\documentclass[aps,prb,floatfix,twocolumn,preprintnumbers,amsmath,amssymb,groupedaddress,showpacs,showkeys]{revtex4-2}

\usepackage{graphicx} 
\usepackage{dcolumn} 
\usepackage{bm} 
\usepackage{color} 
\usepackage{amsmath,amssymb}
\usepackage{mathrsfs}
\usepackage{upgreek}

\usepackage{mathtools}

\usepackage[colorlinks=true,linkcolor=blue,citecolor=blue,urlcolor=blue]{hyperref}

\newcommand{\ua}{\uparrow}
\newcommand{\da}{\downarrow}

\newcommand{\s}{\sigma}

\let\Re\relax
\let\Im\relax
\DeclareMathOperator{\Re}{Re}
\DeclareMathOperator{\Im}{Im}

\usepackage{xr}
\makeatletter
\newcommand*{\addFileDependency}[1]{ 
	\typeout{(#1)}
	\@addtofilelist{#1}
	\IfFileExists{#1}{}{\typeout{No file #1.}}
}
\makeatother

\newcommand*{\myexternaldocument}[2]{%
	\externaldocument{#1/#2}%
	\addFileDependency{#2.tex}%
	\addFileDependency{#1/#2.aux}%
}

\myexternaldocument{aux_files}{sm_revised_v2}

\begin{document}

\title{Spin relaxation, Josephson effect and Yu-Shiba-Rusinov states in superconducting bilayer graphene}

\author{Michael Barth}%
 	\affiliation{Institute for Theoretical Physics, University of Regensburg, 93040 Regensburg, Germany}
 	
\author{Jacob Fuchs}%
 	\affiliation{Institute for Theoretical Physics, University of Regensburg, 93040 Regensburg, Germany}

 	

 	
\author{Denis Kochan}%
\email[Corresponding author: ]{denis.kochan@ur.de}
 	\affiliation{Institute for Theoretical Physics, University of Regensburg, 93040 Regensburg, Germany\\
 	Institute of Physics, Slovak Academy of Sciences, 84511 Bratislava, Slovakia}



\begin{abstract}
    Bilayer graphene has two non-equivalent sublattices and, therefore, the same adatom impurity can manifest in spectrally distinct ways---sharp versus broad resonances near the charge neutrality---depending on the sublattice 
    it adsorbs at. Employing Green's function analytical methods and the numerical \textsc{Kwant} package we investigate the spectral and transport interplay between the resonances and superconducting coherence induced 
    in bilayer graphene by proximity to an s-wave superconductor. Analyzing doping and temperature dependencies of quasi-particle spin-relaxation rates, energies of Yu-Shiba-Rusinov states, Andreev spectra and the supercurrent characteristics of Josephson junctions we find unique superconducting signatures discriminating between resonant and off-resonant regimes. Our findings are in certain aspects going beyond the superconducting bilayer graphene and hold for generic s-wave superconductors functionalized by the resonant magnetic impurities.
\end{abstract}

\date{\today}
\maketitle

\section{Introduction}

Microscopic understanding of spin relaxation is a necessary prerequisite for a proper engineering and functionalization of spintronics devices 
\cite{Zutic2004}. 
Promising candidates for such applications are graphene-based systems \cite{HanKawakamiGmitraFabian_2014,Roche:SpintronicsPerspective2015,Avsar-Colloquium:RMP2020} 
as they offer charge carriers with high mobility, tunable spin-orbit coupling (SOC) and even magnetic-exchange interaction \cite{Zutic:MaterialsToday2018}.
By graphene-based systems we mean graphene and bilayer graphene (BLG) proximitized by layered van der Waals materials, 
such as transition metal dichalcogenides (TMDC) \cite{Podzorov_2004,Shi2015_IonicGatedTMDC-SC,Jo2015_WS2_SC,Navarro-Moratalla2016_TaS2-SC,Costanzo2016_MoS2-SC} or magnetic insulators \cite{Swartz2012:ASCNano,Yang2013:PRL,Mendes2015:PRL,Wei2016:NatMat,Dyrdal2017:2DMat,Hallal2017:2DMat} that offer new possibilities \cite{Sierra:NatNono2021} for exploring (magneto-)transport and (opto-)spintronics phenomena. 
The new functionality in this regards---triggered by the discovery of the superconductivity in twisted BLG \cite{Cao_2018,Yankowitz2019} and by promising perspectives in superconducting spintronics \cite{EschrigPhysToday2011,EschrigRepProgPhys2015,LinderRobinson_NP2015,Guang:APL2021}---is the proximity of graphene and BLG with other low-dimensional superconducting materials. 
Indeed, the proximity induced superconductivity has been experimentally demonstrated in lateral graphene-based Josephson junctions \cite{Heersche2007a, Komatsu2012a,Calado2015,Delagrange_PRL_2018}, alkaline-intercalated graphite~\cite{Li2013,Ludbrook2015,Chapman2016} and also vertical stacks with the interfacial geometries~\cite{Tonnoir2013,DiBernardo2017_p-wave-SCG}. 

Here we 
focus on Bernal stacked BLG in a proximity of an s-wave superconductor whose quasi-particle spin properties can be altered by impurities depending on the sublattice 
they hybridize with. Particularly, we look at light adatoms---like hydrogen, fluorine or copper---and the local magnetic exchange or local SOC interactions that are induced by them. 
Quite generally \cite{Schrieffer-book:1964,Yafet1983,Kochan:PRL_sc_graphene_spin_relaxation}, spin relaxation 
in the s-wave superconductors manifests differently depending on whether the spin-flip scattering is due to SOC (even w.r.t.~time reversal) or magnetic exchange (odd w.r.t.~time reversal). Superconducting coherence enforces composition of the quasi-particle scattering amplitudes in a way that they subtract in the first, and 
sum in the second case what, correspondingly, decreases \cite{Yang_Parkin2010,Hubler:PRL2012,Quay_Aprili_Strunk_2015} or increases \cite{Hebel1957,Poli2008} superconducting spin 
relaxation as compared to the normal phase. 
The enhanced superconducting spin relaxation in the presence of magnetic impurities is known as the Hebel-Slichter-effect \cite{Hebel1957,Hebel1959,Hebel1959-Theory}, and 
the temperature dependence of the ratio of the superconducting rate versus its normal-phase counterpart as the Hebel-Slichter peak. 
The absence of the latter ``serves'' often as a probe of unconventional pairing, however, as scrutinized in \cite{Cavanagh_2021} this can be a red herring. Another reason for the breakdown of the Hebel-Slichter effect are resonances caused by a multiple scattering off the underlying Yu-Shiba-Rusinov (YSR) states \cite{Yu1965,Shiba1968,Rusinov1968,Wehling2008,Lado2016,Brihuega:AdvMaterials2021}---as was shown in detail for the superconducting single layer graphene \cite{Kochan:PRL_sc_graphene_spin_relaxation}. What happens in BLG 
and how different sublattice degrees of freedom enter the game is a subject of the present study.

The main goals of our paper are spin, sublattice and spectral properties of superconducting BLG in the presence of light adatoms that act as magnetic or spin-orbit-coupling resonant scatterers \cite{Wehling2010_ResonantScattering,IrmerPRB2018TopBridgeHollow,PogorelovLoktevKochan:PRB2020}. 
Particularly, 
1) we compare temperature and doping dependencies of spin relaxation rates depending on which sublattice an adatom is hybridizing with, 
2) analyze the subgap spectral properties in terms of the induced YSR states, and 
3) explore critical currents and Andreev bound states (ABS) in the BLG-based superconducting Josephson junctions. 
Though, some of our findings are general---e.g.,~the disappearance of the Hebel-Slichter peak when tuning the chemical potential into resonances---and go beyond BLG specifics. 

The paper is organized as follows; in Sec.~\ref{Sec:Model} we shortly introduce the model Hamiltonian describing BLG and impurities.
The necessary analytical equipment---Green's function, T-matrix and generalities about the YSR spectra---are presented in Sec.~\ref{Sec:GreenFunctions}.
Results and other outcomes from the numerical simulations are extensively summarized and qualitatively discussed in Sec.~\ref{Sec:Result}. 
More technical and \textsc{Kwant} implementation aspects are left for the Supplemental Material  \cite{SM} (see, also, Refs.~\cite{Kwant,Kochan:PRL_sc_graphene_spin_relaxation,Bundesmann_SR_SOC,KatochPRL2018,Denis_MAG_ham,Mashkoori2017ImpurityBS,Octave,Andreev_1966,Kulik_1977,Sauls_2018,JOSEPHSON1962,JOSEPHSON1974,Titov:PhysRevB2006,Munoz:PhysRevB2012,Alidoust:PhysRevB2019,Alidoust:PhysRevResearch2020,ABS_spectral_density_Muralidharan,FURUSAKI1994214,Ostroukh_paper,Akhmerov_supercurrent,McClure:PR1957,Slonczewski:PR1958,KonschuhPRB:2012,McCann_2013,Beenakker_1991,Akhmerov_ABS_smatrix} therein).

\section{Model Hamiltonian}\label{Sec:Model}

\begin{figure}
  \centering
  \includegraphics[width=\columnwidth]{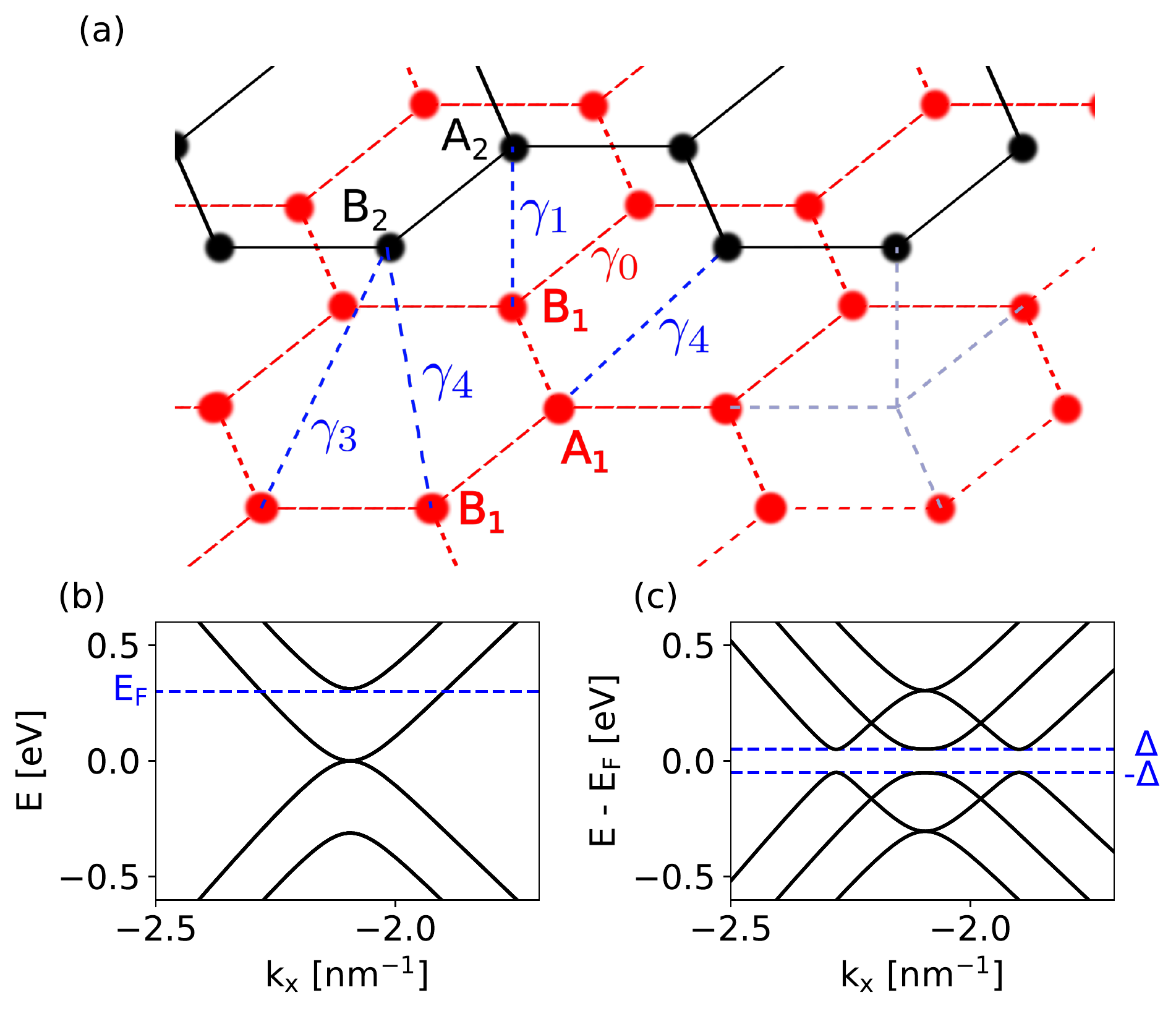}
  \caption{(a) Schematic illustration of the Bernal-stacked BLG with relevant intra- and interlayer orbital hoppings. Carbons $A1$ and $B2$ which are not coupled via the interlayer coupling $\gamma_1$ are conventionally called the low-energy or non-dimer sites, and their counterparts $B1$ and $A2$ 
coupled by $\gamma_1$ as the high-energy or dimer sites; we adopt that terminology in what follows.
  (b) Band structure of BLG around the $K$-point. (c) Quasi-particle band structure of the superconducting BLG at chemical potential $E_F$ shown in panel (b), for the visibility we employed an exaggerated value of 
  $\Delta = 50~\mathrm{meV}$.}
  \label{fig:BLG_lattice}
\end{figure}

We consider superconducting Bernal stacked BLG functionalized with light adatoms that hybridize with carbon $p_z$-orbitals in the
top layer. Such a system is described by the Hamiltonian
\begin{equation}\label{Eq:Hamiltonian}
     H = H_\mathrm{BLG} + H_\mathrm{ada},
\end{equation}
where $H_\mathrm{BLG}$ describes superconducting BLG host, and the An\-der\-son-type Hamiltonian $H_\mathrm{ada}$ takes into account local interactions promoted by the adatom. 
To describe BLG we use the minimal tight-binding Hamiltonian: 
\begin{align}\label{Eq:HamBLG}
    H_\mathrm{BLG}&=
    \sum_{m, n, X, \sigma} (-\gamma_0 \delta_{\langle mn\rangle}-\mu\delta_{mn})
    c^\dagger_{X,m,\sigma} c^{\phantom{\dagger}}_{X,n,\sigma} \nonumber \\ 
         & + \gamma_1 \sum_{m, \sigma} \bigl(c^\dagger_{B1,m,\sigma} c^{\phantom{\dagger}}_{A2,m,\sigma} +c^\dagger_{A2,m,\sigma} c^{\phantom{\dagger}}_{B1,m,\sigma}\bigr)\\
         &+\Delta\sum_{m, X} \bigl(c^\dagger_{X,m,\uparrow} c^\dagger_{X,m,\downarrow}+c^{\phantom{\dagger}}_{X,m,\downarrow} c^{\phantom{\dagger}}_{X,m,\uparrow}\bigr),\nonumber
\end{align}
where $c_{X,m,\sigma}$ and $c_{X,m,\sigma}^\dagger$ are the annihilation and creation operators for an electron with spin $\sigma$, located at lattice site $m$. In order to keep a track 
on the sublattice and layer degrees of freedom we use along $m$ an additional label $X=\{A1,B1,A2,B2\}$, reserving letter $A$~$(B)$ for the A~(B)-sublattice, and number $1$~$(2)$ for the bottom~(top) layer, respectively, see Fig.~\ref{fig:BLG_lattice}. 

The parameters in Eq.~(\ref{Eq:HamBLG}) have the standard meaning; 
$\gamma_0=2.6\,\mathrm{eV}$ describes the intralayer nearest neighbour hopping (mimicked by the symbol $\delta_{\langle m n \rangle}$) along the carbon-carbon bond 
possessing length $\mathrm{a}_{cc} = 0.142\,\mathrm{nm}$,
$\gamma_1 = 0.34\,\mathrm{eV}$ is an interlayer hopping \cite{KonschuhPRB:2012} between the top and bottom carbons separated by a distance $\mathrm{c} = 0.335\,\mathrm{nm}$,
$\mu$ denotes the chemical potential of the system (with zero taken at charge neutrality point of the non-superconducting BLG), 
and, finally, $\Delta$ is the global superconducting s-wave pairing induced by a proximity of BLG with a superconductor. 
In order to capture temperature effects we assume that $\Delta$ follows the conventional BCS dependence well-interpolated by the standard formula:
\begin{equation}\label{Eq:GapDep}
    \Delta(T)=\Delta_0\,\tanh{[1.74 \sqrt{T_{\rm c}/T - 1}]}\,\Theta(T_c-T).
\end{equation}
For concreteness we choose $\Delta_0 = 1\,\mathrm{meV}$, giving us the critical temperature $T_c = \Delta_0/(1.76 \cdot k_{\mathrm{B}}) = 6.953\,\mathrm{K}$~\cite{tinkham2004introduction}. This slightly elevated value of $\Delta_0$ is a compromise between realistic superconducting proximity in layered carbon systems \cite{PhysRevB.98.121411, PhysRevB.101.195405,LeeLee2018}, and a numerical capability to handle transport and spectral calculations \footnote{However, in special cases that involved numerical diagonalization we use even larger  
$\Delta_0 = 50\,\mathrm{meV}$ just to reach convergence and cross-check analytical results.}. 
The system is illustrated in Fig.~\ref{fig:BLG_lattice}, along with its normal and superconducting quasi-particle band structures. 
Two remarks are in order, first, the general BLG Hamiltonian in McClure-Slonczewski-Weiss parameterization~\cite{McClure:PR1957,Slonczewski:PR1958,KonschuhPRB:2012,McCann_2013} involves also additional interlayer orbital hoppings $\gamma_3$ and $\gamma_4$, see Fig.~\ref{fig:BLG_lattice}. We neglect them in what 
follows, although, we checked that they do not bring new qualitative features---for a quantitative comparison of the simple and full models see the Supplemental Material \cite{SM}.
Second, intrinsic SOC of BLG \cite{KonschuhPRB:2012} is two orders of magnitude smaller than a typical local SOC induced 
by adatoms, see \cite{Gmitra2013_SOC-in-H-Graphene,Irmer2015-SOC-F-Graphene,Zollner2016-SOC-Methyl,FrankPRB2017Copper} and Appendix \ref{app:SOC_ham_and_params}, therefore we also neglect in $H_\mathrm{BLG}$ all the intrinsic SOC contributions of the BLG host.

The adatom Hamiltonian $H_\mathrm{ada}$ comprise orbital and spin interactions \cite{Denis_SOC_Ham,Denis_MAG_ham}, i.e.,
\begin{equation}
    H_\mathrm{ada}=V_{o}+V_{s}.
\end{equation}
Assuming the adatom hosts a single electronic orbital governed by the annihilation and creation operators $d$ and $d^\dagger$ the Hamiltonian $V_{o}$ explicitly reads \cite{Gmitra2013_SOC-in-H-Graphene,Irmer2015-SOC-F-Graphene,Zollner2016-SOC-Methyl,FrankPRB2017Copper}: 
\begin{align}
    V_o&=\sum_{\sigma} [(\varepsilon-\mu) d^\dagger_\sigma d^{\phantom{\dagger}}_\sigma +
    \omega (d^\dagger_\sigma c^{\phantom{\dagger}}_{\star\sigma}+ 
    c^{\dagger}_{\star\sigma}d^{\phantom{\dagger}}_\sigma)]\nonumber \\
    &+
    \Delta_{d} (d^\dagger_\uparrow d^\dagger_\downarrow +d^{\phantom{\dagger}}_\downarrow d^{\phantom{\dagger}}_\uparrow), \label{Eq:Vo}
\end{align}
where $c_\star$ and $c_\star^\dagger$ act on the functionalized---dimer or non-dimer---carbon site in the top layer. 
The Anderson-like Hamiltonian $V_o$ is parameterized by the adatom onsite energy $\varepsilon$,
the adatom-carbon hybridization $\omega$, and the adatom-located superconducting pairing $\Delta_d$. Its magnitude is not so crucial for the results presented below and, therefore, for the sake of simplicity we set $\Delta_d$ to the corresponding BLG value $\Delta$, see Eq.~(\ref{Eq:GapDep}). 

For the spin interaction $V_s$ we consider two separate cases: (1) \textit{magnetic exchange} of the adatom $d$-states with a non-itinerant, spin $\tfrac{1}{2}$, magnetic moment $\mathbf{S}$ that effectively develops on the adatom (through the Hubbard interaction, for details see \cite{hewson_1993}) in terms of remaining degrees of freedom dynamically decoupled from $d$-levels, i.e. 
\begin{align}
    V_s^{(1)}=&-J\,\mathbf{s}\cdot\mathbf{S},\label{Eq:Vs}
\end{align}
and (2) \textit{local SOC} Hamiltonian $V_s^{(2)}$, whose explicit, but lengthy expression is provided in Appendix \ref{app:SOC_ham_and_params}.
In the expression for $V_s^{(1)}$, the $\mu$-th component of the itinerant spin operator $\mathbf{s}$ reads, $\mathbf{s}^{\mu}=d^\dagger_{a}\,(\boldsymbol{\sigma}^{\mu})^{\phantom{\dagger}}_{ab}\,d^{\phantom{\dagger}}_b$, 
where $\boldsymbol{\sigma}^\mu$ is $\mu$-th spin Pauli matrix, and $a$ and $b$ run over $\uparrow$ and $\downarrow$ spin-projections of $d$-states.
Spin degrees of freedom of the non-itinerant spin, $\Uparrow$ and $\Downarrow$, are introduced such that $\mathbf{S}$-operator is given as a vector of the Pauli matrices acting 
on these spins. Evaluating the final spin-relaxation rates we trace out $\mathbf{S}$ degrees of freedom, calculation with all details is presented in Ref.~\cite{KochanD2014}.


\section{General considerations: Green's functions and YSR energies}\label{Sec:GreenFunctions}

\subsection{Green's functions}

The starting point for the analytical considerations is the (retarded) Green's resolvent 
\begin{equation}
    \mathbb{G}(z)=(z-H)^{-1},
\end{equation} 
where $H$ denotes the full Hamiltonian of the system, e.g.,~ Eq.~(\ref{Eq:Hamiltonian}), and 
$z=E+i\eta$ the complex energy (with a positive infinitesimal imaginary part) measured with respect to the Fermi level $\mu$.
In what follows we show how to obtain superconducting $\mathbb{G}(z)$ in terms of the normal-phase (and hence simpler) Green's function elements, and how to calculate the corresponding spin-relaxation rates and YSR spectra. To be concrete we stick to the case of BLG with adatoms, but the procedure is in fact general assuming one can split the given 
Hamiltonian $H$ into an unperturbed part (provisionally called $H_\mathrm{BLG}$) and a spatially local but not necessarily a point-like perturbation (in our case $H_\mathrm{ada}$).  

First, defining the Green's resolvent of the unperturbed superconducting system, 
\begin{equation}
    G(z)=(z-H_\mathrm{BLG})^{-1},
\end{equation}
we can express $\mathbb{G}(z)$ in terms of $G(z)$ by means of the Dyson equation, i.e.,
\begin{align}
    \mathbb{G}(z)&=G(z)+G(z)\,H_\mathrm{ada}\,\mathbb{G}(z) \label{Eq:Dyson1}\\
        &=G(z)+G(z)\,\mathbb{T}_\mathrm{ada}(z)\,G(z){\phantom{\bigr)^{-1}}} \label{Eq:Dyson3}\\
        &=\bigl(1-G(z)\,H_\mathrm{ada}\bigr)^{-1}\,G(z)\,.  \label{Eq:Dyson2}
\end{align}
The advantage of the latest expression manifests in the local atomic (tight-binding) basis at which $H_\mathrm{ada}$ becomes 
a matrix with few non-zero rows and columns, and hence its inversion is not a tremendous task. 
In the second equation, Eq.~(\ref{Eq:Dyson3}), we have defined the T-matrix 
\begin{equation}\label{Eq:T-matrix}
   \mathbb{T}_\mathrm{ada}(z)=H_\mathrm{ada}\,\bigl(1-G(z)\,H_\mathrm{ada}\bigr)^{-1}.
\end{equation}

The T-matrix is useful from several points of view. First, inspecting its energy poles within the superconducting gap gives the YSR bound state spectra~\cite{Balatsky:RevModPhys2006}. Second, knowing the T-matrix one can directly access the spin-relaxation~rate~$1/\tau_s$ at a given chemical potential $\mu$, temperature $T$, and the adatom concentration (per number of carbons) $\eta_\mathrm{ada}$, by evaluating the following expression~\cite{Yafet1983,Kochan:PRL_sc_graphene_spin_relaxation}:
\begin{equation}\label{Eq:Yafet}
     \frac{1}{\tau_s}=
      \dfrac{\iint\limits_{\rm{BZ}} \mathrm{d}\mathbf{k}\,\mathrm{d}\mathbf{q}\,\left|\langle\mathbf{k},\ua\hspace{-1mm}|\mathbb{T}|\mathbf{q},\da\rangle\right|^2
      \delta(E_{\mathbf{k}}-E_{\mathbf{q}})\left(-\tfrac{\partial g}{\partial E_\mathbf{k}}\right)}
      {\frac{\hbar\pi}{A_{uc}\eta_\mathrm{ada}}\,\int\limits_{\rm{BZ}} \mathrm{d}\mathbf{k} \left(-\tfrac{\partial g}{\partial E_\mathbf{k}}\right)}.
    \end{equation}
Therein, the integrations are taken over the 1st~Brillouin~zone~(BZ) of BLG; $g(E,T)=[\mathrm{e}^{E/(k_{\mathrm{B}}T)}+1]^{-1}$ is the Fermi-Dirac~distribution whose derivative gives thermal smearing, $A_{uc}$ is the area of the BLG unit~cell, and $E_\mathbf{k}$ and $|\mathbf{k},\sigma\rangle$ are, correspondingly, the quasi-particle eigenenergies and eigenstates (normalized to the BLG unit cell) of $H_\mathrm{BLG}$.

To know the T-matrix, Eq.~(\ref{Eq:T-matrix}), we need the unperturbed Green's resolvent $G(z)$ of the superconducting host. The next step is the evaluation of 
$G(z)$ in terms of $g(z)$---the retarded Green's resolvent of BLG in the normal-phase. To this end we express $H_\mathrm{BLG}$, Eq.~(\ref{Eq:HamBLG}), in the Bogoliubov-de~Gennes form 
(in a basis at which the superconducting pairing becomes a diagonal matrix) 
\begin{equation}\label{Eq:BdG-BLG}
   H_\mathrm{BdG} =
        \left(\begin{array}{cc}
            h_\mathrm{BLG}       &          \Delta \\ 
            \Delta^*  &   -h_\mathrm{BLG}^*
    \end{array}\right)\,,
\end{equation}
where $h_\mathrm{BLG}=h_\mathrm{BLG}^*=H_\mathrm{BLG}(\Delta=0)$ comprises the non-superconducting part of Eq.~(\ref{Eq:HamBLG}), i.e.,~an ordinary BLG Hamiltonian held at chemical potential $\mu$. Because of the spatial homogeneity of the s-wave pairing $\Delta$ (constant diagonal matrix) the direct inversion of $z-H_\mathrm{BdG}$ gives
\begin{equation}\label{Eq:GF}
   G(z) =
        \left(\begin{array}{cc}
            g_{+}(Z)+\frac{z}{Z}g_{-}(Z)      &          -\frac{\Delta}{Z}g_{-}(Z) \\ 
            -\frac{\Delta^*}{Z}g_{-}(Z)  &   -g_{+}(Z)+\frac{z}{Z}g_{-}(Z)
    \end{array}\right),
\end{equation}
where 
\begin{align}
    Z &= \sqrt{z^2-|\Delta|^2},\\
    g_{\pm}(Z) &= \tfrac{1}{2}\bigl(Z-h_\mathrm{BLG}\bigr)^{-1} \pm \tfrac{1}{2}\bigl(-Z-h_\mathrm{BLG}\bigr)^{-1}. \label{Eq:smallGF}
\end{align}
The proper branch of the complex square root should be chosen in such a way that $\mathrm{Im}\,Z>0$. 
So we see that the whole Green's function calculation effectively boils down to an ordinary retarded 
Green's resolvent of the non-superconducting BLG Hamiltonian $h_\mathrm{BLG}$, i.e.,~to $g(\pm Z)=(\pm Z-h_\mathrm{BLG})^{-1}$.

The above equation, Eq.~(\ref{Eq:GF}), is an operator identity in the Bogoliubov-de~Gennes form expressed 
in a basis in which the pairing component $\Delta$ becomes a diagonal matrix.
For the later purposes we would need the matrix elements of $g(z)$ in the local atomic (Wannier) basis, particularly, 
one matrix element involving the $p_z$ orbital $|\Psi_\star\rangle$ located on carbon site $C_\star$ that hosts the adatom impurity. 
Such on-site Green's function element---also known as the locator Green's function---reads
\begin{equation}\label{Eq:LocatorGeneral}
    g_{C_\star}(z)=\langle\Psi_\star| g(z)| \Psi_\star\rangle=\int \mathrm{d}\epsilon\,\frac{\varrho_\star(\epsilon+\mu)}{z-\epsilon},
\end{equation}
where $\varrho_\star$ is the normal-phase DOS of the unperturbed system projected on the atomic site $C_\star$ and the integration runs over the corresponding quasi-particle bandwidth. The projected DOS, 
$\varrho_\star(z)=\sum_{\mathbf{k}}\delta(z-\epsilon_\mathbf{k})|\langle\Psi_\star|\mathbf{k}\rangle|^2$, can be routinely computed from the known eigenvalues, $\epsilon_{\mathbf{k}}$, and eigenvectors, $|\mathbf{k}\rangle$, of $h_\mathrm{BLG}$.

Up to now the discussion was general without any explicit reference to superconducting or normal-phase BLG Hamiltonians $H_\mathrm{BLG}$ and $h_\mathrm{BLG}=H_\mathrm{BLG}(\Delta=0)$, see Eq.~(\ref{Eq:HamBLG}). In what follows we express $g_{C_\star}(z)$ for BLG assuming normal-phase Hamiltonian $h_\mathrm{BLG}$ with only $\gamma_0$ and $\gamma_1$ hoppings. In this case the integral in Eq.~(\ref{Eq:LocatorGeneral}) can be computed analytically, see Ref.~\cite{Denis_MAG_ham}. 
The resulting $g_{C_\star}(z)$ for the dimer and non-dimer $C_\star$ sites are as follows: 
\begin{align}
  g_{C_\star}^{\mathrm{d}}(z)  &=z \left[F(z^2+\gamma_1 z) + F(z^2-\gamma_1 z)\right]\label{Eq:onsite-GF-11},\\
  g_{C_\star}^{\mathrm{nd}}(z) &=g_{C_\star}^{\mathrm{d}}(z) + \gamma_1\bigl[F(z^2+\gamma_1 z)-F(z^2-\gamma_1 z)\bigr]\label{Eq:onsite-GF-22},
\end{align}
where
\begin{align}
  \label{Eq:onsite-integral}
  F(\zeta) &= \frac{A_{uc}}{4\pi\nu_0^2}\left[I\left(\sqrt{\zeta}/{\nu_0}\right)
  +
  I\left(-\sqrt{\zeta}/{\nu_0}\right)\right],\\
  I(p) &= \frac{1}{2}\ln{\left|\frac{\Re^2(p)+\Im^2(p)}{(\Lambda-\Re(p))^2+\Im^2(p)}\right|}\nonumber\\
  &-i\arctan{\frac{\Re(p)}{\Im(p)}}-i\arctan{\frac{\Lambda-\Re(p)}{\Im(p)}}.
\end{align}
In the above expressions $A_{uc}=3 \sqrt{3} \mathrm{a}_{cc}^2 / 2$ is the area of the BLG unit cell, $\nu_0 = 3\mathrm{a}_{cc} \gamma_0/ 2$ and the momentum cut-off
$\Lambda = 2 \sqrt{\sqrt{3} \pi} / (3 \mathrm{a}_{cc})$. Moreover, to keep track on dimensions of different arguments
entering functions $F$ and $I$, we use rather distinct letters, $z$, $\zeta$ and $p$, which have, correspondingly, units of energy, energy square and momentum.

\begin{figure*}
  \centering
  \includegraphics[width=1.99\columnwidth]{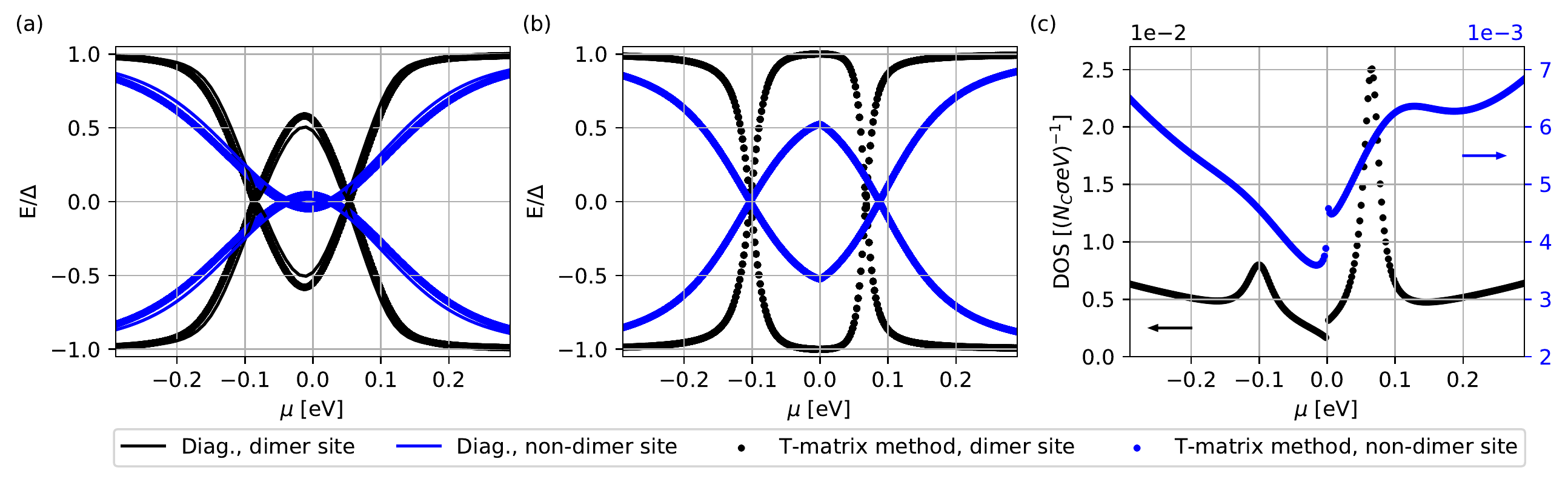}
  \caption{Spectral comparison---superconducting YSR states and normal-phase resonances in BLG: 
  YSRs' energies in units of $\Delta$ vs.~chemical potential $\mu$ for superconducting BLG with the pairing gap 
  $\Delta_0 = 50\,\mathrm{meV}$ [panel~(a)] and $\Delta_0 = 1\,\mathrm{meV}$ [panel~(b)] computed analytically (dots) 
  and numerically (lines) for hydrogen magnetic impurity chemisorbed on the dimer (black) and 
  non-dimer (blue) site; 
  the results of numerical diagonalization displayed in panel (a) were carried on 
  a rectangular flake with the width $W = 601a$ and length $L = 601a$, and hard-wall boundary conditions. 
  Panel (c) shows perturbed DOS in normal BLG (per carbon atom and spin) vs.~chemical potential 
  for concentration $\eta_\mathrm{ada}=0.7\,\%$ of dimer (black, units on left axis) and non-dimer (blue, units on right 
  axis) magnetic impurities. Positions of the resonance peaks in the normal-phase correspond to the minima of YSR states in panels~(a)~and~(b), in accordance with a prediction in Sec.~\ref{Sec:ToyModel}.}
  \label{fig:Comparison_YSRS_energies}
\end{figure*}

\subsection{Yu-Shiba-Rusinov states and resonances---a toy model and its predictions}\label{Sec:ToyModel}

In this section we show under quite general assumptions that resonances caused by magnetic impurities in the non-superconducting systems can trigger---after turning into the superconducting-phase---a formation of YRS states with energies deep inside the superconducting gap. This phenomenon is quite generic and holds for homogeneous s-wave superconductors 
with low concentrations of resonant magnetic impurities---assuming the resonance life-time in the normal-phase is larger 
than the corresponding Larmor precession time, what is the case in single and bilayer graphene. 

It is clear from Eqs.~(\ref{Eq:Dyson1})~and~(\ref{Eq:Dyson2}) and the definition of the Green's resolvent that the eigenenergies of the full Hamiltonian $H$ can be read off from the singularities of $\mathbb{G}(z)=(z-H)^{-1}$ sending $\eta=\mathrm{Im}\,z$ to zero.
We take as a reference some unperturbed superconducting system, e.g.~BLG. 
Let us look at eigenstates of $H=H_\mathrm{BLG}+H^{\prime}_\mathrm{ada}$ that can develop inside the superconducting gap of the unperturbed host due to a coupling with a local perturbation centered on 
a particular atomic site $C_{\star}$:
\begin{equation}\label{Eq:HamPrime}
  H^{\prime}_\mathrm{ada}=\sum\limits_{\sigma} (U+\sigma J) c^\dagger_{\star\sigma}c^{\phantom{\dagger}}_{\star\sigma}. 
\end{equation}
The above Hamiltonian represents a perturbation of the Lifshitz-type \cite{Lifshitz:Book1988} that is parameterized by the on-site energy $U$ and the magnetic interactions $J$ (the term involving chemical potential $\mu$ is in the unperturbed Hamiltonian).
This does not cause a fundamental limitation since in certain regimes the Anderson impurity model given by the adatom Hamiltonian 
\begin{equation}\label{Eq:Ada-magnetic}
    H_\mathrm{ada}=V_o+V_s^{(1)},
\end{equation}
see Eqs.~(\ref{Eq:Vo})~and~(\ref{Eq:Vs}), or even the more general Hubbard impurity model, can be down-folded \cite{hewson_1993} into the form given by Eq.~(\ref{Eq:HamPrime}). 
In what follows we assume that the orbital energy scale dominates over the magnetic one, 
i.e.,~$U^2\gg J^2$.

It is clear from Eq.~(\ref{Eq:Dyson2}) that the in-gap states can be extracted from singularities \cite{Balatsky:RevModPhys2006} of $\bigl(1-G(z)\,H^{\prime}_\mathrm{ada}\bigr)^{-1}$, therefore one needs to inspect energies $|E|<|\Delta|$ at which the ``secular determinant'' of the operator $1-G(z)\,H^{\prime}_\mathrm{ada}$ turns to zero, i.e.,
\begin{equation}\label{Eq:SecDet}
    \det{\bigl[1-(E-H_\mathrm{BLG})^{-1}\,H^{\prime}_\mathrm{ada}\bigr]}=0.
\end{equation}
Since $H^{\prime}_\mathrm{ada}$ is located on the atomic site $C_\star$ we just need the corresponding locator of $G(E)=(E-H_\mathrm{BLG})^{-1}$, i.e.,
\begin{align}\label{Eq:LocatorGF}
    G_{C_\star}(E)&=\langle \Psi_{\star}|G(E)|\Psi_{\star} \rangle\,.
\end{align}
Correspondingly, $\langle \Psi_{\star}|G(E)|\Psi_{\star} \rangle$ is a $2\times 2$ matrix in the reduced particle-hole Nambu space~\footnote{As a comment, while in this toy model we assume no macroscopic spin polarization neither spin-orbit interaction in the unperturbed system we just employ the reduced Nambu formalism, however, one should keep in mind that for any solution with an energy $E$ the full Nambu-space approach will give as a solution also 
the energy $-E$.}. 
Substituting for $G(E)$ in Eq.~(\ref{Eq:LocatorGF}) the corresponding expression from Eq.~(\ref{Eq:GF}), we can rewrite $G_{C_\star}(E)$ in terms of the locators of $g_\pm(E)$, see Eq.~(\ref{Eq:smallGF}), and even further in terms of the locators of the normal-phase resolvents $(\pm Z(E)-h_\mathrm{BLG})^{-1}$, where for the in-gap states $\pm Z(E)=\pm i\sqrt{|\Delta|^2-E^2}$.
The locator that is finally needed to be calculated turns to be the following integral
\begin{equation}\label{Eq:gammaE}
  g_{C_\star}(\pm Z(E))=\int \mathrm{d}x\,\frac{\varrho_{\star}(x)}{\mu\pm i\sqrt{|\Delta|^2-E^2}-x}\equiv \gamma(\pm E)\,,
\end{equation}
see also Eq.~(\ref{Eq:LocatorGeneral}).

Similarly, the perturbation $H^{\prime}_\mathrm{ada}$ turns to be a $2\times 2$ matrix in the particle-hole space with the following Bogoliubov-de~Gennes form:
\begin{align}
    H_{\mathrm{BdG}}^{\prime}&=
    \left(\begin{array}{cc}
         U+J& 0  \\
         0 & -U + J 
    \end{array}\right).
\end{align}
Hence the secular determinant of the operator $1-G(z)\,H^{\prime}_\mathrm{ada}$, Eq.~(\ref{Eq:SecDet}), reduces in the local atomic basis just to a determinant of an ordinary $2\times 2$ matrix. 
So finally, the in-gap energies $|E|<|\Delta|$ of the perturbed problem satisfy the following (integro-algebraic) secular equation: 
\begin{align}
\mathrm{Re}\Bigl\{ \bigl[1 &- (U+J)\,\gamma(E)\bigr] \bigl[1-(U-J)\,\gamma(-E)\bigr] \Bigr\} \nonumber \\
&= J\,\frac{E}{i\sqrt{|\Delta|^2-E^2}}\,\bigl[\gamma(E)-\gamma(-E)\bigr] \label{Eq:SecDet2}.    
\end{align}
Further, using a fact that $\gamma(-E)=\overline{\gamma(E)}$, the left hand side of the above equation can be expressed as 
a sum of two terms: $\bigl\{1- (U+J)\,\mathrm{Re}[\gamma(E)]\bigr\}\bigl\{1- (U-J)\,\mathrm{Re}[\gamma(E)]\bigr\}$
and $(U^2-J^2)\{\mathrm{Im}[\gamma(E)]\}^2$. We will show in a sequel that at resonances the first of them turns 
to zero and, correspondingly, the secular equation, Eq.~(\ref{Eq:SecDet2}), simplifies even more:
\begin{align}\label{Eq:SecDet3}
(U^2-J^2)\,\mathrm{Im}[\gamma(E)] &= 2\frac{E J}{\sqrt{|\Delta|^2-E^2}}.    
\end{align}

Let us recall that the normal-phase \emph{resonant energies} $\mu_\pm$ of the unperturbed host under an action of $H^{\prime}_\mathrm{ada}$ are defined \cite{Lifshitz1956,AndersonPR:1961,Lifshitz1964,Elliott1974,Lifshitz:Book1988} by the following equations:
\begin{equation}
  \lim\limits_{\eta\rightarrow 0}  
  \int \mathrm{d}x\,\frac{(\mu_{\pm}-x) }{(\mu_{\pm}-x)^2+\eta^2}\,\varrho_{\star}(x)=\frac{1}{U\pm J}.
\end{equation}
In practise one relaxes infinitesimality of $\eta$ and uses some fixed value smaller than the corresponding 
resonance width \cite{AndersonPR:1961}  
\begin{equation}
    \Gamma_{c\pm}=\pi\,|U^2-J^2|\,\varrho_{\star}(\mu_\pm),
\end{equation}
which is inversely proportional to the lifetime of the resonance ($\tau_{\text{life}}=\hbar/\Gamma_{c}$). This constraint on the magnitude of $\eta$ implies that the resonance energies $\mu_\pm$ are given with an uncertainty of $\Gamma_{c\pm}$.
Assume we have a superconducting system at the chemical potential $\mu$ close to $\mu_{+}$ or $\mu_{-}$ (within a range of $\Gamma_{c\pm}$) that possesses a superconducting gap $\Delta$, such that $\Gamma_{c\pm} \gtrsim |\Delta| > \sqrt{|\Delta|^2-E^2}$. Taking the real part of Eq.~(\ref{Eq:gammaE}) we can write:
\begin{equation}
  \mathrm{Re}[\gamma(E)]=\int \frac{\mathrm{d}x\,(\mu_{\pm}-x)\,\varrho_{\star}(x)}{(\mu_{\pm}-x)^2+(|\Delta|^2-E^2)}
  \simeq\frac{1}{U\pm J}.
\end{equation}
This guaranties that the term $\bigl\{1- (U+J)\mathrm{Re}[\gamma(E)]\bigr\}\bigl\{1- (U-J)\mathrm{Re}[\gamma(E)]\bigr\}\simeq 0$.
Similarly, for the imaginary part of Eq.~(\ref{Eq:gammaE}) we get:
\begin{align}
    \mathrm{Im}[\gamma(E)]&=-\int \frac{\mathrm{d}x\,\sqrt{|\Delta|^2-E^2}\,\varrho_{\star}(x)}{(\mu_{\pm}-x)^2+(|\Delta|^2-E^2)}\simeq -\pi\varrho_{\star}(\mu_\pm),
\end{align}
where the last equality holds for the unperturbed system with a relatively wide bandwidth and properly varying density of states $\varrho_\star$ on the scale larger than $\Gamma_{c\pm}$. 
Within these assumptions the expression for the secular determinant, Eq.~(\ref{Eq:SecDet3}), finally reads~\footnote{\emph{Private correspondence:} a very similar 
formula (unpublished) was obtained using a different perspective by Dr.~Tom\'{a}\v{s} Novotn\'{y}.}:
\begin{equation}\label{Eq:YSR-new_formula}
    \frac{|E|}{\sqrt{|\Delta|^2-E^2}}\simeq\frac{\Gamma_{c\pm}}{2|J|}=\frac{1}{2}\frac{\hbar/|J|}{\hbar/\Gamma_{c\pm}}=
    \frac{1}{2}\frac{\tau_{\text{Larmor}}}{\tau_{\text{life}}}.
\end{equation}

The above formula gives the energies of YSR states for a superconducting system whose Fermi level $\mu$ is tuned to the vicinity of the normal-phase resonance, i.e. $\mu\simeq\mu_{\pm}$ withing a range of $\Gamma_{c\pm}$. 
Knowing the resonance width $\Gamma_{c\pm}$ and the strength of magnetic exchange $J$, or equivalently, the lifetime 
$\tau_{\text{life}}$ of the normal-phase resonance and the Larmor precession time, $\tau_{\text{Larmor}}=\hbar/|J|$, 
due to magnetic exchange $J$ one can easily get the corresponding YSR energies:
\begin{equation}\label{Eq:YSR-new_formula_1}
    E_{\text{YSR}}
    =\pm\frac{|\Delta|}{\sqrt{1+4 J^2/\Gamma_c^2}}
    =\pm\frac{|\Delta|}{\sqrt{1+4\tau^2_{\text{life}}/\tau^2_{\text{Larmor}}}}.
\end{equation}

Scrutinizing Eqs.~(\ref{Eq:YSR-new_formula})~and~(\ref{Eq:YSR-new_formula_1}) further, we see that whenever the Larmor precession time is substantially smaller than the resonance lifetime the corresponding YSR energies would be very close to the center of the superconducting gap, i.e.,~$|E|\simeq 0$. 
Moreover, having two atomic sites---say dimer and non-dimer in the case of BLG---out of which the first gives rise to a narrower resonance than the second, then for the same magnetic $J$ the corresponding YSR energies would be deeper inside the gap for the first site than for the second. 
Our findings are pointing along the similar lines as those of the recent study \cite{Uldemolins_2021} that investigated formation and coupling of the YSR states to a substrate when tuning the Fermi level into the Van Hove singularity.
 

Based on the above considerations one can already predict what to expect for the quasi-particles' spin relaxation. Quasi-particles occupy energies above the superconducting gap, while the YSR states carrying magnetic moments are inside the gap. The larger is the energy separation between the two groups, the more ``invisible'' these states become for each other. Consequently, we expect substantially weakened quasi-particle spin relaxation at chemical potentials that yield YSR states deep inside the superconducting gap. 
The effect should be more visible when lowering the temperature since there $\Delta$ grows with a lowered $T$ 
according to Eq.~(\ref{Eq:GapDep}). Of course, at too low temperatures the spin relaxation quenches naturally because of the absence of free quasi-particle states which rather pair and enter the BCS condensate. 
\begin{figure*}[hbt]
  \centering
  \includegraphics[width=1.9\columnwidth]{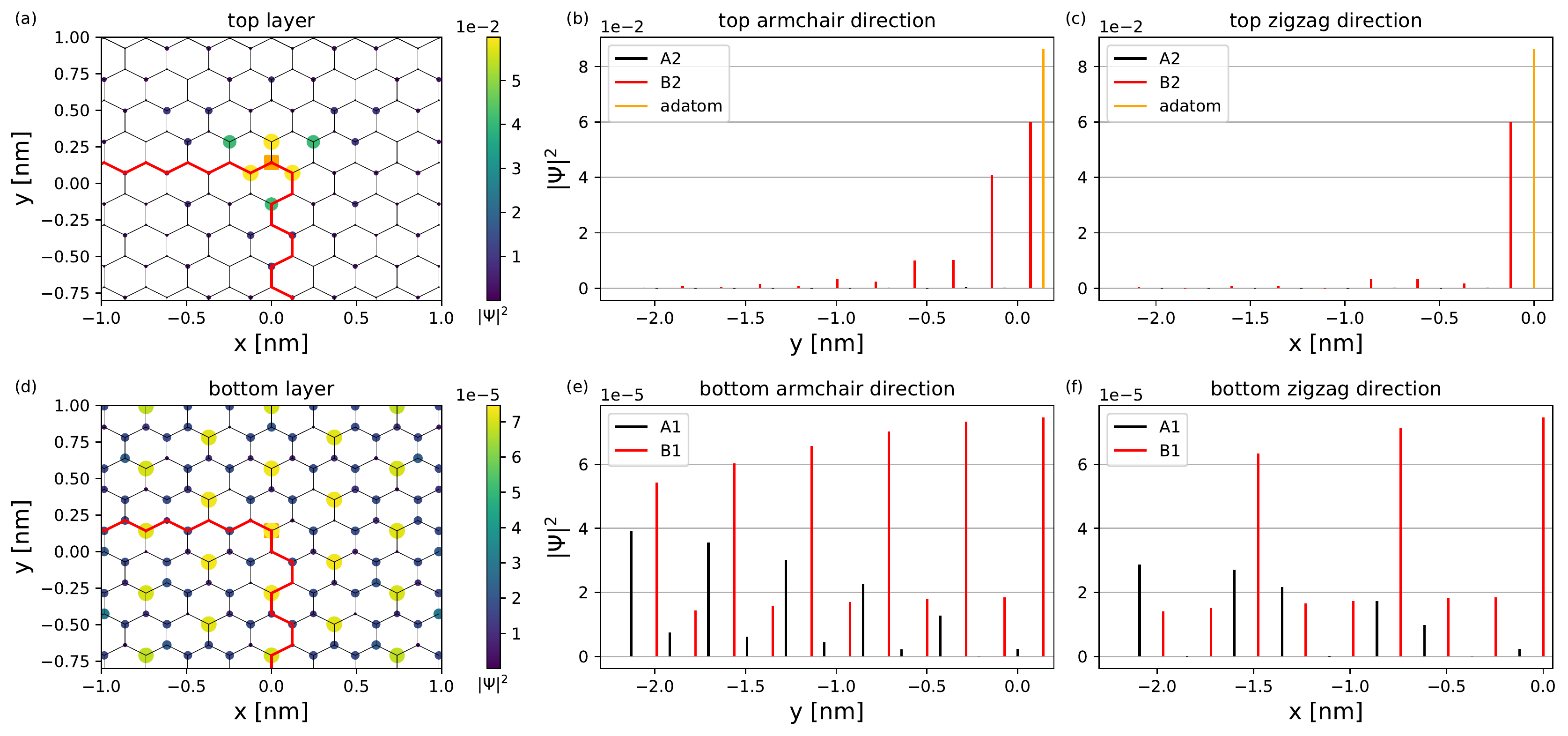}
  \caption{Sub-lattice resolved probabilities of YSR states in top~(a) and bottom~(d) layers originating from a magnetic adatom (hydrogen) chemisorbed at a dimer carbon in the top layer at chemical potential $\mu=-0.1\,\mathrm{eV}$; for the corresponding energy spectrum see Fig.~\ref{fig:Comparison_YSRS_energies}. 
  Side panels show sublattice-resolved probabilities in the corresponding layers for two representative directions, armchair [panels (b)~and~(e)] and zig-zag [panels (c)~and~(f)]. The numerical calculations---exact diagonalization---employed: $\Delta_0 = 50\,\mathrm{meV}$, $W = 601a$ and $L = 601a$.}
  \label{fig:YSRS_probdens_dimer}
\end{figure*}

\section{Results}\label{Sec:Result}

We implemented Hamiltonian $H_\mathrm{BLG}+H_\mathrm{ada}$, Eq.~(\ref{Eq:Hamiltonian}), for the hydrogen functionalized superconducting BLG in \textsc{Kwant}, and calculated its various transport, relaxation and spectral properties. 
The very detailed numerical implementation scheme is provided in the Supplemental Material \cite{SM} for readers willing to adopt it to 
other materials or further spintronics applications. 
We have chosen hydrogen, since it is the most probable and natural atomic contaminant coming from organic solvents 
used in a sample-fabrication process, and also, because it acts as a resonant magnetic scatterer \cite{Kochan2014_PRL-SR-Graphene, Denis_MAG_ham}.
Of course, methodology as developed can be used for any adatom species that are well described by Hamiltonian 
$H_\mathrm{ada}$.

Discussing the results, we start from spectral and spatial properties of YSR states, continue with spin relaxation, and end up with the Andreev spectra and critical currents of the BLG-based Josephson junctions. Moreover, we assume dilute adatom concentrations that do not affect the magnitude of the proximity-induced superconducting gap $|\Delta|$, neither giving 
it pronounced local spatial variations on the length scale shorter than the coherence length. To make fully self-consistent approach is beyond the scope of the present paper.

\subsection{Yu-Shiba-Rusinov states}\label{Sec:ResultYSR}

Figure~\ref{fig:Comparison_YSRS_energies}~(a) compares the YSR spectra for hydrogenated superconducting BLG 
versus chemical potential computed analytically---solutions of Eq.~(\ref{Eq:SecDet}) for the adatom Hamiltonian $H_\mathrm{ada}$, Eq.~(\ref{Eq:Ada-magnetic})---and by direct numerical diagonalization. 
The obtained spectra by both methods match quantitatively very well for $\Delta_0=50$\,meV up to a very tiny 
offset stemming from finite size effects and fixed convergence tolerance of $10^{-5}$ in the numerical 
diagonalization procedure.  
The main features of the YSR spectra for the gap of $50$\,meV are clearly visible in Fig.~\ref{fig:Comparison_YSRS_energies}~(a), and are also reproduced for a smaller gap of $1$\,meV displayed in Fig.~\ref{fig:Comparison_YSRS_energies}~(b).
The magnetic impurity on the dimer site exhibits two, well-separated, doping regions---around $\mu = -0.1\,\mathrm{eV}$ and $\mu = 0.08\,\mathrm{eV}$---hosting YSR states with energies deep inside the gap, while the non-dimer site supports the low energy YSR states over a much broader doping region. 
As derived in Sec.~\ref{Sec:ToyModel}, deep lying YSR states should form in resonances, therefore in Fig.~\ref{fig:Comparison_YSRS_energies}~(c) we show the analytical DOS for BLG in the normal-phase perturbed by $0.7\,\%$ of resonant magnetic impurities---resonance peaks in the DOS match perfectly 
with ``(almost) zero energy'' YSR states.  

Seeing the YSRs' energies and DOS induced by adatoms at dimer and non-dimer sites we can expect certain spectral differences in the corresponding spin relaxations. As mentioned above, the extended quasi-particle states occupy energies over the superconducting gap, while the localized YSR states are inside the 
gap. The larger is their energy separation the more ``ineffective'' is their mutual interaction and hence substantially weakened would be scattering and spin relaxation. 

\begin{figure*}[hbt]
	\centering
	\includegraphics[width=1.9\columnwidth]{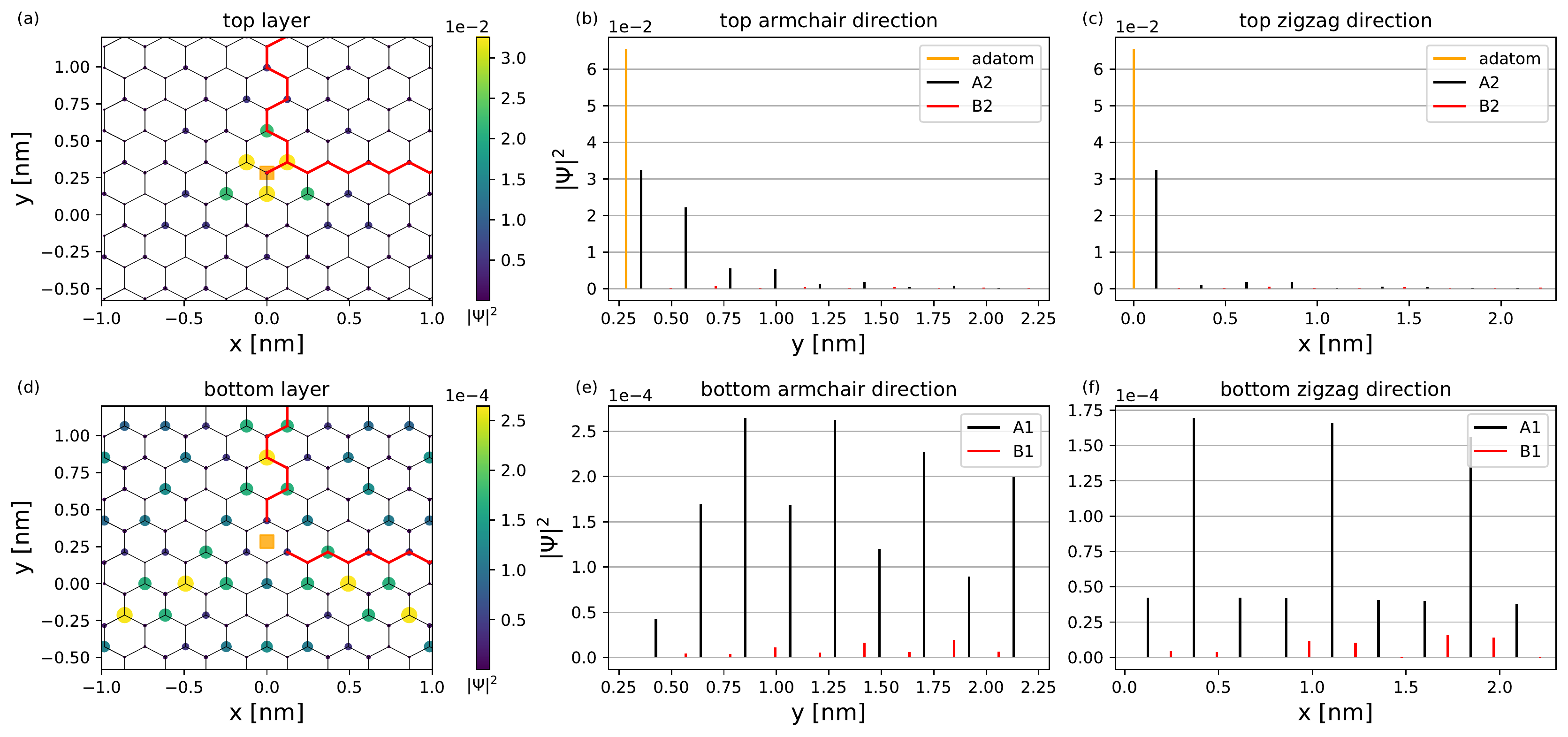}
	\caption{Sub-lattice resolved probabilities of YSR states in top~(a) and bottom~(d) layers originating from a magnetic adatom (hydrogen) chemisorbed at a non-dimer carbon in the top layer at chemical potential $\mu=-0.1\,\mathrm{eV}$; for the corresponding energy spectrum see Fig.~\ref{fig:Comparison_YSRS_energies}. 
		Side panels show sublattice-resolved probabilities in the corresponding layers for two representative directions, armchair [panels (b)~and~(e)] and zig-zag [panels (c)~and~(f)]. The numerical calculations---exact diagonalization---employed: $\Delta_0 = 50\,\mathrm{meV}$, $W = 601a$ and $L = 601a$.}
	\label{fig:YSRS_probdens_non_dimer}
\end{figure*}

This is quite a general statement irrespective of BLG that is based on the energy overlap argument. 
However, in the case of BLG what would matter on top of this, is the spatial overlap between the localized YSR states and propagating quasi-particle modes within BLG. 
Figures~\ref{fig:YSRS_probdens_dimer}~and~\ref{fig:YSRS_probdens_non_dimer} show, correspondingly, the sublattice resolved YSR probabilities originating from hydrogen magnetic impurities chemisorbed at dimer and non-dimer carbon sites. The plotted eigenstates' probabilities correspond to the YSR spectra in Fig.~\ref{fig:Comparison_YSRS_energies} for the particular chemical potential of $\mu=-0.1\,\mathrm{eV}$ [value at which one of the dimer resonances
in the normal system appears, see Fig.~\ref{fig:Comparison_YSRS_energies}~(c)]. 
Inspecting Figs.~\ref{fig:YSRS_probdens_dimer}~and~\ref{fig:YSRS_probdens_non_dimer} we see that for the magnetic impurity chemisorbed on the dimer (non-dimer) carbon site in the top layer, the corresponding YSR states dominantly occupy the opposite---non-dimer (dimer) top sublattice---of BLG. 
The spatial profiles of the YSR probability densities with their threefold symmetry matches with the results of the recent study of YSR states in twisted BLG \cite{Lado:PhysRevMaterials2019}.

Moreover, diagonalizing $H_\mathrm{BdG}$, Eq.~(\ref{Eq:BdG-BLG}), for $\mu$ in $[-\gamma_1,+\gamma_1]$, one sees 
that the BLG quasi-particle states are built primarily on $p_z$ orbitals belonging to the low-energy $B2$ and $A1$ carbons, i.e., they propagate mainly through the non-dimer sublattice of BLG, see Fig.~\ref{fig:BLG_lattice}.  
Thus from a pure geometrical point of view, there is a substantially larger (smaller) spatial overlap between 
these low-energy BLG states and YSR states originating from the dimer (non-dimer) impurities, since the latter 
spread over the non-dimer (dimer) sublattice. 
Therefore for $\mu$ in $[-\gamma_1,+\gamma_1]$, we expect a stronger spin relaxation for magnetic impurities at dimer 
than non-dimer sites.
%

\subsection{Spin relaxation}

In s-wave superconductors the quasi-particle spin relaxation by non-resonant magnetic impurities follows the conventional Hebel-Slichter picture~\cite{Hebel1957,Hebel1959,Hebel1959-Theory}. That is, when entering from 
the normal into the superconducting phase the spin relaxation rate initially increases due to the superconducting coherence; lowering temperature further it starts to saturate, and by approaching a milli-Kelvin regime the spin relaxation quenches completely due to the lack of quasi-particle excitations. 

\begin{figure*}
  \centering
  \includegraphics[width=1.9\columnwidth]{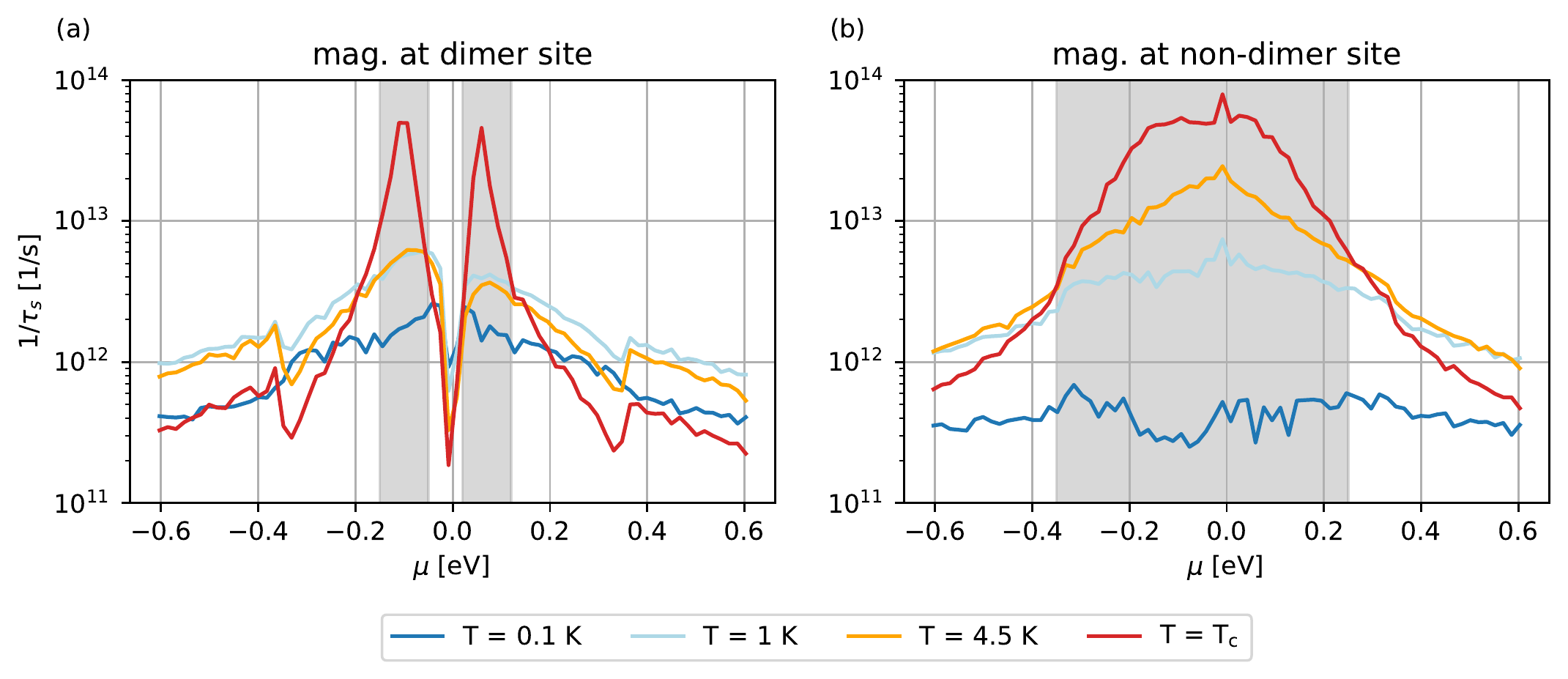}
  \caption{Quasi-particle spin-relaxation rates in superconducting BLG with $\Delta_0=1$\,meV for different temperatures $T$ (different colors) as functions 
of chemical potential $\mu$, the grey shaded backgrounds mark the doping regions at which the system behaves resonantly. Panels~(a)~and~(b)~display, correspondingly, spin-relaxation rate for hydrogen magnetic impurity chemisorbed on dimer and non-dimer 
site employing the rate formula given by Eq.~(\ref{Eq:spin-flip-probability7}). Calculation was performed for BLG Hamiltonian with all $\gamma$-hoppings involved, the spatial dimensions of the scattering region were fixed to $W=131\,a$ and $L=4\,a$ and the phase averaging counted 20 equally spaced values 
  of $k_{\mathrm{trans}}W$ in the interval $[0;2\pi)$. The impurity concentration for this configuration corresponds to $\eta_\mathrm{ada} = 0.0413~\mathrm{\%}$. }
  \label{fig:Rates_mag}
\end{figure*}

Figures~\ref{fig:Rates_mag}~(a)~and~(b) show temperature and doping dependencies of spin relaxation in hydrogenated superconducting BLG with the pairing gap
$\Delta_0=1$\,meV. Obviously, we see that the spin relaxation due to resonant magnetic impurities does not follow the Hebel-Slichter picture over the whole ranges of doping. Passing into the superconducting phase the spin relaxation in BLG drops down substantially with lowered $T$ at doping regions around the resonances---particularly, in the dimer case for $\mu\in[-0.2,0.2]\,\mathrm{eV}$ and  in the non-dimer one for $\mu\in[-0.4,0.4]$\,eV---and enhances at doping levels away from them. The reason for the drop was first elucidated in Ref.~\cite{Kochan:PRL_sc_graphene_spin_relaxation}, and counts the formation of YSR states lying deep inside the superconducting gap. The latter energetically decouple from the quasi-particle ranges, as explained in Sec.~\ref{Sec:ToyModel} and documented in Fig.~\ref{fig:Comparison_YSRS_energies}. Consequently, the reduced energy overlap between the two groups of states---which gets more pronounced when 
lowering $T$ and raising $\Delta(T)$ in accordance with Eq.~(\ref{Eq:GapDep})---implies the lowered spin-relaxation. 
In the regions far away from the resonances, the YSR states are close to the gap edges, and the spin relaxation follows the conventional Hebel-Slichter scenario.
For the moderate temperatures (above 1\,K) the cross-over from the resonant to Hebel-Slichter picture in BLG appears around $|\mu|\simeq 0.2$\,eV in the dimer case and $|\mu|\simeq 0.4$\,eV in the non-dimer one. 

Impurity spectral features---positions of the resonance peaks and their widths, see Fig.~\ref{fig:Comparison_YSRS_energies}~(c)---affect doping dependencies of
the spin-relaxation rates already in the normal-phase \cite{Denis_MAG_ham}. The spin-relaxation rate for the spectrally narrow dimer impurity shows two pronounced shoulders in $1/\tau_s$, see Fig.~\ref{fig:Rates_mag}~(a), while the spectrally wide non-dimer resonance washes out the sub-peak structure producing a single wide hump in Fig.~\ref{fig:Rates_mag}~(b), of course this depends on the mutual strengths of the exchange $V_s^{(1)}$ and orbital interaction $V_o$, for the extended discussion see \cite{Denis_MAG_ham}. 
In reality, the spin-relaxation rate would be broadened due to other effects, like electron-hole puddles, variations of orbital parameters with doping and temperature, spatial separation of impurities etc., so the final rate gets effectively smeared out and its internal shoulder-like structure is not necessarily observed directly \cite{Han:PRL2011,Yang:PRL2011,Ingla-Aynes:PRB2015,Avsar:NPG2016}. In the case of dimer impurity, we see that around $|\mu|\simeq \gamma_1=0.3$\,eV the spin-relaxation rate slightly jumps up. This is because around this energy the electronic states from high energy carbons $A2$ and $B1$, see Fig.~\ref{fig:BLG_lattice}, enter the transport and the number of scattering channels raises. 
It is worth to compare the magnitudes of spin relaxation rates in Figs.~\ref{fig:Rates_mag} (a) and (b) for impurities at dimer and non-dimer sites when passing from the normal-phase at $T_c=6.953$\,K to the superconducting-phase at milli-Kelvin range around $T=0.1$\,K. We see that the dimer impurity relaxes quasi-particles' spins 
faster than the non-dimer one at very-low $T$ superconducting-phase, but this turns approaching $T_c$ and going into the normal-phase. 
Again, this is the consequence of the wave function overlaps between the extended low-energy quasi-particle BLG modes and the localized YSR states as displayed in Figs.~\ref{fig:YSRS_probdens_dimer}~and~\ref{fig:YSRS_probdens_non_dimer}.

In Appendix \ref{app:SOC_ham_and_params} we also show results for the spin-relaxation rates in the case of SOC active hydrogen impurities. As expected, the rates exhibit a strong decrease over the whole doping range when lowering the temperature for both impurity configurations. These findings are consistent with the calculations in single layer graphene \cite{Kochan:PRL_sc_graphene_spin_relaxation}.

\subsection{Critical current of the BLG-based Josephson junction}

\begin{figure}
  \centering
  \includegraphics[width=\columnwidth]{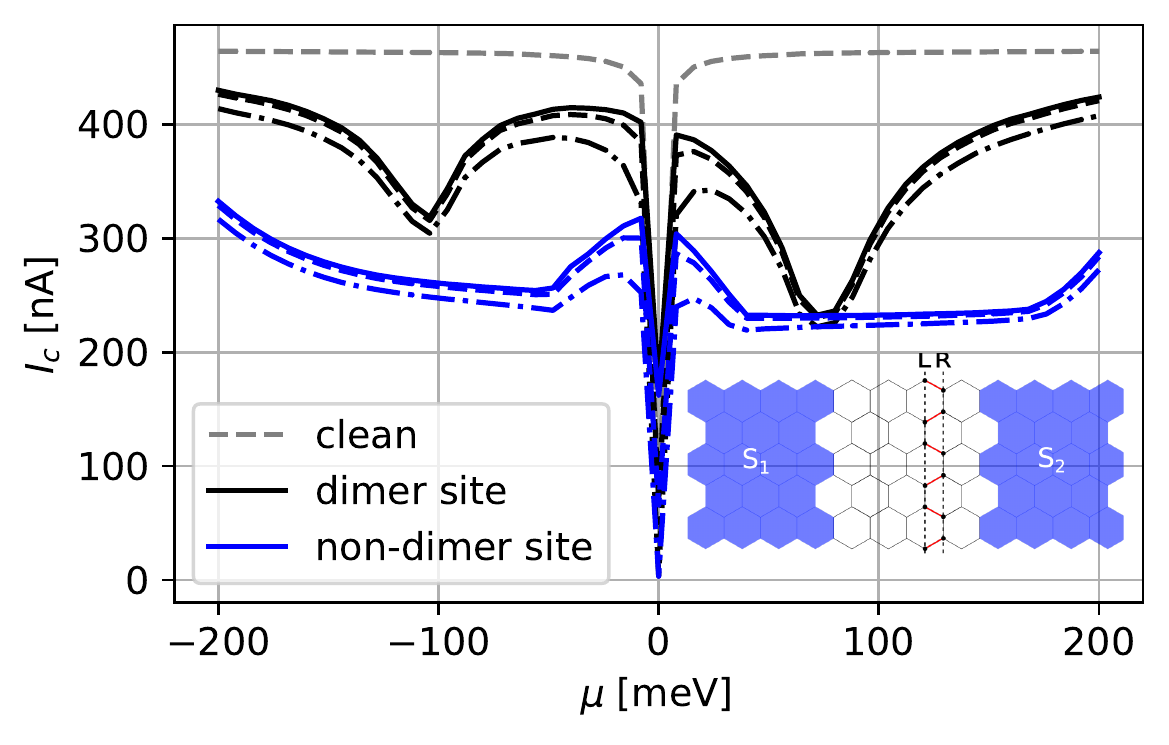}
  \caption{Critical currents of the BLG-based Josephson junctions (see inset for schematics) with and without resonant magnetic impurities calculated by Eq.~(\ref{eq:furusaki}). 
  Data for different lengths are displayed by different styles---the full lines corresponds to length $L=10a$, the dashed lines to $L=20a$ 
  and the dashed-dotted lines to $L=60a$. Different colors code different impurity contents---the clean junction result is displayed by grey, and 
  data for a junction with a single hydrogen in its center at dimer/non-dimer site by black/blue.
  The width of the junctions is fixed to $W=40a$, and the pairing gap $\Delta_0$ in the BLG-based superconducting leads equals $1$\,meV.}
  \label{fig:JJ_crit_current}
\end{figure}
Figure~\ref{fig:JJ_crit_current} illustrates the critical currents of the BLG-based Josephson junctions as functions of chemical potential for different lengths and different hydrogen positions.
We study junctions functionalized with the di\-mer/non-di\-mer resonant magnetic impurities (data displayed by black/blue), as well as a junction without them (data in grey), for a junction schematic see the inset in Fig.~\ref{fig:JJ_crit_current}. 
In the latter benchmark case we just plot the critical current for $L=20a$, as the length dependence is not affecting the magnitude of $I_c$ too strongly. 
Comparing the scaling of the critical current $I_c$ with the chemical potential $\mu$ for the dimer and non-dimer impurity cases, we see that $I_c$ drops its value at those doping levels where the BLG system in the normal-phase hits its resonances, for comparison see DOS in Fig.~\ref{fig:Comparison_YSRS_energies}(c). 
The effect is more pronounced for the narrow resonance in the dimer case, but also the non-dimer impurity displays a wide plateau in $I_c$ that is spreading over 
its resonance width. 

We see that $I_c$ for the non-dimer case is lower than $I_c$ for the dimer one, implying the system is more perturbed 
by the resonant scattering off magnetic impurities chemisorbed on the non-dimer sites.   
For the same reason the spin relaxation rate $1/\tau_s$ for the non-dimer position 
in the normal phase ($T=T_c$) is larger than the corresponding quantity for the dimer one, see Figs.~\ref{fig:Rates_mag}. 
The explanation why this happens was given in \cite{Denis_MAG_ham}: an impurity chemisorbed at the non-dimer (dimer) site gives rise to a resonant impurity state in the normal phase that is located on the dimer (non-dimer) sublattice---similarly as for the YSR states, see Figs.~\ref{fig:YSRS_probdens_dimer}~and~\ref{fig:YSRS_probdens_non_dimer}. However, there is one substantial difference compared with the localized YSR states. The resonant levels are virtually bound, meaning, their spatial probability falls off with a distance polynomially, 
\cite{Pereira:PRL2006,Castro:PRL2010}. Since the top dimer sublattice of BLG couples via $\gamma_1$-hopping with the bottom layer, the effect of resonant state living on the dimer top sublattice would be felt also on the bottom layer. While the both layers are affected by the resonance the scattering is more damaging---this is the reason for the larger $1/\tau_s$ and smaller $I_c$ in the normal-phase for the impurity located on the non-dimer site. 
Contrary, the resonant state due to the dimer impurity, which keeps located on the non-dimer top sub-lattice of BLG would only weakly protrude into the bottom layer---non-dimer carbons do not hybridize via $\gamma_1$---and hence an electron propagating in BLG is effectively less scattered off the dimer impurities since the bottom layer gives it a green light to move freely.     

So we believe that the Josephson current spectroscopy can serve as another sensible probe for discriminating between different resonant impurities reflecting their spectral and resonant features.

\begin{figure*}
  \centering
  \includegraphics[width=1.9\columnwidth]{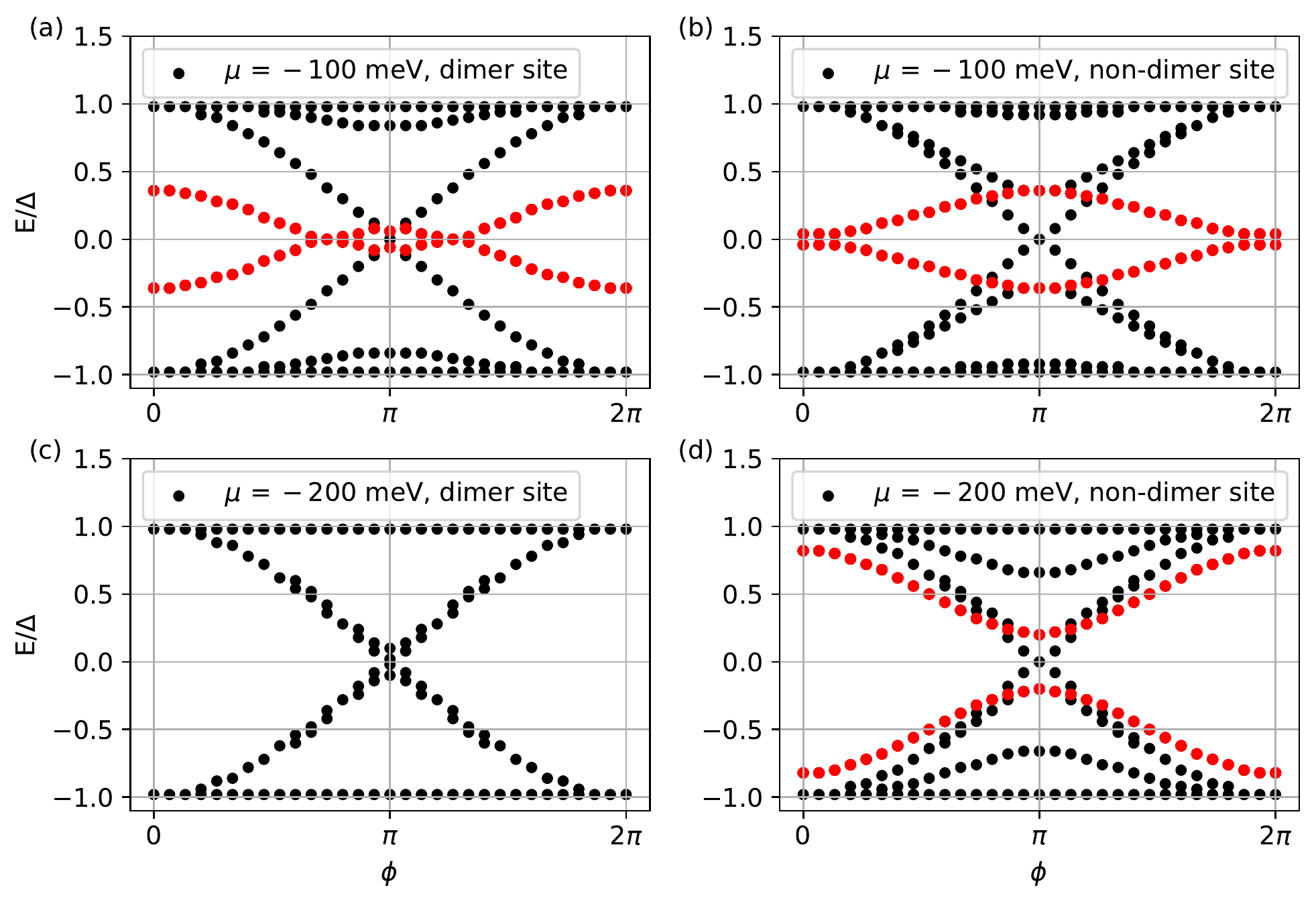}
  \caption{ABS spectra of BLG-based Josephson junction functionalized by resonant magnetic impurities. Panels~(a)~and~(b) show correspondingly the ABS spectra for a hydrogen impurity at dimer and non-dimer site at chemical potential $\mu=-100\,\mathrm{meV}$ (resonance), while panels~(c)~and~(d) correspond to the same configuration at doping level $\mu=-200\,\mathrm{meV}$ (off-resonance for the dimer impurity). Bound state energies are obtained by sampling maximas of $d(E)$, Eq.~(\ref{eq:spectral_density}), for energies $E$ inside the superconducting gap. The size of the normal spacer has width $W=40a$ and length $L=10a$ and contains a single impurity positioned in the center. Bands shown in red correspond to ABS that result from the interplay of the magnetic exchange and the superconducting pairing and share certain spectral similarities with the YSR states.}
  \label{fig:ABS_w40a_L10a}
\end{figure*}

\subsection{Andreev bound state spectrum of BLG-based Josephson junction}

Figure~\ref{fig:ABS_w40a_L10a} shows the Andreev in-gap spectra for the BLG-based Josephson junction functionalized by dimer/non-dimer hydrogen impurities as functions of the phase difference $\phi$. We consider the same geometry and system sizes (width $W=40a$ and length $L=10a$) as were used for the calculation of 
the critical currents in Fig.~\ref{fig:JJ_crit_current}. Moreover, we calculate the ABS spectra for the two representative chemical potentials, 
$\mu=-100\,\mathrm{meV}$ and $\mu=-200\,\mathrm{meV}$, that set different resonant regimes. 

Panels~\ref{fig:ABS_w40a_L10a}~(a)~and~(b)~display the corresponding ABS energies for $\mu=-100\,\mathrm{meV}$ at which both chemisorption positions host resonances in the normal phase. The first remarkable feature in the spectra for both impurity positions is the presence of the ABS bands (shown in red) that are detached from the continuum spectrum and spread around the center of the gap. 
This is very similar to the YSR spectra shown in Fig.~\ref{fig:Comparison_YSRS_energies}, where at the same doping level $\mu$ develops the YSR bound states whose energies are located close to the center of the gap. 
Because of this spectral similarity one can consider the red ABS as Josephson-junction descendants of the corresponding YSR states, despite, strictly speaking, the YSR states being defined for impurities embedded directly inside a superconductor and not inside the normal-spacer of the Josephson junction. 
Comparing closely the dimer [panel (a)] and non-dimer [panel (b)] parts we see also the black dotted ABS with the typical Andreev $\phi$-dispersions determined mainly 
by the junction length $L$ and the S/N-interface transparency \cite{Kulik1970, Costa2018}.
We assume a transparent junction realized, for example, on a flake of BLG that is proximitized by two superconductors with different phases which are separated by a non-proximitized normal region. Contrasting the slopes of red and black ABS branches for both chemisorption positions we see that in the non-dimer case the slopes of the red and black bands are mostly opposite implying a suppression of the critical 
current since $I(\phi) \propto \sum_{\mathrm{ABS}}\partial E^{\mathrm{ABS}}(\phi) / \partial \phi$.

Next, let us change the chemical potential to the lower value of $\mu=-200\,\mathrm{meV}$, such that the dimer site is already out of the resonance, while the 
non-dimer one is still ``in a mild shadow'' of it, see the DOS features in Fig.~\ref{fig:Comparison_YSRS_energies}~(c). 
The corresponding ABS spectra are displayed in panels~\ref{fig:ABS_w40a_L10a}~(c)~and~(d). In contrast to the previous cases, the ABS ``resembling'' 
the YSR states are absent (more precisely overlying with other branches) for the dimer case, but are still optically visible for the non-dimer 
one---again displayed in red, although now spreading energetically more away the center of the gap. 
The remaining bound state energies---displayed by black---resemble the standard ABS dispersions. So off resonances the magnetic impurities in the normal-spacer act on the formation of the ABS as non-magnetic scatterers.
In the supplemental material we also provide a comparison to a different calculation approach with switched off magnetic moments in order to cross-check the employed numerics.

\section{Conclusions}\label{Sec:Conclusions}

In summary, we have shown that the superconducting BLG in the presence of resonant magnetic impurities experiences interesting spin phenomena that are 
manifested in 
1) an unusual doping and temperature dependency of spin-relaxation rates, 
2) subgap spectra hosting deep-lying YSR states, 
3) magnitudes of critical currents and 
4) Andreev bound states in the BLG-based Josephson junctions. 
BLG has two non-equivalent sublattices, hence, the same magnetic adatom hybridizing with BLG can show differing superconducting behaviour. Our secondary aim 
was to trace these features in detail and understand their origins from the point of view of resonant scattering in the normal BLG phase. 

Coming to the spin relaxation, we have convincingly demonstrated by implementing an S-matrix approach
that it can depart from the conventional Hebel-Slichter scenario when taking into account the multiple scattering processes. Meaning, the quasi-particle spin-relaxation rates can substantially decrease once the system is turned into the superconducting phase. 
Furthermore, the detailed numerical implementation scheme we have developed using the existing \textsc{Kwant} functionalities, see the Supplemental Material \cite{SM}, represents \emph{per-se} an important taking home message. It
allows us to simulate spin relaxation, as well, other spectral characteristics including the YSR and Andreev bound states. 

Beyond the BLG, we have demonstrated under quite general conditions that at doping levels that are tuned to the normal-state resonances, the corresponding 
YSR states separate from the quasi-particle coherence peaks and immerse deep in the center of the gap, or even cross there. Such zero energy YSR states have a profound impact on the topological nature of the underlying superconducting ground-state with practical applications for the YSR \cite{Menard2015a,Heinrich:ProgSurfScience2018,WangWiesendanger:PRL2021,Brihuega:AdvMaterials2021} 
and Josephson spectroscopy \cite{Kuster:NatCom2021},
as well on the Shiba-band engineering. Particularly in a connection with topological quantum-phase transitions and parity-changing of the condensate wave function \cite{Sakurai:ProgTheorPhys1970,SauDemler:PRB2013,Pientka:PhysScripta2015,Costa2018}. 
We derived a formula, Eq.~(\ref{Eq:YSR-new_formula_1}), for the YSR energies assuming the system is doped in resonance.
Knowing the resonant width of the modified DOS and the strength of the exchange coupling, one can predict with the help of Eq.~(\ref{Eq:YSR-new_formula_1}) the YSR energies, or vice-versa, knowing the width from normal-phase transport measurements and the YSR energies from the STM one can estimate a magnitude of the exchange strength between itinerant electrons and localized magnetic moments. 

We are not aware of any experiments probing spin relaxation in superconducting graphene neither BLG, but we believe that our results can trigger some, or can shed some light on the similar super-spintronics phenomena explored in other low-dimensional superconductors.

\paragraph*{Acknowledgements.}
This work was supported by Deutsche Forschungsge\-meinschaft (DFG, German Research Foundation) within Project-ID 314695032-SFB 1277 (project A07) and the Elitenetzwerk Bayern Doktorandenkolleg “Topological Insulators”. D.K. acknowledges a partial support from the project SUPERSPIN funded by Slovak Academy of Sciences via the initiative IMPULZ 2021.
We thank Dr.~Marco Aprili, Dr.~Andreas Costa, Dr.~Ferdinand Evers, Dr.~Jaroslav Fabian, Dr.~Richard Hlubina, Dr.~Tom\'{a}\v{s} Novotn\'{y} and Dr.~Klaus Richter for useful discussions.

\clearpage
\appendix

\section{Model parameters, local SOC Hamiltonian and the corresponding spin relaxation}\label{app:SOC_ham_and_params}

An external impurity hybridizing with BLG modifies apart of the orbital degrees of freedom, Hamiltonian $V_o$, also the local SOC environment. To investigate 
an impact of the local SOC on the quasi-particle spin relaxation we use the following tight-binding Hamiltonian: 
\begin{align*}
    V_s^{(2)}=&
            \frac{i\lambda_I^A}{3\sqrt{3}} \sum\limits_{m \in \rm{C_{nnn}} \atop {}}\sum\limits_{\sigma} c^\dagger_{0\s} \left(\hat{s}_{z}\right)_{\s\s} c^{\phantom{\dagger}}_{m\s}+ \mathrm{h.c.} \nonumber\\
             +& \frac{i\lambda_I^B}{3\sqrt{3}} \sum\limits_{{m,n\in\mathrm{C_{nn}} \atop m\neq n}}\sum\limits_{\sigma} c^\dagger_{m\s}\,\nu^{\phantom{\dagger}}_{mn} \left(\hat{s}_{z}\right)_{\s\s} c^{\phantom{\dagger}}_{n\s} \label{Eq:H_soc}\nonumber\\
             +& \frac{2i\lambda_R}{3}        \sum\limits_{m \in \rm{C_{nn}} \atop {}}\sum\limits_{\sigma\neq\sigma'} c^\dagger_{0\s}\left(\bm{\hat{s}}\times\bm{d}_{0m}\right)^{\phantom{\dagger}}_{z,\s\s'} c^{\phantom{\dagger}}_{m\s'} + \mathrm{h.c.} \nonumber \\
             +& \frac{2i\lambda_\mathrm{PIA}^A}{3}        \sum\limits_{m\in\mathrm{C_{nnn}} \atop {}}\sum\limits_{\s\neq\s'} c^\dagger_{0\s}\left(\bm{d}_{0m}\times\bm{\hat{s}}\right)^{\phantom{\dagger}}_{z,\s\s'} c^{\phantom{\dagger}}_{m\s'} + \mathrm{h.c.}\nonumber\\
             +& \frac{2i\lambda_\mathrm{PIA}^B}{3}        \sum\limits_{{m,n\in\mathrm{C_{nn}} \atop m\neq n}}\sum\limits_{\s\neq\s'} c^\dagger_{m\s}\left(\bm{d}_{mn}\times\bm{\hat{s}}\right)^{\phantom{\dagger}}_{z,\s\s'} c^{\phantom{\dagger}}_{n\s'}\,,
\end{align*}
for details see Ref.~\cite{Denis_SOC_Ham}.

\begin{figure}
  \centering
  \includegraphics[width=\columnwidth]{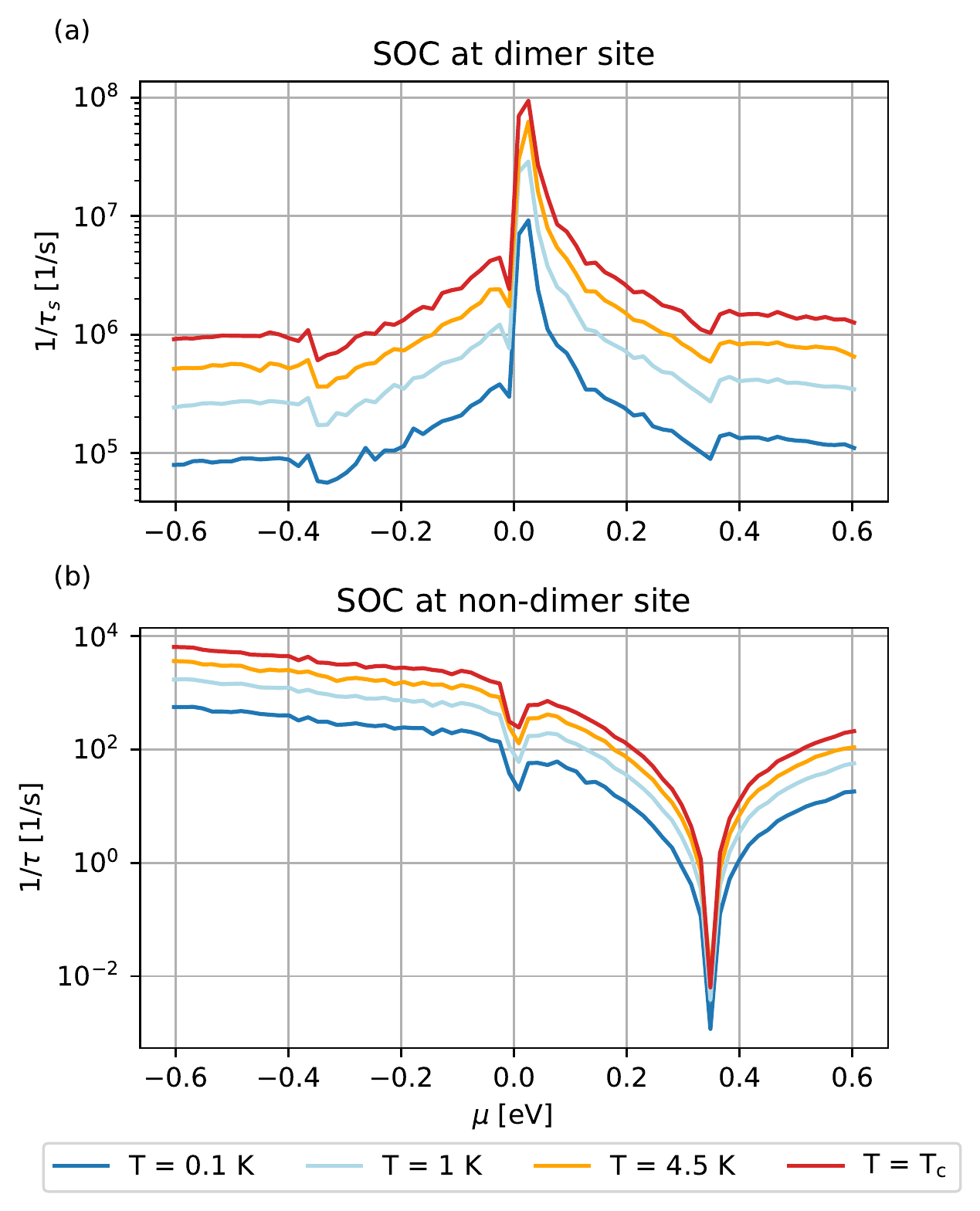}
  \caption{Temperature dependence of the quasi-particle spin-relaxation rates versus doping for superconducting BLG in the presence of a spin-orbit active hydrogen impurity at dimer and non-dimer site. Results are obtained from \textsc{Kwant} simulations with the help of Eq.~(\ref{Eq:spin-flip-probability7}). The system size was fixed to $\mathrm{W}=131a$ and $\mathrm{L=4a}$, giving $\eta_\mathrm{ada} = 0.0413~\mathrm{\%}$. The phase averaging was performed for 20 equally spaced values of $k_{\mathrm{trans}}$ in the interval $[0;2\pi]$.}
  \label{fig:Rates_soc}
\end{figure}

The parameters entering Hamiltonians $V_o$, $V_s^{(1)}$ and $V_s^{(2)}$ that are used in this study correspond to hydrogen impurity, the values are obtained from fitting DFT calculations~\cite{Kochan2014_PRL-SR-Graphene,Gmitra2013_SOC-in-H-Graphene} and are summarized in table~\ref{table:params}.
\begin{table}[h!]
\begin{center}
    \begin{tabular}{ c|c|c } 
     hydrogen          & dimer [eV]               & non-dimer [eV]       \\  \hline
     $\epsilon$        & 0.25                     & 0.35                 \\
     $\omega$          & 6.5                      & 5.5                  \\
     $J$                 & -0.4                     & -0.4                 \\
     $\lambda_I^A$     & $-0.21\cdot 10^{-3}$     & $-0.21\cdot 10^{-3}$ \\
     $\lambda_I^B$     & 0                        & 0                    \\
     $\lambda_R$       & $0.33\cdot 10^{-3}$      & $0.33\cdot 10^{-3}$  \\
     $\lambda_\mathrm{PIA}^A$ & 0                        & 0                    \\
     $\lambda_\mathrm{PIA}^B$ & $0.77\cdot 10^{-3}$      & $0.77\cdot 10^{-3}$  \\
    \end{tabular}
\end{center}
\caption{Hamiltonian parameters}
\label{table:params}
\end{table}
Figure~\ref{fig:Rates_soc} shows quasi-particle spin-relaxation rates versus doping for a spin-orbit active hydrogen impurity, again for several representative temperatures going from the critical $T_c$ down to zero. 
The relaxation rate shows clear differences for the dimer, panel (a), and the non-dimer, panel (b) positions. 
While the dimer case displays a strong enhancement of the rate around $\mu=0$, the rate is heavily suppressed in the non-dimer case for $\mu=0.34\,\mathrm{eV}$. These features remain 
insensitive to the variation of temperature and transcend also into the superconducting-phase. Passing from the normal to superconducting regime, we observe a global reduction of the spin-relaxation rate by an order of magnitude. This observations match with the results obtained for superconducting single layer graphene~\cite{Kochan:PRL_sc_graphene_spin_relaxation}.

\newpage
\bibliography{literatur}

\begin{thebibliography}{117}%
\makeatletter
\providecommand \@ifxundefined [1]{%
 \@ifx{#1\undefined}
}%
\providecommand \@ifnum [1]{%
 \ifnum #1\expandafter \@firstoftwo
 \else \expandafter \@secondoftwo
 \fi
}%
\providecommand \@ifx [1]{%
 \ifx #1\expandafter \@firstoftwo
 \else \expandafter \@secondoftwo
 \fi
}%
\providecommand \natexlab [1]{#1}%
\providecommand \enquote  [1]{``#1''}%
\providecommand \bibnamefont  [1]{#1}%
\providecommand \bibfnamefont [1]{#1}%
\providecommand \citenamefont [1]{#1}%
\providecommand \href@noop [0]{\@secondoftwo}%
\providecommand \href [0]{\begingroup \@sanitize@url \@href}%
\providecommand \@href[1]{\@@startlink{#1}\@@href}%
\providecommand \@@href[1]{\endgroup#1\@@endlink}%
\providecommand \@sanitize@url [0]{\catcode `\\12\catcode `\$12\catcode
  `\&12\catcode `\#12\catcode `\^12\catcode `\_12\catcode `\%12\relax}%
\providecommand \@@startlink[1]{}%
\providecommand \@@endlink[0]{}%
\providecommand \url  [0]{\begingroup\@sanitize@url \@url }%
\providecommand \@url [1]{\endgroup\@href {#1}{\urlprefix }}%
\providecommand \urlprefix  [0]{URL }%
\providecommand \Eprint [0]{\href }%
\providecommand \doibase [0]{https://doi.org/}%
\providecommand \selectlanguage [0]{\@gobble}%
\providecommand \bibinfo  [0]{\@secondoftwo}%
\providecommand \bibfield  [0]{\@secondoftwo}%
\providecommand \translation [1]{[#1]}%
\providecommand \BibitemOpen [0]{}%
\providecommand \bibitemStop [0]{}%
\providecommand \bibitemNoStop [0]{.\EOS\space}%
\providecommand \EOS [0]{\spacefactor3000\relax}%
\providecommand \BibitemShut  [1]{\csname bibitem#1\endcsname}%
\let\auto@bib@innerbib\@empty
\bibitem [{\citenamefont {{\v{Z}}uti{\'{c}}}\ \emph {et~al.}(2004)\citenamefont
  {{\v{Z}}uti{\'{c}}}, \citenamefont {Fabian},\ and\ \citenamefont
  {Das~Sarma}}]{Zutic2004}%
  \BibitemOpen
  \bibfield  {author} {\bibinfo {author} {\bibfnamefont {I.}~\bibnamefont
  {{\v{Z}}uti{\'{c}}}}, \bibinfo {author} {\bibfnamefont {J.}~\bibnamefont
  {Fabian}},\ and\ \bibinfo {author} {\bibfnamefont {S.}~\bibnamefont
  {Das~Sarma}},\ }\bibfield  {title} {\bibinfo {title} {{Spintronics:
  Fundamentals and applications}},\ }\href
  {https://doi.org/10.1103/RevModPhys.76.323} {\bibfield  {journal} {\bibinfo
  {journal} {Reviews of Modern Physics}\ }\textbf {\bibinfo {volume} {76}},\
  \bibinfo {pages} {323} (\bibinfo {year} {2004})}\BibitemShut {NoStop}%
\bibitem [{\citenamefont {Han}\ \emph {et~al.}(2014)\citenamefont {Han},
  \citenamefont {Kawakami}, \citenamefont {Gmitra},\ and\ \citenamefont
  {Fabian}}]{HanKawakamiGmitraFabian_2014}%
  \BibitemOpen
  \bibfield  {author} {\bibinfo {author} {\bibfnamefont {W.}~\bibnamefont
  {Han}}, \bibinfo {author} {\bibfnamefont {R.~K.}\ \bibnamefont {Kawakami}},
  \bibinfo {author} {\bibfnamefont {M.}~\bibnamefont {Gmitra}},\ and\ \bibinfo
  {author} {\bibfnamefont {J.}~\bibnamefont {Fabian}},\ }\bibfield  {title}
  {\bibinfo {title} {{Graphene spintronics}},\ }\href
  {https://doi.org/10.1038/nnano.2014.214} {\bibfield  {journal} {\bibinfo
  {journal} {Nature Nanotechnology}\ }\textbf {\bibinfo {volume} {9}},\
  \bibinfo {pages} {794} (\bibinfo {year} {2014})}\BibitemShut {NoStop}%
\bibitem [{\citenamefont {Roche}\ \emph {et~al.}(2015)\citenamefont {Roche},
  \citenamefont {{\AA}kerman}, \citenamefont {Beschoten}, \citenamefont
  {Charlier}, \citenamefont {Chshiev}, \citenamefont {Prasad~Dash},
  \citenamefont {Dlubak}, \citenamefont {Fabian}, \citenamefont {Fert},
  \citenamefont {Guimar{\~{a}}es}, \citenamefont {Guinea}, \citenamefont
  {Grigorieva}, \citenamefont {Sch{\"{o}}nenberger}, \citenamefont {Seneor},
  \citenamefont {Stampfer}, \citenamefont {Valenzuela}, \citenamefont
  {Waintal},\ and\ \citenamefont {van
  Wees}}]{Roche:SpintronicsPerspective2015}%
  \BibitemOpen
  \bibfield  {author} {\bibinfo {author} {\bibfnamefont {S.}~\bibnamefont
  {Roche}}, \bibinfo {author} {\bibfnamefont {J.}~\bibnamefont {{\AA}kerman}},
  \bibinfo {author} {\bibfnamefont {B.}~\bibnamefont {Beschoten}}, \bibinfo
  {author} {\bibfnamefont {J.-C.}\ \bibnamefont {Charlier}}, \bibinfo {author}
  {\bibfnamefont {M.}~\bibnamefont {Chshiev}}, \bibinfo {author} {\bibfnamefont
  {S.}~\bibnamefont {Prasad~Dash}}, \bibinfo {author} {\bibfnamefont
  {B.}~\bibnamefont {Dlubak}}, \bibinfo {author} {\bibfnamefont
  {J.}~\bibnamefont {Fabian}}, \bibinfo {author} {\bibfnamefont
  {A.}~\bibnamefont {Fert}}, \bibinfo {author} {\bibfnamefont {M.}~\bibnamefont
  {Guimar{\~{a}}es}}, \bibinfo {author} {\bibfnamefont {F.}~\bibnamefont
  {Guinea}}, \bibinfo {author} {\bibfnamefont {I.}~\bibnamefont {Grigorieva}},
  \bibinfo {author} {\bibfnamefont {C.}~\bibnamefont {Sch{\"{o}}nenberger}},
  \bibinfo {author} {\bibfnamefont {P.}~\bibnamefont {Seneor}}, \bibinfo
  {author} {\bibfnamefont {C.}~\bibnamefont {Stampfer}}, \bibinfo {author}
  {\bibfnamefont {S.~O.}\ \bibnamefont {Valenzuela}}, \bibinfo {author}
  {\bibfnamefont {X.}~\bibnamefont {Waintal}},\ and\ \bibinfo {author}
  {\bibfnamefont {B.}~\bibnamefont {van Wees}},\ }\bibfield  {title} {\bibinfo
  {title} {{Graphene spintronics: the European Flagship perspective}},\ }\href
  {https://doi.org/10.1088/2053-1583/2/3/030202} {\bibfield  {journal}
  {\bibinfo  {journal} {2D Materials}\ }\textbf {\bibinfo {volume} {2}},\
  \bibinfo {pages} {030202} (\bibinfo {year} {2015})}\BibitemShut {NoStop}%
\bibitem [{\citenamefont {Avsar}\ \emph {et~al.}(2020)\citenamefont {Avsar},
  \citenamefont {Ochoa}, \citenamefont {Guinea}, \citenamefont {\"Ozyilmaz},
  \citenamefont {van Wees},\ and\ \citenamefont
  {Vera-Marun}}]{Avsar-Colloquium:RMP2020}%
  \BibitemOpen
  \bibfield  {author} {\bibinfo {author} {\bibfnamefont {A.}~\bibnamefont
  {Avsar}}, \bibinfo {author} {\bibfnamefont {H.}~\bibnamefont {Ochoa}},
  \bibinfo {author} {\bibfnamefont {F.}~\bibnamefont {Guinea}}, \bibinfo
  {author} {\bibfnamefont {B.}~\bibnamefont {\"Ozyilmaz}}, \bibinfo {author}
  {\bibfnamefont {B.~J.}\ \bibnamefont {van Wees}},\ and\ \bibinfo {author}
  {\bibfnamefont {I.~J.}\ \bibnamefont {Vera-Marun}},\ }\bibfield  {title}
  {\bibinfo {title} {{Colloquium: Spintronics in graphene and other
  two-dimensional materials}},\ }\href
  {https://doi.org/10.1103/RevModPhys.92.021003} {\bibfield  {journal}
  {\bibinfo  {journal} {Rev. Mod. Phys.}\ }\textbf {\bibinfo {volume} {92}},\
  \bibinfo {pages} {021003} (\bibinfo {year} {2020})}\BibitemShut {NoStop}%
\bibitem [{\citenamefont {{\v{Z}}uti{\'{c}}}\ \emph {et~al.}(2019)\citenamefont
  {{\v{Z}}uti{\'{c}}}, \citenamefont {Matos-Abiague}, \citenamefont {Scharf},
  \citenamefont {Dery},\ and\ \citenamefont
  {Belashchenko}}]{Zutic:MaterialsToday2018}%
  \BibitemOpen
  \bibfield  {author} {\bibinfo {author} {\bibfnamefont {I.}~\bibnamefont
  {{\v{Z}}uti{\'{c}}}}, \bibinfo {author} {\bibfnamefont {A.}~\bibnamefont
  {Matos-Abiague}}, \bibinfo {author} {\bibfnamefont {B.}~\bibnamefont
  {Scharf}}, \bibinfo {author} {\bibfnamefont {H.}~\bibnamefont {Dery}},\ and\
  \bibinfo {author} {\bibfnamefont {K.}~\bibnamefont {Belashchenko}},\
  }\bibfield  {title} {\bibinfo {title} {{Proximitized materials}},\ }\href
  {https://doi.org/10.1016/j.mattod.2018.05.003} {\bibfield  {journal}
  {\bibinfo  {journal} {Materials Today}\ }\textbf {\bibinfo {volume} {22}},\
  \bibinfo {pages} {85} (\bibinfo {year} {2019})}\BibitemShut {NoStop}%
\bibitem [{\citenamefont {Podzorov}\ \emph {et~al.}(2004)\citenamefont
  {Podzorov}, \citenamefont {Gershenson}, \citenamefont {Kloc}, \citenamefont
  {Zeis},\ and\ \citenamefont {Bucher}}]{Podzorov_2004}%
  \BibitemOpen
  \bibfield  {author} {\bibinfo {author} {\bibfnamefont {V.}~\bibnamefont
  {Podzorov}}, \bibinfo {author} {\bibfnamefont {M.~E.}\ \bibnamefont
  {Gershenson}}, \bibinfo {author} {\bibfnamefont {C.}~\bibnamefont {Kloc}},
  \bibinfo {author} {\bibfnamefont {R.}~\bibnamefont {Zeis}},\ and\ \bibinfo
  {author} {\bibfnamefont {E.}~\bibnamefont {Bucher}},\ }\bibfield  {title}
  {\bibinfo {title} {High-mobility field-effect transistors based on transition
  metal dichalcogenides},\ }\href {https://doi.org/10.1063/1.1723695}
  {\bibfield  {journal} {\bibinfo  {journal} {Applied Physics Letters}\
  }\textbf {\bibinfo {volume} {84}},\ \bibinfo {pages} {3301} (\bibinfo {year}
  {2004})}\BibitemShut {NoStop}%
\bibitem [{\citenamefont {Shi}\ \emph {et~al.}(2015)\citenamefont {Shi},
  \citenamefont {Ye}, \citenamefont {Zhang}, \citenamefont {Suzuki},
  \citenamefont {Yoshida}, \citenamefont {Miyazaki}, \citenamefont {Inoue},
  \citenamefont {Saito},\ and\ \citenamefont
  {Iwasa}}]{Shi2015_IonicGatedTMDC-SC}%
  \BibitemOpen
  \bibfield  {author} {\bibinfo {author} {\bibfnamefont {W.}~\bibnamefont
  {Shi}}, \bibinfo {author} {\bibfnamefont {J.}~\bibnamefont {Ye}}, \bibinfo
  {author} {\bibfnamefont {Y.}~\bibnamefont {Zhang}}, \bibinfo {author}
  {\bibfnamefont {R.}~\bibnamefont {Suzuki}}, \bibinfo {author} {\bibfnamefont
  {M.}~\bibnamefont {Yoshida}}, \bibinfo {author} {\bibfnamefont
  {J.}~\bibnamefont {Miyazaki}}, \bibinfo {author} {\bibfnamefont
  {N.}~\bibnamefont {Inoue}}, \bibinfo {author} {\bibfnamefont
  {Y.}~\bibnamefont {Saito}},\ and\ \bibinfo {author} {\bibfnamefont
  {Y.}~\bibnamefont {Iwasa}},\ }\bibfield  {title} {\bibinfo {title}
  {{Superconductivity Series in Transition Metal Dichalcogenides by Ionic
  Gating}},\ }\href {https://doi.org/10.1038/srep12534} {\bibfield  {journal}
  {\bibinfo  {journal} {Scientific Reports}\ }\textbf {\bibinfo {volume} {5}},\
  \bibinfo {pages} {12534} (\bibinfo {year} {2015})}\BibitemShut {NoStop}%
\bibitem [{\citenamefont {Jo}\ \emph {et~al.}(2015)\citenamefont {Jo},
  \citenamefont {Costanzo}, \citenamefont {Berger},\ and\ \citenamefont
  {Morpurgo}}]{Jo2015_WS2_SC}%
  \BibitemOpen
  \bibfield  {author} {\bibinfo {author} {\bibfnamefont {S.}~\bibnamefont
  {Jo}}, \bibinfo {author} {\bibfnamefont {D.}~\bibnamefont {Costanzo}},
  \bibinfo {author} {\bibfnamefont {H.}~\bibnamefont {Berger}},\ and\ \bibinfo
  {author} {\bibfnamefont {A.~F.}\ \bibnamefont {Morpurgo}},\ }\bibfield
  {title} {\bibinfo {title} {{Electrostatically Induced Superconductivity at
  the Surface of WS 2}},\ }\href {https://doi.org/10.1021/nl504314c} {\bibfield
   {journal} {\bibinfo  {journal} {Nano Letters}\ }\textbf {\bibinfo {volume}
  {15}},\ \bibinfo {pages} {1197} (\bibinfo {year} {2015})}\BibitemShut
  {NoStop}%
\bibitem [{\citenamefont {Navarro-Moratalla}\ \emph {et~al.}(2016)\citenamefont
  {Navarro-Moratalla}, \citenamefont {Island}, \citenamefont
  {Ma{\~{n}}as-Valero}, \citenamefont {Pinilla-Cienfuegos}, \citenamefont
  {Castellanos-Gomez}, \citenamefont {Quereda}, \citenamefont
  {Rubio-Bollinger}, \citenamefont {Chirolli}, \citenamefont
  {Silva-Guill{\'{e}}n}, \citenamefont {Agra{\"{i}}t}, \citenamefont {Steele},
  \citenamefont {Guinea}, \citenamefont {van~der Zant},\ and\ \citenamefont
  {Coronado}}]{Navarro-Moratalla2016_TaS2-SC}%
  \BibitemOpen
  \bibfield  {author} {\bibinfo {author} {\bibfnamefont {E.}~\bibnamefont
  {Navarro-Moratalla}}, \bibinfo {author} {\bibfnamefont {J.~O.}\ \bibnamefont
  {Island}}, \bibinfo {author} {\bibfnamefont {S.}~\bibnamefont
  {Ma{\~{n}}as-Valero}}, \bibinfo {author} {\bibfnamefont {E.}~\bibnamefont
  {Pinilla-Cienfuegos}}, \bibinfo {author} {\bibfnamefont {A.}~\bibnamefont
  {Castellanos-Gomez}}, \bibinfo {author} {\bibfnamefont {J.}~\bibnamefont
  {Quereda}}, \bibinfo {author} {\bibfnamefont {G.}~\bibnamefont
  {Rubio-Bollinger}}, \bibinfo {author} {\bibfnamefont {L.}~\bibnamefont
  {Chirolli}}, \bibinfo {author} {\bibfnamefont {J.~A.}\ \bibnamefont
  {Silva-Guill{\'{e}}n}}, \bibinfo {author} {\bibfnamefont {N.}~\bibnamefont
  {Agra{\"{i}}t}}, \bibinfo {author} {\bibfnamefont {G.~A.}\ \bibnamefont
  {Steele}}, \bibinfo {author} {\bibfnamefont {F.}~\bibnamefont {Guinea}},
  \bibinfo {author} {\bibfnamefont {H.~S.~J.}\ \bibnamefont {van~der Zant}},\
  and\ \bibinfo {author} {\bibfnamefont {E.}~\bibnamefont {Coronado}},\
  }\bibfield  {title} {\bibinfo {title} {{Enhanced superconductivity in
  atomically thin TaS2}},\ }\href {https://doi.org/10.1038/ncomms11043}
  {\bibfield  {journal} {\bibinfo  {journal} {Nature Communications}\ }\textbf
  {\bibinfo {volume} {7}},\ \bibinfo {pages} {11043} (\bibinfo {year}
  {2016})}\BibitemShut {NoStop}%
\bibitem [{\citenamefont {Costanzo}\ \emph {et~al.}(2016)\citenamefont
  {Costanzo}, \citenamefont {Jo}, \citenamefont {Berger},\ and\ \citenamefont
  {Morpurgo}}]{Costanzo2016_MoS2-SC}%
  \BibitemOpen
  \bibfield  {author} {\bibinfo {author} {\bibfnamefont {D.}~\bibnamefont
  {Costanzo}}, \bibinfo {author} {\bibfnamefont {S.}~\bibnamefont {Jo}},
  \bibinfo {author} {\bibfnamefont {H.}~\bibnamefont {Berger}},\ and\ \bibinfo
  {author} {\bibfnamefont {A.~F.}\ \bibnamefont {Morpurgo}},\ }\bibfield
  {title} {\bibinfo {title} {{Gate-induced superconductivity in atomically thin
  MoS2 crystals}},\ }\href {https://doi.org/10.1038/nnano.2015.314} {\bibfield
  {journal} {\bibinfo  {journal} {Nature Nanotechnology}\ }\textbf {\bibinfo
  {volume} {11}},\ \bibinfo {pages} {339} (\bibinfo {year} {2016})}\BibitemShut
  {NoStop}%
\bibitem [{\citenamefont {Swartz}\ \emph {et~al.}(2012)\citenamefont {Swartz},
  \citenamefont {Odenthal}, \citenamefont {Hao}, \citenamefont {Ruoff},\ and\
  \citenamefont {Kawakami}}]{Swartz2012:ASCNano}%
  \BibitemOpen
  \bibfield  {author} {\bibinfo {author} {\bibfnamefont {A.~G.}\ \bibnamefont
  {Swartz}}, \bibinfo {author} {\bibfnamefont {P.~M.}\ \bibnamefont
  {Odenthal}}, \bibinfo {author} {\bibfnamefont {Y.}~\bibnamefont {Hao}},
  \bibinfo {author} {\bibfnamefont {R.~S.}\ \bibnamefont {Ruoff}},\ and\
  \bibinfo {author} {\bibfnamefont {R.~K.}\ \bibnamefont {Kawakami}},\
  }\bibfield  {title} {\bibinfo {title} {{Integration of the Ferromagnetic
  Insulator EuO onto Graphene}},\ }\href {https://doi.org/10.1021/nn303771f}
  {\bibfield  {journal} {\bibinfo  {journal} {ACS Nano}\ }\textbf {\bibinfo
  {volume} {6}},\ \bibinfo {pages} {10063} (\bibinfo {year}
  {2012})}\BibitemShut {NoStop}%
\bibitem [{\citenamefont {Yang}\ \emph {et~al.}(2013)\citenamefont {Yang},
  \citenamefont {Hallal}, \citenamefont {Terrade}, \citenamefont {Waintal},
  \citenamefont {Roche},\ and\ \citenamefont {Chshiev}}]{Yang2013:PRL}%
  \BibitemOpen
  \bibfield  {author} {\bibinfo {author} {\bibfnamefont {H.~X.}\ \bibnamefont
  {Yang}}, \bibinfo {author} {\bibfnamefont {A.}~\bibnamefont {Hallal}},
  \bibinfo {author} {\bibfnamefont {D.}~\bibnamefont {Terrade}}, \bibinfo
  {author} {\bibfnamefont {X.}~\bibnamefont {Waintal}}, \bibinfo {author}
  {\bibfnamefont {S.}~\bibnamefont {Roche}},\ and\ \bibinfo {author}
  {\bibfnamefont {M.}~\bibnamefont {Chshiev}},\ }\bibfield  {title} {\bibinfo
  {title} {{Proximity Effects Induced in Graphene by Magnetic Insulators:
  First-Principles Calculations on Spin Filtering and Exchange-Splitting
  Gaps}},\ }\href {https://doi.org/10.1103/PhysRevLett.110.046603} {\bibfield
  {journal} {\bibinfo  {journal} {Phys. Rev. Lett.}\ }\textbf {\bibinfo
  {volume} {110}},\ \bibinfo {pages} {046603} (\bibinfo {year}
  {2013})}\BibitemShut {NoStop}%
\bibitem [{\citenamefont {Mendes}\ \emph {et~al.}(2015)\citenamefont {Mendes},
  \citenamefont {Alves~Santos}, \citenamefont {Meireles}, \citenamefont
  {Lacerda}, \citenamefont {Vilela-Le\~ao}, \citenamefont {Machado},
  \citenamefont {Rodr\'{\i}guez-Su\'arez}, \citenamefont {Azevedo},\ and\
  \citenamefont {Rezende}}]{Mendes2015:PRL}%
  \BibitemOpen
  \bibfield  {author} {\bibinfo {author} {\bibfnamefont {J.~B.~S.}\
  \bibnamefont {Mendes}}, \bibinfo {author} {\bibfnamefont {O.}~\bibnamefont
  {Alves~Santos}}, \bibinfo {author} {\bibfnamefont {L.~M.}\ \bibnamefont
  {Meireles}}, \bibinfo {author} {\bibfnamefont {R.~G.}\ \bibnamefont
  {Lacerda}}, \bibinfo {author} {\bibfnamefont {L.~H.}\ \bibnamefont
  {Vilela-Le\~ao}}, \bibinfo {author} {\bibfnamefont {F.~L.~A.}\ \bibnamefont
  {Machado}}, \bibinfo {author} {\bibfnamefont {R.~L.}\ \bibnamefont
  {Rodr\'{\i}guez-Su\'arez}}, \bibinfo {author} {\bibfnamefont
  {A.}~\bibnamefont {Azevedo}},\ and\ \bibinfo {author} {\bibfnamefont {S.~M.}\
  \bibnamefont {Rezende}},\ }\bibfield  {title} {\bibinfo {title}
  {{Spin-Current to Charge-Current Conversion and Magnetoresistance in a Hybrid
  Structure of Graphene and Yttrium Iron Garnet}},\ }\href
  {https://doi.org/10.1103/PhysRevLett.115.226601} {\bibfield  {journal}
  {\bibinfo  {journal} {Phys. Rev. Lett.}\ }\textbf {\bibinfo {volume} {115}},\
  \bibinfo {pages} {226601} (\bibinfo {year} {2015})}\BibitemShut {NoStop}%
\bibitem [{\citenamefont {Wei}\ \emph {et~al.}(2016)\citenamefont {Wei},
  \citenamefont {Lee}, \citenamefont {Lemaitre}, \citenamefont {Pinel},
  \citenamefont {Cutaia}, \citenamefont {Cha}, \citenamefont {Katmis},
  \citenamefont {Zhu}, \citenamefont {Heiman}, \citenamefont {Hone},
  \citenamefont {Moodera},\ and\ \citenamefont {Chen}}]{Wei2016:NatMat}%
  \BibitemOpen
  \bibfield  {author} {\bibinfo {author} {\bibfnamefont {P.}~\bibnamefont
  {Wei}}, \bibinfo {author} {\bibfnamefont {S.}~\bibnamefont {Lee}}, \bibinfo
  {author} {\bibfnamefont {F.}~\bibnamefont {Lemaitre}}, \bibinfo {author}
  {\bibfnamefont {L.}~\bibnamefont {Pinel}}, \bibinfo {author} {\bibfnamefont
  {D.}~\bibnamefont {Cutaia}}, \bibinfo {author} {\bibfnamefont
  {W.}~\bibnamefont {Cha}}, \bibinfo {author} {\bibfnamefont {F.}~\bibnamefont
  {Katmis}}, \bibinfo {author} {\bibfnamefont {Y.}~\bibnamefont {Zhu}},
  \bibinfo {author} {\bibfnamefont {D.}~\bibnamefont {Heiman}}, \bibinfo
  {author} {\bibfnamefont {J.}~\bibnamefont {Hone}}, \bibinfo {author}
  {\bibfnamefont {J.~S.}\ \bibnamefont {Moodera}},\ and\ \bibinfo {author}
  {\bibfnamefont {C.-T.}\ \bibnamefont {Chen}},\ }\bibfield  {title} {\bibinfo
  {title} {{Strong interfacial exchange field in the graphene/EuS
  heterostructure}},\ }\href {https://doi.org/10.1038/nmat4603} {\bibfield
  {journal} {\bibinfo  {journal} {Nature Materials}\ }\textbf {\bibinfo
  {volume} {15}},\ \bibinfo {pages} {711} (\bibinfo {year} {2016})}\BibitemShut
  {NoStop}%
\bibitem [{\citenamefont {Dyrdał}\ and\ \citenamefont
  {Barnaś}(2017)}]{Dyrdal2017:2DMat}%
  \BibitemOpen
  \bibfield  {author} {\bibinfo {author} {\bibfnamefont {A.}~\bibnamefont
  {Dyrdał}}\ and\ \bibinfo {author} {\bibfnamefont {J.}~\bibnamefont
  {Barnaś}},\ }\bibfield  {title} {\bibinfo {title} {{Anomalous, spin, and
  valley Hall effects in graphene deposited on ferromagnetic substrates}},\
  }\href {https://doi.org/10.1088/2053-1583/aa7bac} {\bibfield  {journal}
  {\bibinfo  {journal} {2D Materials}\ }\textbf {\bibinfo {volume} {4}},\
  \bibinfo {pages} {034003} (\bibinfo {year} {2017})}\BibitemShut {NoStop}%
\bibitem [{\citenamefont {Hallal}\ \emph {et~al.}(2017)\citenamefont {Hallal},
  \citenamefont {Ibrahim}, \citenamefont {Yang}, \citenamefont {Roche},\ and\
  \citenamefont {Chshiev}}]{Hallal2017:2DMat}%
  \BibitemOpen
  \bibfield  {author} {\bibinfo {author} {\bibfnamefont {A.}~\bibnamefont
  {Hallal}}, \bibinfo {author} {\bibfnamefont {F.}~\bibnamefont {Ibrahim}},
  \bibinfo {author} {\bibfnamefont {H.}~\bibnamefont {Yang}}, \bibinfo {author}
  {\bibfnamefont {S.}~\bibnamefont {Roche}},\ and\ \bibinfo {author}
  {\bibfnamefont {M.}~\bibnamefont {Chshiev}},\ }\bibfield  {title} {\bibinfo
  {title} {Tailoring magnetic insulator proximity effects in graphene:
  first-principles calculations},\ }\href
  {https://doi.org/10.1088/2053-1583/aa6663} {\bibfield  {journal} {\bibinfo
  {journal} {2D Materials}\ }\textbf {\bibinfo {volume} {4}},\ \bibinfo {pages}
  {025074} (\bibinfo {year} {2017})}\BibitemShut {NoStop}%
\bibitem [{\citenamefont {Sierra}\ \emph {et~al.}(2021)\citenamefont {Sierra},
  \citenamefont {Fabian}, \citenamefont {Kawakami}, \citenamefont {Roche},\
  and\ \citenamefont {Valenzuela}}]{Sierra:NatNono2021}%
  \BibitemOpen
  \bibfield  {author} {\bibinfo {author} {\bibfnamefont {J.~F.}\ \bibnamefont
  {Sierra}}, \bibinfo {author} {\bibfnamefont {J.}~\bibnamefont {Fabian}},
  \bibinfo {author} {\bibfnamefont {R.~K.}\ \bibnamefont {Kawakami}}, \bibinfo
  {author} {\bibfnamefont {S.}~\bibnamefont {Roche}},\ and\ \bibinfo {author}
  {\bibfnamefont {S.~O.}\ \bibnamefont {Valenzuela}},\ }\bibfield  {title}
  {\bibinfo {title} {{Van der Waals heterostructures for spintronics and
  opto-spintronics}},\ }\bibfield  {journal} {\bibinfo  {journal} {Nature
  Nanotechnology}\ }\href {https://doi.org/10.1038/s41565-021-00936-x}
  {10.1038/s41565-021-00936-x} (\bibinfo {year} {2021})\BibitemShut {NoStop}%
\bibitem [{\citenamefont {Cao}\ \emph {et~al.}(2018)\citenamefont {Cao},
  \citenamefont {Fatemi}, \citenamefont {Fang}, \citenamefont {Watanabe},
  \citenamefont {Taniguchi}, \citenamefont {Kaxiras},\ and\ \citenamefont
  {Jarillo-Herrero}}]{Cao_2018}%
  \BibitemOpen
  \bibfield  {author} {\bibinfo {author} {\bibfnamefont {Y.}~\bibnamefont
  {Cao}}, \bibinfo {author} {\bibfnamefont {V.}~\bibnamefont {Fatemi}},
  \bibinfo {author} {\bibfnamefont {S.}~\bibnamefont {Fang}}, \bibinfo {author}
  {\bibfnamefont {K.}~\bibnamefont {Watanabe}}, \bibinfo {author}
  {\bibfnamefont {T.}~\bibnamefont {Taniguchi}}, \bibinfo {author}
  {\bibfnamefont {E.}~\bibnamefont {Kaxiras}},\ and\ \bibinfo {author}
  {\bibfnamefont {P.}~\bibnamefont {Jarillo-Herrero}},\ }\bibfield  {title}
  {\bibinfo {title} {Unconventional superconductivity in magic-angle graphene
  superlattices},\ }\href {https://doi.org/10.1038/nature26160} {\bibfield
  {journal} {\bibinfo  {journal} {Nature}\ }\textbf {\bibinfo {volume} {556}},\
  \bibinfo {pages} {43–50} (\bibinfo {year} {2018})}\BibitemShut {NoStop}%
\bibitem [{\citenamefont {Yankowitz}\ \emph {et~al.}(2019)\citenamefont
  {Yankowitz}, \citenamefont {Chen}, \citenamefont {Polshyn}, \citenamefont
  {Zhang}, \citenamefont {Watanabe}, \citenamefont {Taniguchi}, \citenamefont
  {Graf}, \citenamefont {Young},\ and\ \citenamefont {Dean}}]{Yankowitz2019}%
  \BibitemOpen
  \bibfield  {author} {\bibinfo {author} {\bibfnamefont {M.}~\bibnamefont
  {Yankowitz}}, \bibinfo {author} {\bibfnamefont {S.}~\bibnamefont {Chen}},
  \bibinfo {author} {\bibfnamefont {H.}~\bibnamefont {Polshyn}}, \bibinfo
  {author} {\bibfnamefont {Y.}~\bibnamefont {Zhang}}, \bibinfo {author}
  {\bibfnamefont {K.}~\bibnamefont {Watanabe}}, \bibinfo {author}
  {\bibfnamefont {T.}~\bibnamefont {Taniguchi}}, \bibinfo {author}
  {\bibfnamefont {D.}~\bibnamefont {Graf}}, \bibinfo {author} {\bibfnamefont
  {A.~F.}\ \bibnamefont {Young}},\ and\ \bibinfo {author} {\bibfnamefont
  {C.~R.}\ \bibnamefont {Dean}},\ }\bibfield  {title} {\bibinfo {title}
  {{Tuning superconductivity in twisted bilayer graphene}},\ }\href
  {https://doi.org/10.1126/science.aav1910} {\bibfield  {journal} {\bibinfo
  {journal} {Science}\ }\textbf {\bibinfo {volume} {363}},\ \bibinfo {pages}
  {1059} (\bibinfo {year} {2019})}\BibitemShut {NoStop}%
\bibitem [{\citenamefont {Eschrig}(2011)}]{EschrigPhysToday2011}%
  \BibitemOpen
  \bibfield  {author} {\bibinfo {author} {\bibfnamefont {M.}~\bibnamefont
  {Eschrig}},\ }\bibfield  {title} {\bibinfo {title} {{Spin-polarized
  supercurrents for spintronics}},\ }\href {https://doi.org/10.1063/1.3541944}
  {\bibfield  {journal} {\bibinfo  {journal} {Physics Today}\ }\textbf
  {\bibinfo {volume} {64}},\ \bibinfo {pages} {43} (\bibinfo {year}
  {2011})}\BibitemShut {NoStop}%
\bibitem [{\citenamefont {Eschrig}(2015)}]{EschrigRepProgPhys2015}%
  \BibitemOpen
  \bibfield  {author} {\bibinfo {author} {\bibfnamefont {M.}~\bibnamefont
  {Eschrig}},\ }\bibfield  {title} {\bibinfo {title} {{Spin-polarized
  supercurrents for spintronics: a review of current progress}},\ }\href
  {https://doi.org/10.1088/0034-4885/78/10/104501} {\bibfield  {journal}
  {\bibinfo  {journal} {Reports on Progress in Physics}\ }\textbf {\bibinfo
  {volume} {78}},\ \bibinfo {pages} {104501} (\bibinfo {year}
  {2015})}\BibitemShut {NoStop}%
\bibitem [{\citenamefont {Linder}\ and\ \citenamefont
  {Robinson}(2015)}]{LinderRobinson_NP2015}%
  \BibitemOpen
  \bibfield  {author} {\bibinfo {author} {\bibfnamefont {J.}~\bibnamefont
  {Linder}}\ and\ \bibinfo {author} {\bibfnamefont {J.~W.~A.}\ \bibnamefont
  {Robinson}},\ }\bibfield  {title} {\bibinfo {title} {{Superconducting
  spintronics}},\ }\href {https://doi.org/10.1038/nphys3242} {\bibfield
  {journal} {\bibinfo  {journal} {Nature Physics}\ }\textbf {\bibinfo {volume}
  {11}},\ \bibinfo {pages} {307} (\bibinfo {year} {2015})}\BibitemShut
  {NoStop}%
\bibitem [{\citenamefont {Yang}\ \emph {et~al.}(2021)\citenamefont {Yang},
  \citenamefont {Ciccarelli},\ and\ \citenamefont {Robinson}}]{Guang:APL2021}%
  \BibitemOpen
  \bibfield  {author} {\bibinfo {author} {\bibfnamefont {G.}~\bibnamefont
  {Yang}}, \bibinfo {author} {\bibfnamefont {C.}~\bibnamefont {Ciccarelli}},\
  and\ \bibinfo {author} {\bibfnamefont {J.~W.~A.}\ \bibnamefont {Robinson}},\
  }\bibfield  {title} {\bibinfo {title} {{Boosting spintronics with
  superconductivity}},\ }\href {https://doi.org/10.1063/5.0048904} {\bibfield
  {journal} {\bibinfo  {journal} {APL Materials}\ }\textbf {\bibinfo {volume}
  {9}},\ \bibinfo {pages} {050703} (\bibinfo {year} {2021})}\BibitemShut
  {NoStop}%
\bibitem [{\citenamefont {Heersche}\ \emph {et~al.}(2007)\citenamefont
  {Heersche}, \citenamefont {Jarillo-Herrero}, \citenamefont {Oostinga},
  \citenamefont {Vandersypen},\ and\ \citenamefont {Morpurgo}}]{Heersche2007a}%
  \BibitemOpen
  \bibfield  {author} {\bibinfo {author} {\bibfnamefont {H.~B.}\ \bibnamefont
  {Heersche}}, \bibinfo {author} {\bibfnamefont {P.}~\bibnamefont
  {Jarillo-Herrero}}, \bibinfo {author} {\bibfnamefont {J.~B.}\ \bibnamefont
  {Oostinga}}, \bibinfo {author} {\bibfnamefont {L.~M.~K.}\ \bibnamefont
  {Vandersypen}},\ and\ \bibinfo {author} {\bibfnamefont {A.~F.}\ \bibnamefont
  {Morpurgo}},\ }\bibfield  {title} {\bibinfo {title} {{Bipolar supercurrent in
  graphene}},\ }\href {https://doi.org/10.1038/nature05555} {\bibfield
  {journal} {\bibinfo  {journal} {Nature}\ }\textbf {\bibinfo {volume} {446}},\
  \bibinfo {pages} {56} (\bibinfo {year} {2007})}\BibitemShut {NoStop}%
\bibitem [{\citenamefont {Komatsu}\ \emph {et~al.}(2012)\citenamefont
  {Komatsu}, \citenamefont {Li}, \citenamefont {Autier-Laurent}, \citenamefont
  {Bouchiat},\ and\ \citenamefont {Gu{\'{e}}ron}}]{Komatsu2012a}%
  \BibitemOpen
  \bibfield  {author} {\bibinfo {author} {\bibfnamefont {K.}~\bibnamefont
  {Komatsu}}, \bibinfo {author} {\bibfnamefont {C.}~\bibnamefont {Li}},
  \bibinfo {author} {\bibfnamefont {S.}~\bibnamefont {Autier-Laurent}},
  \bibinfo {author} {\bibfnamefont {H.}~\bibnamefont {Bouchiat}},\ and\
  \bibinfo {author} {\bibfnamefont {S.}~\bibnamefont {Gu{\'{e}}ron}},\
  }\bibfield  {title} {\bibinfo {title} {{Superconducting proximity effect in
  long superconductor/graphene/superconductor junctions: From specular Andreev
  reflection at zero field to the quantum Hall regime}},\ }\href
  {https://doi.org/10.1103/PhysRevB.86.115412} {\bibfield  {journal} {\bibinfo
  {journal} {Physical Review B}\ }\textbf {\bibinfo {volume} {86}},\ \bibinfo
  {pages} {115412} (\bibinfo {year} {2012})}\BibitemShut {NoStop}%
\bibitem [{\citenamefont {Calado}\ \emph {et~al.}(2015)\citenamefont {Calado},
  \citenamefont {Goswami}, \citenamefont {Nanda}, \citenamefont {Diez},
  \citenamefont {Akhmerov}, \citenamefont {Watanabe}, \citenamefont
  {Taniguchi}, \citenamefont {Klapwijk},\ and\ \citenamefont
  {Vandersypen}}]{Calado2015}%
  \BibitemOpen
  \bibfield  {author} {\bibinfo {author} {\bibfnamefont {V.~E.}\ \bibnamefont
  {Calado}}, \bibinfo {author} {\bibfnamefont {S.}~\bibnamefont {Goswami}},
  \bibinfo {author} {\bibfnamefont {G.}~\bibnamefont {Nanda}}, \bibinfo
  {author} {\bibfnamefont {M.}~\bibnamefont {Diez}}, \bibinfo {author}
  {\bibfnamefont {A.~R.}\ \bibnamefont {Akhmerov}}, \bibinfo {author}
  {\bibfnamefont {K.}~\bibnamefont {Watanabe}}, \bibinfo {author}
  {\bibfnamefont {T.}~\bibnamefont {Taniguchi}}, \bibinfo {author}
  {\bibfnamefont {T.~M.}\ \bibnamefont {Klapwijk}},\ and\ \bibinfo {author}
  {\bibfnamefont {L.~M.~K.}\ \bibnamefont {Vandersypen}},\ }\bibfield  {title}
  {\bibinfo {title} {{Ballistic Josephson junctions in edge-contacted
  graphene}},\ }\href {https://doi.org/10.1038/nnano.2015.156} {\bibfield
  {journal} {\bibinfo  {journal} {Nature Nanotechnology}\ }\textbf {\bibinfo
  {volume} {10}},\ \bibinfo {pages} {761} (\bibinfo {year} {2015})}\BibitemShut
  {NoStop}%
\bibitem [{\citenamefont {Indolese}\ \emph {et~al.}(2018)\citenamefont
  {Indolese}, \citenamefont {Delagrange}, \citenamefont {Makk}, \citenamefont
  {Wallbank}, \citenamefont {Wanatabe}, \citenamefont {Taniguchi},\ and\
  \citenamefont {Sch{\"{o}}nenberger}}]{Delagrange_PRL_2018}%
  \BibitemOpen
  \bibfield  {author} {\bibinfo {author} {\bibfnamefont {D.~I.}\ \bibnamefont
  {Indolese}}, \bibinfo {author} {\bibfnamefont {R.}~\bibnamefont
  {Delagrange}}, \bibinfo {author} {\bibfnamefont {P.}~\bibnamefont {Makk}},
  \bibinfo {author} {\bibfnamefont {J.~R.}\ \bibnamefont {Wallbank}}, \bibinfo
  {author} {\bibfnamefont {K.}~\bibnamefont {Wanatabe}}, \bibinfo {author}
  {\bibfnamefont {T.}~\bibnamefont {Taniguchi}},\ and\ \bibinfo {author}
  {\bibfnamefont {C.}~\bibnamefont {Sch{\"{o}}nenberger}},\ }\bibfield  {title}
  {\bibinfo {title} {{Signatures of van Hove Singularities Probed by the
  Supercurrent in a Graphene-hBN Superlattice}},\ }\href
  {https://doi.org/10.1103/PhysRevLett.121.137701} {\bibfield  {journal}
  {\bibinfo  {journal} {Physical Review Letters}\ }\textbf {\bibinfo {volume}
  {121}},\ \bibinfo {pages} {137701} (\bibinfo {year} {2018})}\BibitemShut
  {NoStop}%
\bibitem [{\citenamefont {Li}\ \emph {et~al.}(2013)\citenamefont {Li},
  \citenamefont {Feng}, \citenamefont {Zhang}, \citenamefont {Ou},
  \citenamefont {Chen}, \citenamefont {He}, \citenamefont {Wang}, \citenamefont
  {Guo}, \citenamefont {Liu}, \citenamefont {Xue},\ and\ \citenamefont
  {Ma}}]{Li2013}%
  \BibitemOpen
  \bibfield  {author} {\bibinfo {author} {\bibfnamefont {K.}~\bibnamefont
  {Li}}, \bibinfo {author} {\bibfnamefont {X.}~\bibnamefont {Feng}}, \bibinfo
  {author} {\bibfnamefont {W.}~\bibnamefont {Zhang}}, \bibinfo {author}
  {\bibfnamefont {Y.}~\bibnamefont {Ou}}, \bibinfo {author} {\bibfnamefont
  {L.}~\bibnamefont {Chen}}, \bibinfo {author} {\bibfnamefont {K.}~\bibnamefont
  {He}}, \bibinfo {author} {\bibfnamefont {L.-L.}\ \bibnamefont {Wang}},
  \bibinfo {author} {\bibfnamefont {L.}~\bibnamefont {Guo}}, \bibinfo {author}
  {\bibfnamefont {G.}~\bibnamefont {Liu}}, \bibinfo {author} {\bibfnamefont
  {Q.-K.}\ \bibnamefont {Xue}},\ and\ \bibinfo {author} {\bibfnamefont
  {X.}~\bibnamefont {Ma}},\ }\bibfield  {title} {\bibinfo {title}
  {{Superconductivity in Ca-intercalated epitaxial graphene on silicon
  carbide}},\ }\href {https://doi.org/10.1063/1.4817781} {\bibfield  {journal}
  {\bibinfo  {journal} {Applied Physics Letters}\ }\textbf {\bibinfo {volume}
  {103}},\ \bibinfo {pages} {062601} (\bibinfo {year} {2013})}\BibitemShut
  {NoStop}%
\bibitem [{\citenamefont {Ludbrook}\ \emph {et~al.}(2015)\citenamefont
  {Ludbrook}, \citenamefont {Levy}, \citenamefont {Nigge}, \citenamefont
  {Zonno}, \citenamefont {Schneider}, \citenamefont {Dvorak}, \citenamefont
  {Veenstra}, \citenamefont {Zhdanovich}, \citenamefont {Wong}, \citenamefont
  {Dosanjh}, \citenamefont {Stra{\ss}er}, \citenamefont {St{\"{o}}hr},
  \citenamefont {Forti}, \citenamefont {Ast}, \citenamefont {Starke},\ and\
  \citenamefont {Damascelli}}]{Ludbrook2015}%
  \BibitemOpen
  \bibfield  {author} {\bibinfo {author} {\bibfnamefont {B.~M.}\ \bibnamefont
  {Ludbrook}}, \bibinfo {author} {\bibfnamefont {G.}~\bibnamefont {Levy}},
  \bibinfo {author} {\bibfnamefont {P.}~\bibnamefont {Nigge}}, \bibinfo
  {author} {\bibfnamefont {M.}~\bibnamefont {Zonno}}, \bibinfo {author}
  {\bibfnamefont {M.}~\bibnamefont {Schneider}}, \bibinfo {author}
  {\bibfnamefont {D.~J.}\ \bibnamefont {Dvorak}}, \bibinfo {author}
  {\bibfnamefont {C.~N.}\ \bibnamefont {Veenstra}}, \bibinfo {author}
  {\bibfnamefont {S.}~\bibnamefont {Zhdanovich}}, \bibinfo {author}
  {\bibfnamefont {D.}~\bibnamefont {Wong}}, \bibinfo {author} {\bibfnamefont
  {P.}~\bibnamefont {Dosanjh}}, \bibinfo {author} {\bibfnamefont
  {C.}~\bibnamefont {Stra{\ss}er}}, \bibinfo {author} {\bibfnamefont
  {A.}~\bibnamefont {St{\"{o}}hr}}, \bibinfo {author} {\bibfnamefont
  {S.}~\bibnamefont {Forti}}, \bibinfo {author} {\bibfnamefont {C.~R.}\
  \bibnamefont {Ast}}, \bibinfo {author} {\bibfnamefont {U.}~\bibnamefont
  {Starke}},\ and\ \bibinfo {author} {\bibfnamefont {A.}~\bibnamefont
  {Damascelli}},\ }\bibfield  {title} {\bibinfo {title} {{Evidence for
  superconductivity in Li-decorated monolayer graphene}},\ }\href
  {https://doi.org/10.1073/pnas.1510435112} {\bibfield  {journal} {\bibinfo
  {journal} {Proceedings of the National Academy of Sciences}\ }\textbf
  {\bibinfo {volume} {112}},\ \bibinfo {pages} {11795} (\bibinfo {year}
  {2015})}\BibitemShut {NoStop}%
\bibitem [{\citenamefont {Chapman}\ \emph {et~al.}(2016)\citenamefont
  {Chapman}, \citenamefont {Su}, \citenamefont {Howard}, \citenamefont
  {Kundys}, \citenamefont {Grigorenko}, \citenamefont {Guinea}, \citenamefont
  {Geim}, \citenamefont {Grigorieva},\ and\ \citenamefont
  {Nair}}]{Chapman2016}%
  \BibitemOpen
  \bibfield  {author} {\bibinfo {author} {\bibfnamefont {J.}~\bibnamefont
  {Chapman}}, \bibinfo {author} {\bibfnamefont {Y.}~\bibnamefont {Su}},
  \bibinfo {author} {\bibfnamefont {C.~A.}\ \bibnamefont {Howard}}, \bibinfo
  {author} {\bibfnamefont {D.}~\bibnamefont {Kundys}}, \bibinfo {author}
  {\bibfnamefont {A.~N.}\ \bibnamefont {Grigorenko}}, \bibinfo {author}
  {\bibfnamefont {F.}~\bibnamefont {Guinea}}, \bibinfo {author} {\bibfnamefont
  {A.~K.}\ \bibnamefont {Geim}}, \bibinfo {author} {\bibfnamefont {I.~V.}\
  \bibnamefont {Grigorieva}},\ and\ \bibinfo {author} {\bibfnamefont {R.~R.}\
  \bibnamefont {Nair}},\ }\bibfield  {title} {\bibinfo {title}
  {{Superconductivity in Ca-doped graphene laminates}},\ }\href
  {https://doi.org/10.1038/srep23254} {\bibfield  {journal} {\bibinfo
  {journal} {Scientific Reports}\ }\textbf {\bibinfo {volume} {6}},\ \bibinfo
  {pages} {23254} (\bibinfo {year} {2016})}\BibitemShut {NoStop}%
\bibitem [{\citenamefont {Tonnoir}\ \emph {et~al.}(2013)\citenamefont
  {Tonnoir}, \citenamefont {Kimouche}, \citenamefont {Coraux}, \citenamefont
  {Magaud}, \citenamefont {Delsol}, \citenamefont {Gilles},\ and\ \citenamefont
  {Chapelier}}]{Tonnoir2013}%
  \BibitemOpen
  \bibfield  {author} {\bibinfo {author} {\bibfnamefont {C.}~\bibnamefont
  {Tonnoir}}, \bibinfo {author} {\bibfnamefont {A.}~\bibnamefont {Kimouche}},
  \bibinfo {author} {\bibfnamefont {J.}~\bibnamefont {Coraux}}, \bibinfo
  {author} {\bibfnamefont {L.}~\bibnamefont {Magaud}}, \bibinfo {author}
  {\bibfnamefont {B.}~\bibnamefont {Delsol}}, \bibinfo {author} {\bibfnamefont
  {B.}~\bibnamefont {Gilles}},\ and\ \bibinfo {author} {\bibfnamefont
  {C.}~\bibnamefont {Chapelier}},\ }\bibfield  {title} {\bibinfo {title}
  {{Induced Superconductivity in Graphene Grown on Rhenium}},\ }\href
  {https://doi.org/10.1103/PhysRevLett.111.246805} {\bibfield  {journal}
  {\bibinfo  {journal} {Physical Review Letters}\ }\textbf {\bibinfo {volume}
  {111}},\ \bibinfo {pages} {246805} (\bibinfo {year} {2013})}\BibitemShut
  {NoStop}%
\bibitem [{\citenamefont {Di~Bernardo}\ \emph {et~al.}(2017)\citenamefont
  {Di~Bernardo}, \citenamefont {Millo}, \citenamefont {Barbone}, \citenamefont
  {Alpern}, \citenamefont {Kalcheim}, \citenamefont {Sassi}, \citenamefont
  {Ott}, \citenamefont {De~Fazio}, \citenamefont {Yoon}, \citenamefont {Amado},
  \citenamefont {Ferrari}, \citenamefont {Linder},\ and\ \citenamefont
  {Robinson}}]{DiBernardo2017_p-wave-SCG}%
  \BibitemOpen
  \bibfield  {author} {\bibinfo {author} {\bibfnamefont {A.}~\bibnamefont
  {Di~Bernardo}}, \bibinfo {author} {\bibfnamefont {O.}~\bibnamefont {Millo}},
  \bibinfo {author} {\bibfnamefont {M.}~\bibnamefont {Barbone}}, \bibinfo
  {author} {\bibfnamefont {H.}~\bibnamefont {Alpern}}, \bibinfo {author}
  {\bibfnamefont {Y.}~\bibnamefont {Kalcheim}}, \bibinfo {author}
  {\bibfnamefont {U.}~\bibnamefont {Sassi}}, \bibinfo {author} {\bibfnamefont
  {A.~K.}\ \bibnamefont {Ott}}, \bibinfo {author} {\bibfnamefont
  {D.}~\bibnamefont {De~Fazio}}, \bibinfo {author} {\bibfnamefont
  {D.}~\bibnamefont {Yoon}}, \bibinfo {author} {\bibfnamefont {M.}~\bibnamefont
  {Amado}}, \bibinfo {author} {\bibfnamefont {A.~C.}\ \bibnamefont {Ferrari}},
  \bibinfo {author} {\bibfnamefont {J.}~\bibnamefont {Linder}},\ and\ \bibinfo
  {author} {\bibfnamefont {J.~W.~A.}\ \bibnamefont {Robinson}},\ }\bibfield
  {title} {\bibinfo {title} {{p-wave triggered superconductivity in
  single-layer graphene on an electron-doped oxide superconductor}},\ }\href
  {https://doi.org/10.1038/ncomms14024} {\bibfield  {journal} {\bibinfo
  {journal} {Nature Communications}\ }\textbf {\bibinfo {volume} {8}},\
  \bibinfo {pages} {14024} (\bibinfo {year} {2017})}\BibitemShut {NoStop}%
\bibitem [{\citenamefont {Schrieffer}(1964)}]{Schrieffer-book:1964}%
  \BibitemOpen
  \bibfield  {author} {\bibinfo {author} {\bibfnamefont {J.~R.}\ \bibnamefont
  {Schrieffer}},\ }\href@noop {} {\emph {\bibinfo {title} {{Theory of
  Superconductivity}}}}\ (\bibinfo  {publisher} {Benjamin},\ \bibinfo {address}
  {New York},\ \bibinfo {year} {1964})\BibitemShut {NoStop}%
\bibitem [{\citenamefont {Yafet}(1983)}]{Yafet1983}%
  \BibitemOpen
  \bibfield  {author} {\bibinfo {author} {\bibfnamefont {Y.}~\bibnamefont
  {Yafet}},\ }\bibfield  {title} {\bibinfo {title} {{Conduction electron spin
  relaxation in the superconducting state}},\ }\href
  {https://doi.org/10.1016/0375-9601(83)90874-5} {\bibfield  {journal}
  {\bibinfo  {journal} {Physics Letters A}\ }\textbf {\bibinfo {volume} {98}},\
  \bibinfo {pages} {287} (\bibinfo {year} {1983})}\BibitemShut {NoStop}%
\bibitem [{\citenamefont {Kochan}\ \emph {et~al.}(2020)\citenamefont {Kochan},
  \citenamefont {Barth}, \citenamefont {Costa}, \citenamefont {Richter},\ and\
  \citenamefont {Fabian}}]{Kochan:PRL_sc_graphene_spin_relaxation}%
  \BibitemOpen
  \bibfield  {author} {\bibinfo {author} {\bibfnamefont {D.}~\bibnamefont
  {Kochan}}, \bibinfo {author} {\bibfnamefont {M.}~\bibnamefont {Barth}},
  \bibinfo {author} {\bibfnamefont {A.}~\bibnamefont {Costa}}, \bibinfo
  {author} {\bibfnamefont {K.}~\bibnamefont {Richter}},\ and\ \bibinfo {author}
  {\bibfnamefont {J.}~\bibnamefont {Fabian}},\ }\bibfield  {title} {\bibinfo
  {title} {{Spin Relaxation in $s$-Wave Superconductors in the Presence of
  Resonant Spin-Flip Scatterers}},\ }\href
  {https://doi.org/10.1103/PhysRevLett.125.087001} {\bibfield  {journal}
  {\bibinfo  {journal} {Phys. Rev. Lett.}\ }\textbf {\bibinfo {volume} {125}},\
  \bibinfo {pages} {087001} (\bibinfo {year} {2020})}\BibitemShut {NoStop}%
\bibitem [{\citenamefont {Yang}\ \emph {et~al.}(2010)\citenamefont {Yang},
  \citenamefont {Yang}, \citenamefont {Takahashi}, \citenamefont {Maekawa},\
  and\ \citenamefont {Parkin}}]{Yang_Parkin2010}%
  \BibitemOpen
  \bibfield  {author} {\bibinfo {author} {\bibfnamefont {H.}~\bibnamefont
  {Yang}}, \bibinfo {author} {\bibfnamefont {S.-H.}\ \bibnamefont {Yang}},
  \bibinfo {author} {\bibfnamefont {S.}~\bibnamefont {Takahashi}}, \bibinfo
  {author} {\bibfnamefont {S.}~\bibnamefont {Maekawa}},\ and\ \bibinfo {author}
  {\bibfnamefont {S.~S.~P.}\ \bibnamefont {Parkin}},\ }\bibfield  {title}
  {\bibinfo {title} {{Extremely long quasiparticle spin lifetimes in
  superconducting aluminium using MgO tunnel spin injectors}},\ }\href
  {https://doi.org/10.1038/nmat2781} {\bibfield  {journal} {\bibinfo  {journal}
  {Nature Materials}\ }\textbf {\bibinfo {volume} {9}},\ \bibinfo {pages} {586}
  (\bibinfo {year} {2010})}\BibitemShut {NoStop}%
\bibitem [{\citenamefont {H{\"{u}}bler}\ \emph {et~al.}(2012)\citenamefont
  {H{\"{u}}bler}, \citenamefont {Wolf}, \citenamefont {Beckmann},\ and\
  \citenamefont {V.~L{\"{o}}hneysen}}]{Hubler:PRL2012}%
  \BibitemOpen
  \bibfield  {author} {\bibinfo {author} {\bibfnamefont {F.}~\bibnamefont
  {H{\"{u}}bler}}, \bibinfo {author} {\bibfnamefont {M.~J.}\ \bibnamefont
  {Wolf}}, \bibinfo {author} {\bibfnamefont {D.}~\bibnamefont {Beckmann}},\
  and\ \bibinfo {author} {\bibfnamefont {H.}~\bibnamefont
  {V.~L{\"{o}}hneysen}},\ }\bibfield  {title} {\bibinfo {title} {{Long-range
  spin-polarized quasiparticle transport in mesoscopic al superconductors with
  a zeeman splitting}},\ }\href
  {https://doi.org/10.1103/PhysRevLett.109.207001} {\bibfield  {journal}
  {\bibinfo  {journal} {Physical Review Letters}\ }\textbf {\bibinfo {volume}
  {109}},\ \bibinfo {pages} {207001} (\bibinfo {year} {2012})}\BibitemShut
  {NoStop}%
\bibitem [{\citenamefont {Quay}\ \emph {et~al.}(2015)\citenamefont {Quay},
  \citenamefont {Weideneder}, \citenamefont {Chiffaudel}, \citenamefont
  {Strunk},\ and\ \citenamefont {Aprili}}]{Quay_Aprili_Strunk_2015}%
  \BibitemOpen
  \bibfield  {author} {\bibinfo {author} {\bibfnamefont {C.~H.~L.}\
  \bibnamefont {Quay}}, \bibinfo {author} {\bibfnamefont {M.}~\bibnamefont
  {Weideneder}}, \bibinfo {author} {\bibfnamefont {Y.}~\bibnamefont
  {Chiffaudel}}, \bibinfo {author} {\bibfnamefont {C.}~\bibnamefont {Strunk}},\
  and\ \bibinfo {author} {\bibfnamefont {M.}~\bibnamefont {Aprili}},\
  }\bibfield  {title} {\bibinfo {title} {{Quasiparticle spin resonance and
  coherence in superconducting aluminium}},\ }\href
  {https://doi.org/10.1038/ncomms9660} {\bibfield  {journal} {\bibinfo
  {journal} {Nature Communications}\ }\textbf {\bibinfo {volume} {6}},\
  \bibinfo {pages} {8660} (\bibinfo {year} {2015})}\BibitemShut {NoStop}%
\bibitem [{\citenamefont {Hebel}\ and\ \citenamefont
  {Slichter}(1957)}]{Hebel1957}%
  \BibitemOpen
  \bibfield  {author} {\bibinfo {author} {\bibfnamefont {L.~C.}\ \bibnamefont
  {Hebel}}\ and\ \bibinfo {author} {\bibfnamefont {C.~P.}\ \bibnamefont
  {Slichter}},\ }\bibfield  {title} {\bibinfo {title} {{Nuclear Relaxation in
  Superconducting Aluminum}},\ }\href {https://doi.org/10.1103/PhysRev.107.901}
  {\bibfield  {journal} {\bibinfo  {journal} {Physical Review}\ }\textbf
  {\bibinfo {volume} {107}},\ \bibinfo {pages} {901} (\bibinfo {year}
  {1957})}\BibitemShut {NoStop}%
\bibitem [{\citenamefont {Poli}\ \emph {et~al.}(2008)\citenamefont {Poli},
  \citenamefont {Morten}, \citenamefont {Urech}, \citenamefont {Brataas},
  \citenamefont {Haviland},\ and\ \citenamefont {Korenivski}}]{Poli2008}%
  \BibitemOpen
  \bibfield  {author} {\bibinfo {author} {\bibfnamefont {N.}~\bibnamefont
  {Poli}}, \bibinfo {author} {\bibfnamefont {J.~P.}\ \bibnamefont {Morten}},
  \bibinfo {author} {\bibfnamefont {M.}~\bibnamefont {Urech}}, \bibinfo
  {author} {\bibfnamefont {A.}~\bibnamefont {Brataas}}, \bibinfo {author}
  {\bibfnamefont {D.~B.}\ \bibnamefont {Haviland}},\ and\ \bibinfo {author}
  {\bibfnamefont {V.}~\bibnamefont {Korenivski}},\ }\bibfield  {title}
  {\bibinfo {title} {{Spin Injection and Relaxation in a Mesoscopic
  Superconductor}},\ }\href {https://doi.org/10.1103/PhysRevLett.100.136601}
  {\bibfield  {journal} {\bibinfo  {journal} {Physical Review Letters}\
  }\textbf {\bibinfo {volume} {100}},\ \bibinfo {pages} {136601} (\bibinfo
  {year} {2008})}\BibitemShut {NoStop}%
\bibitem [{\citenamefont {Hebel}\ and\ \citenamefont
  {Slichter}(1959)}]{Hebel1959}%
  \BibitemOpen
  \bibfield  {author} {\bibinfo {author} {\bibfnamefont {L.~C.}\ \bibnamefont
  {Hebel}}\ and\ \bibinfo {author} {\bibfnamefont {C.~P.}\ \bibnamefont
  {Slichter}},\ }\bibfield  {title} {\bibinfo {title} {{Nuclear Spin Relaxation
  in Normal and Superconducting Aluminum}},\ }\href
  {https://doi.org/10.1103/PhysRev.113.1504} {\bibfield  {journal} {\bibinfo
  {journal} {Physical Review}\ }\textbf {\bibinfo {volume} {113}},\ \bibinfo
  {pages} {1504} (\bibinfo {year} {1959})}\BibitemShut {NoStop}%
\bibitem [{\citenamefont {Hebel}(1959)}]{Hebel1959-Theory}%
  \BibitemOpen
  \bibfield  {author} {\bibinfo {author} {\bibfnamefont {L.~C.}\ \bibnamefont
  {Hebel}},\ }\bibfield  {title} {\bibinfo {title} {{Theory of Nuclear Spin
  Relaxation in Superconductors}},\ }\href
  {https://doi.org/10.1103/PhysRev.116.79} {\bibfield  {journal} {\bibinfo
  {journal} {Physical Review}\ }\textbf {\bibinfo {volume} {116}},\ \bibinfo
  {pages} {79} (\bibinfo {year} {1959})}\BibitemShut {NoStop}%
\bibitem [{\citenamefont {Cavanagh}\ and\ \citenamefont
  {Powell}(2021)}]{Cavanagh_2021}%
  \BibitemOpen
  \bibfield  {author} {\bibinfo {author} {\bibfnamefont {D.~C.}\ \bibnamefont
  {Cavanagh}}\ and\ \bibinfo {author} {\bibfnamefont {B.~J.}\ \bibnamefont
  {Powell}},\ }\bibfield  {title} {\bibinfo {title} {{Fate of the
  Hebel-Slichter peak in superconductors with strong antiferromagnetic
  fluctuations}},\ }\href {https://doi.org/10.1103/PhysRevResearch.3.013241}
  {\bibfield  {journal} {\bibinfo  {journal} {Phys. Rev. Research}\ }\textbf
  {\bibinfo {volume} {3}},\ \bibinfo {pages} {013241} (\bibinfo {year}
  {2021})}\BibitemShut {NoStop}%
\bibitem [{\citenamefont {Yu}(1965)}]{Yu1965}%
  \BibitemOpen
  \bibfield  {author} {\bibinfo {author} {\bibfnamefont {L.}~\bibnamefont
  {Yu}},\ }\bibfield  {title} {\bibinfo {title} {{Bound State in
  Superconductors with Paramagnetic Impurities}},\ }\href@noop {} {\bibfield
  {journal} {\bibinfo  {journal} {Acta Physica Sinica}\ }\textbf {\bibinfo
  {volume} {21}},\ \bibinfo {pages} {75} (\bibinfo {year} {1965})}\BibitemShut
  {NoStop}%
\bibitem [{\citenamefont {Shiba}(1968)}]{Shiba1968}%
  \BibitemOpen
  \bibfield  {author} {\bibinfo {author} {\bibfnamefont {H.}~\bibnamefont
  {Shiba}},\ }\bibfield  {title} {\bibinfo {title} {{Classical Spins in
  Superconductors}},\ }\href {https://doi.org/10.1143/PTP.40.435} {\bibfield
  {journal} {\bibinfo  {journal} {Progress of Theoretical Physics}\ }\textbf
  {\bibinfo {volume} {40}},\ \bibinfo {pages} {435} (\bibinfo {year}
  {1968})}\BibitemShut {NoStop}%
\bibitem [{\citenamefont {Rusinov}(1968)}]{Rusinov1968}%
  \BibitemOpen
  \bibfield  {author} {\bibinfo {author} {\bibfnamefont {A.~I.}\ \bibnamefont
  {Rusinov}},\ }\bibfield  {title} {\bibinfo {title} {{Superconductivity near a
  paramagnetic impurity}},\ }\href@noop {} {\bibfield  {journal} {\bibinfo
  {journal} {Zh. Eksp. Teor. Fiz.}\ }\textbf {\bibinfo {volume} {9}},\ \bibinfo
  {pages} {146} (\bibinfo {year} {1968})}\BibitemShut {NoStop}%
\bibitem [{\citenamefont {Wehling}\ \emph {et~al.}(2008)\citenamefont
  {Wehling}, \citenamefont {Dahal}, \citenamefont {Lichtenstein},\ and\
  \citenamefont {Balatsky}}]{Wehling2008}%
  \BibitemOpen
  \bibfield  {author} {\bibinfo {author} {\bibfnamefont {T.~O.}\ \bibnamefont
  {Wehling}}, \bibinfo {author} {\bibfnamefont {H.~P.}\ \bibnamefont {Dahal}},
  \bibinfo {author} {\bibfnamefont {A.~I.}\ \bibnamefont {Lichtenstein}},\ and\
  \bibinfo {author} {\bibfnamefont {A.~V.}\ \bibnamefont {Balatsky}},\
  }\bibfield  {title} {\bibinfo {title} {{Local impurity effects in
  superconducting graphene}},\ }\href
  {https://doi.org/10.1103/PhysRevB.78.035414} {\bibfield  {journal} {\bibinfo
  {journal} {Physical Review B}\ }\textbf {\bibinfo {volume} {78}},\ \bibinfo
  {pages} {035414} (\bibinfo {year} {2008})}\BibitemShut {NoStop}%
\bibitem [{\citenamefont {Lado}\ and\ \citenamefont
  {Fern{\'{a}}ndez-Rossier}(2016)}]{Lado2016}%
  \BibitemOpen
  \bibfield  {author} {\bibinfo {author} {\bibfnamefont {J.~L.}\ \bibnamefont
  {Lado}}\ and\ \bibinfo {author} {\bibfnamefont {J.}~\bibnamefont
  {Fern{\'{a}}ndez-Rossier}},\ }\bibfield  {title} {\bibinfo {title}
  {{Unconventional Yu-Shiba-Rusinov states in hydrogenated graphene}},\ }\href
  {https://doi.org/10.1088/2053-1583/3/2/025001} {\bibfield  {journal}
  {\bibinfo  {journal} {2D Materials}\ }\textbf {\bibinfo {volume} {3}},\
  \bibinfo {pages} {0} (\bibinfo {year} {2016})}\BibitemShut {NoStop}%
\bibitem [{\citenamefont {Cortés-del Río}\ \emph {et~al.}(2021)\citenamefont
  {Cortés-del Río}, \citenamefont {Lado}, \citenamefont {Cherkez},
  \citenamefont {Mallet}, \citenamefont {Veuillen}, \citenamefont {Cuevas},
  \citenamefont {Gómez-Rodríguez}, \citenamefont {Fernández-Rossier},\ and\
  \citenamefont {Brihuega}}]{Brihuega:AdvMaterials2021}%
  \BibitemOpen
  \bibfield  {author} {\bibinfo {author} {\bibfnamefont {E.}~\bibnamefont
  {Cortés-del Río}}, \bibinfo {author} {\bibfnamefont {J.~L.}\ \bibnamefont
  {Lado}}, \bibinfo {author} {\bibfnamefont {V.}~\bibnamefont {Cherkez}},
  \bibinfo {author} {\bibfnamefont {P.}~\bibnamefont {Mallet}}, \bibinfo
  {author} {\bibfnamefont {J.-Y.}\ \bibnamefont {Veuillen}}, \bibinfo {author}
  {\bibfnamefont {J.~C.}\ \bibnamefont {Cuevas}}, \bibinfo {author}
  {\bibfnamefont {J.~M.}\ \bibnamefont {Gómez-Rodríguez}}, \bibinfo {author}
  {\bibfnamefont {J.}~\bibnamefont {Fernández-Rossier}},\ and\ \bibinfo
  {author} {\bibfnamefont {I.}~\bibnamefont {Brihuega}},\ }\bibfield  {title}
  {\bibinfo {title} {{Observation of Yu–Shiba–Rusinov States in
  Superconducting Graphene}},\ }\href
  {https://doi.org/https://doi.org/10.1002/adma.202008113} {\bibfield
  {journal} {\bibinfo  {journal} {Advanced Materials}\ }\textbf {\bibinfo
  {volume} {33}},\ \bibinfo {pages} {2008113} (\bibinfo {year}
  {2021})}\BibitemShut {NoStop}%
\bibitem [{\citenamefont {Wehling}\ \emph {et~al.}(2010)\citenamefont
  {Wehling}, \citenamefont {Yuan}, \citenamefont {Lichtenstein}, \citenamefont
  {Geim},\ and\ \citenamefont {Katsnelson}}]{Wehling2010_ResonantScattering}%
  \BibitemOpen
  \bibfield  {author} {\bibinfo {author} {\bibfnamefont {T.~O.}\ \bibnamefont
  {Wehling}}, \bibinfo {author} {\bibfnamefont {S.}~\bibnamefont {Yuan}},
  \bibinfo {author} {\bibfnamefont {A.~I.}\ \bibnamefont {Lichtenstein}},
  \bibinfo {author} {\bibfnamefont {A.~K.}\ \bibnamefont {Geim}},\ and\
  \bibinfo {author} {\bibfnamefont {M.~I.}\ \bibnamefont {Katsnelson}},\
  }\bibfield  {title} {\bibinfo {title} {{Resonant Scattering by Realistic
  Impurities in Graphene}},\ }\href
  {https://doi.org/10.1103/PhysRevLett.105.056802} {\bibfield  {journal}
  {\bibinfo  {journal} {Physical Review Letters}\ }\textbf {\bibinfo {volume}
  {105}},\ \bibinfo {pages} {056802} (\bibinfo {year} {2010})}\BibitemShut
  {NoStop}%
\bibitem [{\citenamefont {Irmer}\ \emph {et~al.}(2018)\citenamefont {Irmer},
  \citenamefont {Kochan}, \citenamefont {Lee},\ and\ \citenamefont
  {Fabian}}]{IrmerPRB2018TopBridgeHollow}%
  \BibitemOpen
  \bibfield  {author} {\bibinfo {author} {\bibfnamefont {S.}~\bibnamefont
  {Irmer}}, \bibinfo {author} {\bibfnamefont {D.}~\bibnamefont {Kochan}},
  \bibinfo {author} {\bibfnamefont {J.}~\bibnamefont {Lee}},\ and\ \bibinfo
  {author} {\bibfnamefont {J.}~\bibnamefont {Fabian}},\ }\bibfield  {title}
  {\bibinfo {title} {{Resonant scattering due to adatoms in graphene: Top,
  bridge, and hollow positions}},\ }\href
  {https://doi.org/10.1103/PhysRevB.97.075417} {\bibfield  {journal} {\bibinfo
  {journal} {Physical Review B}\ }\textbf {\bibinfo {volume} {97}},\ \bibinfo
  {pages} {075417} (\bibinfo {year} {2018})}\BibitemShut {NoStop}%
\bibitem [{\citenamefont {Pogorelov}\ \emph {et~al.}(2020)\citenamefont
  {Pogorelov}, \citenamefont {Loktev},\ and\ \citenamefont
  {Kochan}}]{PogorelovLoktevKochan:PRB2020}%
  \BibitemOpen
  \bibfield  {author} {\bibinfo {author} {\bibfnamefont {Y.~G.}\ \bibnamefont
  {Pogorelov}}, \bibinfo {author} {\bibfnamefont {V.~M.}\ \bibnamefont
  {Loktev}},\ and\ \bibinfo {author} {\bibfnamefont {D.}~\bibnamefont
  {Kochan}},\ }\bibfield  {title} {\bibinfo {title} {Impurity resonance effects
  in graphene versus impurity location, concentration, and sublattice
  occupation},\ }\href {https://doi.org/10.1103/PhysRevB.102.155414} {\bibfield
   {journal} {\bibinfo  {journal} {Phys. Rev. B}\ }\textbf {\bibinfo {volume}
  {102}},\ \bibinfo {pages} {155414} (\bibinfo {year} {2020})}\BibitemShut
  {NoStop}%
\bibitem [{SM()}]{SM}%
  \BibitemOpen
  \href@noop {} {\bibinfo  {journal} {See Supplemental Material at~[link]
  including
  Refs.~\cite{Kwant,Kochan:PRL_sc_graphene_spin_relaxation,Bundesmann_SR_SOC,KatochPRL2018,Kochan2015_PRL-SR-BLG,Mashkoori2017ImpurityBS,Octave,Andreev_1966,Kulik_1977,Sauls_2018,JOSEPHSON1962,JOSEPHSON1974,Titov:PhysRevB2006,Munoz:PhysRevB2012,Alidoust:PhysRevB2019,Alidoust:PhysRevResearch2020,ABS_spectral_density_Muralidharan,FURUSAKI1994214,Ostroukh_paper,Akhmerov_supercurrent,McClure:PR1957,Slonczewski:PR1958,KonschuhPRB:2012,McCann_2013,Beenakker_1991,Akhmerov_ABS_smatrix}
  for detailed explanations of the numerical calculations and additional
  information regarding the employed model}\ }\BibitemShut {NoStop}%
\bibitem [{\citenamefont {Groth}\ \emph {et~al.}(2014)\citenamefont {Groth},
  \citenamefont {Wimmer}, \citenamefont {Akhmerov},\ and\ \citenamefont
  {Waintal}}]{Kwant}%
  \BibitemOpen
\bibfield  {journal} {  }\bibfield  {author} {\bibinfo {author} {\bibfnamefont
  {C.~W.}\ \bibnamefont {Groth}}, \bibinfo {author} {\bibfnamefont
  {M.}~\bibnamefont {Wimmer}}, \bibinfo {author} {\bibfnamefont {A.~R.}\
  \bibnamefont {Akhmerov}},\ and\ \bibinfo {author} {\bibfnamefont
  {X.}~\bibnamefont {Waintal}},\ }\bibfield  {title} {\bibinfo {title} {{Kwant:
  a software package for quantum transport}},\ }\href
  {https://doi.org/10.1088/1367-2630/16/6/063065} {\bibfield  {journal}
  {\bibinfo  {journal} {New Journal of Physics}\ }\textbf {\bibinfo {volume}
  {16}},\ \bibinfo {pages} {063065} (\bibinfo {year} {2014})}\BibitemShut
  {NoStop}%
\bibitem [{\citenamefont {Bundesmann}\ \emph {et~al.}(2015)\citenamefont
  {Bundesmann}, \citenamefont {Kochan}, \citenamefont {Tkatschenko},
  \citenamefont {Fabian},\ and\ \citenamefont {Richter}}]{Bundesmann_SR_SOC}%
  \BibitemOpen
  \bibfield  {author} {\bibinfo {author} {\bibfnamefont {J.}~\bibnamefont
  {Bundesmann}}, \bibinfo {author} {\bibfnamefont {D.}~\bibnamefont {Kochan}},
  \bibinfo {author} {\bibfnamefont {F.}~\bibnamefont {Tkatschenko}}, \bibinfo
  {author} {\bibfnamefont {J.}~\bibnamefont {Fabian}},\ and\ \bibinfo {author}
  {\bibfnamefont {K.}~\bibnamefont {Richter}},\ }\bibfield  {title} {\bibinfo
  {title} {{Theory of spin-orbit-induced spin relaxation in functionalized
  graphene}},\ }\href {https://doi.org/10.1103/PhysRevB.92.081403} {\bibfield
  {journal} {\bibinfo  {journal} {Physical Review B}\ }\textbf {\bibinfo
  {volume} {92}},\ \bibinfo {pages} {081403(R)} (\bibinfo {year}
  {2015})}\BibitemShut {NoStop}%
\bibitem [{\citenamefont {Katoch}\ \emph {et~al.}(2018)\citenamefont {Katoch},
  \citenamefont {Zhu}, \citenamefont {Kochan}, \citenamefont {Singh},
  \citenamefont {Fabian},\ and\ \citenamefont {Kawakami}}]{KatochPRL2018}%
  \BibitemOpen
  \bibfield  {author} {\bibinfo {author} {\bibfnamefont {J.}~\bibnamefont
  {Katoch}}, \bibinfo {author} {\bibfnamefont {T.}~\bibnamefont {Zhu}},
  \bibinfo {author} {\bibfnamefont {D.}~\bibnamefont {Kochan}}, \bibinfo
  {author} {\bibfnamefont {S.}~\bibnamefont {Singh}}, \bibinfo {author}
  {\bibfnamefont {J.}~\bibnamefont {Fabian}},\ and\ \bibinfo {author}
  {\bibfnamefont {R.~K.~R.}\ \bibnamefont {Kawakami}},\ }\bibfield  {title}
  {\bibinfo {title} {{Transport Spectroscopy of Sublattice-Resolved Resonant
  Scattering in Hydrogen-Doped Bilayer Graphene}},\ }\href
  {https://doi.org/10.1103/PhysRevLett.121.136801} {\bibfield  {journal}
  {\bibinfo  {journal} {Physical Review Letters}\ }\textbf {\bibinfo {volume}
  {121}},\ \bibinfo {pages} {136801} (\bibinfo {year} {2018})}\BibitemShut
  {NoStop}%
\bibitem [{\citenamefont {Kochan}\ \emph
  {et~al.}(2014{\natexlab{a}})\citenamefont {Kochan}, \citenamefont {Gmitra},\
  and\ \citenamefont {Fabian}}]{Denis_MAG_ham}%
  \BibitemOpen
  \bibfield  {author} {\bibinfo {author} {\bibfnamefont {D.}~\bibnamefont
  {Kochan}}, \bibinfo {author} {\bibfnamefont {M.}~\bibnamefont {Gmitra}},\
  and\ \bibinfo {author} {\bibfnamefont {J.}~\bibnamefont {Fabian}},\
  }\bibfield  {title} {\bibinfo {title} {{Spin Relaxation Mechanism in
  Graphene: Resonant Scattering by Magnetic Impurities}},\ }\href
  {https://doi.org/10.1103/PhysRevLett.112.116602} {\bibfield  {journal}
  {\bibinfo  {journal} {Phys. Rev. Lett.}\ }\textbf {\bibinfo {volume} {112}},\
  \bibinfo {pages} {116602} (\bibinfo {year} {2014}{\natexlab{a}})}\BibitemShut
  {NoStop}%
\bibitem [{\citenamefont {Mashkoori}\ \emph {et~al.}(2017)\citenamefont
  {Mashkoori}, \citenamefont {Bj{\"o}rnson},\ and\ \citenamefont
  {Black-Schaffer}}]{Mashkoori2017ImpurityBS}%
  \BibitemOpen
  \bibfield  {author} {\bibinfo {author} {\bibfnamefont {M.}~\bibnamefont
  {Mashkoori}}, \bibinfo {author} {\bibfnamefont {K.}~\bibnamefont
  {Bj{\"o}rnson}},\ and\ \bibinfo {author} {\bibfnamefont {A.}~\bibnamefont
  {Black-Schaffer}},\ }\bibfield  {title} {\bibinfo {title} {Impurity bound
  states in fully gapped d-wave superconductors with subdominant order
  parameters},\ }\href@noop {} {\bibfield  {journal} {\bibinfo  {journal}
  {Scientific Reports}\ }\textbf {\bibinfo {volume} {7}} (\bibinfo {year}
  {2017})}\BibitemShut {NoStop}%
\bibitem [{\citenamefont {Eaton}\ \emph {et~al.}(2020)\citenamefont {Eaton},
  \citenamefont {Bateman}, \citenamefont {Hauberg},\ and\ \citenamefont
  {Wehbring}}]{Octave}%
  \BibitemOpen
  \bibfield  {author} {\bibinfo {author} {\bibfnamefont {J.~W.}\ \bibnamefont
  {Eaton}}, \bibinfo {author} {\bibfnamefont {D.}~\bibnamefont {Bateman}},
  \bibinfo {author} {\bibfnamefont {S.}~\bibnamefont {Hauberg}},\ and\ \bibinfo
  {author} {\bibfnamefont {R.}~\bibnamefont {Wehbring}},\ }\href
  {https://www.gnu.org/software/octave/doc/v6.1.0/} {\emph {\bibinfo {title}
  {{GNU Octave} version 6.1.0 manual: a high-level interactive language for
  numerical computations}}} (\bibinfo {year} {2020})\BibitemShut {NoStop}%
\bibitem [{\citenamefont {{Andreev}}(1966)}]{Andreev_1966}%
  \BibitemOpen
  \bibfield  {author} {\bibinfo {author} {\bibfnamefont {A.~F.}\ \bibnamefont
  {{Andreev}}},\ }\bibfield  {title} {\bibinfo {title} {{Electron Spectrum of
  the Intermediate State of Superconductors}},\ }\href@noop {} {\bibfield
  {journal} {\bibinfo  {journal} {Soviet Journal of Experimental and
  Theoretical Physics}\ }\textbf {\bibinfo {volume} {22}},\ \bibinfo {pages}
  {455} (\bibinfo {year} {1966})}\BibitemShut {NoStop}%
\bibitem [{\citenamefont {Kulik}\ and\ \citenamefont
  {Omel'yanchuk}(1977)}]{Kulik_1977}%
  \BibitemOpen
  \bibfield  {author} {\bibinfo {author} {\bibfnamefont {I.~O.}\ \bibnamefont
  {Kulik}}\ and\ \bibinfo {author} {\bibfnamefont {A.~N.}\ \bibnamefont
  {Omel'yanchuk}},\ }\bibfield  {title} {\bibinfo {title} {Properties of
  superconducting microbridges in the pure limit},\ }\bibfield  {journal}
  {\bibinfo  {journal} {Sov. J. Low Temp. Phys. (Engl. Transl.); (United
  States)}\ }\textbf {\bibinfo {volume} {3}},\ \href
  {https://www.osti.gov/biblio/6927703} {} (\bibinfo {year} {1977})\BibitemShut
  {NoStop}%
\bibitem [{\citenamefont {Sauls}(2018)}]{Sauls_2018}%
  \BibitemOpen
  \bibfield  {author} {\bibinfo {author} {\bibfnamefont {J.~A.}\ \bibnamefont
  {Sauls}},\ }\bibfield  {title} {\bibinfo {title} {Andreev bound states and
  their signatures},\ }\href {https://doi.org/10.1098/rsta.2018.0140}
  {\bibfield  {journal} {\bibinfo  {journal} {Philosophical Transactions of the
  Royal Society A: Mathematical, Physical and Engineering Sciences}\ }\textbf
  {\bibinfo {volume} {376}},\ \bibinfo {pages} {20180140} (\bibinfo {year}
  {2018})}\BibitemShut {NoStop}%
\bibitem [{\citenamefont {Josephson}(1962)}]{JOSEPHSON1962}%
  \BibitemOpen
  \bibfield  {author} {\bibinfo {author} {\bibfnamefont {B.}~\bibnamefont
  {Josephson}},\ }\bibfield  {title} {\bibinfo {title} {Possible new effects in
  superconductive tunnelling},\ }\href
  {https://doi.org/https://doi.org/10.1016/0031-9163(62)91369-0} {\bibfield
  {journal} {\bibinfo  {journal} {Physics Letters}\ }\textbf {\bibinfo {volume}
  {1}},\ \bibinfo {pages} {251} (\bibinfo {year} {1962})}\BibitemShut {NoStop}%
\bibitem [{\citenamefont {Josephson}(1974)}]{JOSEPHSON1974}%
  \BibitemOpen
  \bibfield  {author} {\bibinfo {author} {\bibfnamefont {B.~D.}\ \bibnamefont
  {Josephson}},\ }\bibfield  {title} {\bibinfo {title} {The discovery of
  tunnelling supercurrents},\ }\href
  {https://doi.org/10.1103/RevModPhys.46.251} {\bibfield  {journal} {\bibinfo
  {journal} {Rev. Mod. Phys.}\ }\textbf {\bibinfo {volume} {46}},\ \bibinfo
  {pages} {251} (\bibinfo {year} {1974})}\BibitemShut {NoStop}%
\bibitem [{\citenamefont {Titov}\ and\ \citenamefont
  {Beenakker}(2006)}]{Titov:PhysRevB2006}%
  \BibitemOpen
  \bibfield  {author} {\bibinfo {author} {\bibfnamefont {M.}~\bibnamefont
  {Titov}}\ and\ \bibinfo {author} {\bibfnamefont {C.~W.~J.}\ \bibnamefont
  {Beenakker}},\ }\bibfield  {title} {\bibinfo {title} {Josephson effect in
  ballistic graphene},\ }\href {https://doi.org/10.1103/PhysRevB.74.041401}
  {\bibfield  {journal} {\bibinfo  {journal} {Phys. Rev. B}\ }\textbf {\bibinfo
  {volume} {74}},\ \bibinfo {pages} {041401(R)} (\bibinfo {year}
  {2006})}\BibitemShut {NoStop}%
\bibitem [{\citenamefont {Mu\~noz}\ \emph {et~al.}(2012)\citenamefont
  {Mu\~noz}, \citenamefont {Covaci},\ and\ \citenamefont
  {Peeters}}]{Munoz:PhysRevB2012}%
  \BibitemOpen
  \bibfield  {author} {\bibinfo {author} {\bibfnamefont {W.~A.}\ \bibnamefont
  {Mu\~noz}}, \bibinfo {author} {\bibfnamefont {L.}~\bibnamefont {Covaci}},\
  and\ \bibinfo {author} {\bibfnamefont {F.~M.}\ \bibnamefont {Peeters}},\
  }\bibfield  {title} {\bibinfo {title} {Tight-binding study of bilayer
  graphene josephson junctions},\ }\href
  {https://doi.org/10.1103/PhysRevB.86.184505} {\bibfield  {journal} {\bibinfo
  {journal} {Phys. Rev. B}\ }\textbf {\bibinfo {volume} {86}},\ \bibinfo
  {pages} {184505} (\bibinfo {year} {2012})}\BibitemShut {NoStop}%
\bibitem [{\citenamefont {Alidoust}\ \emph {et~al.}(2019)\citenamefont
  {Alidoust}, \citenamefont {Willatzen},\ and\ \citenamefont
  {Jauho}}]{Alidoust:PhysRevB2019}%
  \BibitemOpen
  \bibfield  {author} {\bibinfo {author} {\bibfnamefont {M.}~\bibnamefont
  {Alidoust}}, \bibinfo {author} {\bibfnamefont {M.}~\bibnamefont
  {Willatzen}},\ and\ \bibinfo {author} {\bibfnamefont {A.-P.}\ \bibnamefont
  {Jauho}},\ }\bibfield  {title} {\bibinfo {title} {Symmetry of superconducting
  correlations in displaced bilayers of graphene},\ }\href
  {https://doi.org/10.1103/PhysRevB.99.155413} {\bibfield  {journal} {\bibinfo
  {journal} {Phys. Rev. B}\ }\textbf {\bibinfo {volume} {99}},\ \bibinfo
  {pages} {155413} (\bibinfo {year} {2019})}\BibitemShut {NoStop}%
\bibitem [{\citenamefont {Alidoust}\ \emph {et~al.}(2020)\citenamefont
  {Alidoust}, \citenamefont {Jauho},\ and\ \citenamefont
  {Akola}}]{Alidoust:PhysRevResearch2020}%
  \BibitemOpen
  \bibfield  {author} {\bibinfo {author} {\bibfnamefont {M.}~\bibnamefont
  {Alidoust}}, \bibinfo {author} {\bibfnamefont {A.-P.}\ \bibnamefont
  {Jauho}},\ and\ \bibinfo {author} {\bibfnamefont {J.}~\bibnamefont {Akola}},\
  }\bibfield  {title} {\bibinfo {title} {Josephson effect in graphene bilayers
  with adjustable relative displacement},\ }\href
  {https://doi.org/10.1103/PhysRevResearch.2.032074} {\bibfield  {journal}
  {\bibinfo  {journal} {Phys. Rev. Research}\ }\textbf {\bibinfo {volume}
  {2}},\ \bibinfo {pages} {032074(R)} (\bibinfo {year} {2020})}\BibitemShut
  {NoStop}%
\bibitem [{\citenamefont {Sriram}\ \emph {et~al.}(2019)\citenamefont {Sriram},
  \citenamefont {Kalantre}, \citenamefont {Gharavi}, \citenamefont {Baugh},\
  and\ \citenamefont {Muralidharan}}]{ABS_spectral_density_Muralidharan}%
  \BibitemOpen
  \bibfield  {author} {\bibinfo {author} {\bibfnamefont {P.}~\bibnamefont
  {Sriram}}, \bibinfo {author} {\bibfnamefont {S.~S.}\ \bibnamefont
  {Kalantre}}, \bibinfo {author} {\bibfnamefont {K.}~\bibnamefont {Gharavi}},
  \bibinfo {author} {\bibfnamefont {J.}~\bibnamefont {Baugh}},\ and\ \bibinfo
  {author} {\bibfnamefont {B.}~\bibnamefont {Muralidharan}},\ }\bibfield
  {title} {\bibinfo {title} {{Supercurrent interference in semiconductor
  nanowire Josephson junctions}},\ }\href
  {https://doi.org/10.1103/PhysRevB.100.155431} {\bibfield  {journal} {\bibinfo
   {journal} {Phys. Rev. B}\ }\textbf {\bibinfo {volume} {100}},\ \bibinfo
  {pages} {155431} (\bibinfo {year} {2019})}\BibitemShut {NoStop}%
\bibitem [{\citenamefont {Furusaki}(1994)}]{FURUSAKI1994214}%
  \BibitemOpen
  \bibfield  {author} {\bibinfo {author} {\bibfnamefont {A.}~\bibnamefont
  {Furusaki}},\ }\bibfield  {title} {\bibinfo {title} {{DC Josephson effect in
  dirty SNS junctions: Numerical study}},\ }\href
  {https://doi.org/https://doi.org/10.1016/0921-4526(94)90061-2} {\bibfield
  {journal} {\bibinfo  {journal} {Physica B: Condensed Matter}\ }\textbf
  {\bibinfo {volume} {203}},\ \bibinfo {pages} {214} (\bibinfo {year}
  {1994})}\BibitemShut {NoStop}%
\bibitem [{\citenamefont {Ostroukh}\ \emph {et~al.}(2016)\citenamefont
  {Ostroukh}, \citenamefont {Baxevanis}, \citenamefont {Akhmerov},\ and\
  \citenamefont {Beenakker}}]{Ostroukh_paper}%
  \BibitemOpen
  \bibfield  {author} {\bibinfo {author} {\bibfnamefont {V.~P.}\ \bibnamefont
  {Ostroukh}}, \bibinfo {author} {\bibfnamefont {B.}~\bibnamefont {Baxevanis}},
  \bibinfo {author} {\bibfnamefont {A.~R.}\ \bibnamefont {Akhmerov}},\ and\
  \bibinfo {author} {\bibfnamefont {C.~W.~J.}\ \bibnamefont {Beenakker}},\
  }\bibfield  {title} {\bibinfo {title} {{Two-dimensional Josephson vortex
  lattice and anomalously slow decay of the Fraunhofer oscillations in a
  ballistic SNS junction with a warped Fermi surface}},\ }\href
  {https://doi.org/10.1103/PhysRevB.94.094514} {\bibfield  {journal} {\bibinfo
  {journal} {Phys. Rev. B}\ }\textbf {\bibinfo {volume} {94}},\ \bibinfo
  {pages} {094514} (\bibinfo {year} {2016})}\BibitemShut {NoStop}%
\bibitem [{\citenamefont {Zuo}\ \emph {et~al.}(2017)\citenamefont {Zuo},
  \citenamefont {Mourik}, \citenamefont {Szombati}, \citenamefont {Nijholt},
  \citenamefont {van Woerkom}, \citenamefont {Geresdi}, \citenamefont {Chen},
  \citenamefont {Ostroukh}, \citenamefont {Akhmerov}, \citenamefont {Plissard},
  \citenamefont {Car}, \citenamefont {Bakkers}, \citenamefont {Pikulin},
  \citenamefont {Kouwenhoven},\ and\ \citenamefont
  {Frolov}}]{Akhmerov_supercurrent}%
  \BibitemOpen
  \bibfield  {author} {\bibinfo {author} {\bibfnamefont {K.}~\bibnamefont
  {Zuo}}, \bibinfo {author} {\bibfnamefont {V.}~\bibnamefont {Mourik}},
  \bibinfo {author} {\bibfnamefont {D.~B.}\ \bibnamefont {Szombati}}, \bibinfo
  {author} {\bibfnamefont {B.}~\bibnamefont {Nijholt}}, \bibinfo {author}
  {\bibfnamefont {D.~J.}\ \bibnamefont {van Woerkom}}, \bibinfo {author}
  {\bibfnamefont {A.}~\bibnamefont {Geresdi}}, \bibinfo {author} {\bibfnamefont
  {J.}~\bibnamefont {Chen}}, \bibinfo {author} {\bibfnamefont {V.~P.}\
  \bibnamefont {Ostroukh}}, \bibinfo {author} {\bibfnamefont {A.~R.}\
  \bibnamefont {Akhmerov}}, \bibinfo {author} {\bibfnamefont {S.~R.}\
  \bibnamefont {Plissard}}, \bibinfo {author} {\bibfnamefont {D.}~\bibnamefont
  {Car}}, \bibinfo {author} {\bibfnamefont {E.~P. A.~M.}\ \bibnamefont
  {Bakkers}}, \bibinfo {author} {\bibfnamefont {D.~I.}\ \bibnamefont
  {Pikulin}}, \bibinfo {author} {\bibfnamefont {L.~P.}\ \bibnamefont
  {Kouwenhoven}},\ and\ \bibinfo {author} {\bibfnamefont {S.~M.}\ \bibnamefont
  {Frolov}},\ }\bibfield  {title} {\bibinfo {title} {{Supercurrent Interference
  in Few-Mode Nanowire Josephson Junctions}},\ }\href
  {https://doi.org/10.1103/PhysRevLett.119.187704} {\bibfield  {journal}
  {\bibinfo  {journal} {Phys. Rev. Lett.}\ }\textbf {\bibinfo {volume} {119}},\
  \bibinfo {pages} {187704} (\bibinfo {year} {2017})}\BibitemShut {NoStop}%
\bibitem [{\citenamefont {McClure}(1957)}]{McClure:PR1957}%
  \BibitemOpen
  \bibfield  {author} {\bibinfo {author} {\bibfnamefont {J.~W.}\ \bibnamefont
  {McClure}},\ }\bibfield  {title} {\bibinfo {title} {{Band Structure of
  Graphite and de Haas-van Alphen Effect}},\ }\href
  {https://doi.org/10.1103/PhysRev.108.612} {\bibfield  {journal} {\bibinfo
  {journal} {Phys. Rev.}\ }\textbf {\bibinfo {volume} {108}},\ \bibinfo {pages}
  {612} (\bibinfo {year} {1957})}\BibitemShut {NoStop}%
\bibitem [{\citenamefont {Slonczewski}\ and\ \citenamefont
  {Weiss}(1958)}]{Slonczewski:PR1958}%
  \BibitemOpen
  \bibfield  {author} {\bibinfo {author} {\bibfnamefont {J.~C.}\ \bibnamefont
  {Slonczewski}}\ and\ \bibinfo {author} {\bibfnamefont {P.~R.}\ \bibnamefont
  {Weiss}},\ }\bibfield  {title} {\bibinfo {title} {{Band Structure of
  Graphite}},\ }\href {https://doi.org/10.1103/PhysRev.109.272} {\bibfield
  {journal} {\bibinfo  {journal} {Phys. Rev.}\ }\textbf {\bibinfo {volume}
  {109}},\ \bibinfo {pages} {272} (\bibinfo {year} {1958})}\BibitemShut
  {NoStop}%
\bibitem [{\citenamefont {Konschuh}\ \emph {et~al.}(2012)\citenamefont
  {Konschuh}, \citenamefont {Gmitra}, \citenamefont {Kochan},\ and\
  \citenamefont {Fabian}}]{KonschuhPRB:2012}%
  \BibitemOpen
  \bibfield  {author} {\bibinfo {author} {\bibfnamefont {S.}~\bibnamefont
  {Konschuh}}, \bibinfo {author} {\bibfnamefont {M.}~\bibnamefont {Gmitra}},
  \bibinfo {author} {\bibfnamefont {D.}~\bibnamefont {Kochan}},\ and\ \bibinfo
  {author} {\bibfnamefont {J.}~\bibnamefont {Fabian}},\ }\bibfield  {title}
  {\bibinfo {title} {Theory of spin-orbit coupling in bilayer graphene},\
  }\href {https://doi.org/10.1103/PhysRevB.85.115423} {\bibfield  {journal}
  {\bibinfo  {journal} {Phys. Rev. B}\ }\textbf {\bibinfo {volume} {85}},\
  \bibinfo {pages} {115423} (\bibinfo {year} {2012})}\BibitemShut {NoStop}%
\bibitem [{\citenamefont {McCann}\ and\ \citenamefont
  {Koshino}(2013)}]{McCann_2013}%
  \BibitemOpen
  \bibfield  {author} {\bibinfo {author} {\bibfnamefont {E.}~\bibnamefont
  {McCann}}\ and\ \bibinfo {author} {\bibfnamefont {M.}~\bibnamefont
  {Koshino}},\ }\bibfield  {title} {\bibinfo {title} {The electronic properties
  of bilayer graphene},\ }\href {https://doi.org/10.1088/0034-4885/76/5/056503}
  {\bibfield  {journal} {\bibinfo  {journal} {Reports on Progress in Physics}\
  }\textbf {\bibinfo {volume} {76}},\ \bibinfo {pages} {056503} (\bibinfo
  {year} {2013})}\BibitemShut {NoStop}%
\bibitem [{\citenamefont {Beenakker}(1991)}]{Beenakker_1991}%
  \BibitemOpen
  \bibfield  {author} {\bibinfo {author} {\bibfnamefont {C.~W.~J.}\
  \bibnamefont {Beenakker}},\ }\bibfield  {title} {\bibinfo {title} {{Universal
  limit of critical-current fluctuations in mesoscopic Josephson junctions}},\
  }\href {https://doi.org/10.1103/PhysRevLett.67.3836} {\bibfield  {journal}
  {\bibinfo  {journal} {Phys. Rev. Lett.}\ }\textbf {\bibinfo {volume} {67}},\
  \bibinfo {pages} {3836} (\bibinfo {year} {1991})}\BibitemShut {NoStop}%
\bibitem [{\citenamefont {van Heck}\ \emph {et~al.}(2014)\citenamefont {van
  Heck}, \citenamefont {Mi},\ and\ \citenamefont
  {Akhmerov}}]{Akhmerov_ABS_smatrix}%
  \BibitemOpen
  \bibfield  {author} {\bibinfo {author} {\bibfnamefont {B.}~\bibnamefont {van
  Heck}}, \bibinfo {author} {\bibfnamefont {S.}~\bibnamefont {Mi}},\ and\
  \bibinfo {author} {\bibfnamefont {A.~R.}\ \bibnamefont {Akhmerov}},\
  }\bibfield  {title} {\bibinfo {title} {{Single fermion manipulation via
  superconducting phase differences in multiterminal Josephson junctions}},\
  }\href {https://doi.org/10.1103/PhysRevB.90.155450} {\bibfield  {journal}
  {\bibinfo  {journal} {Phys. Rev. B}\ }\textbf {\bibinfo {volume} {90}},\
  \bibinfo {pages} {155450} (\bibinfo {year} {2014})}\BibitemShut {NoStop}%
\bibitem [{\citenamefont {Tinkham}(2004)}]{tinkham2004introduction}%
  \BibitemOpen
  \bibfield  {author} {\bibinfo {author} {\bibfnamefont {M.}~\bibnamefont
  {Tinkham}},\ }\href {https://books.google.de/books?id=k6AO9nRYbioC} {\emph
  {\bibinfo {title} {{Introduction to Superconductivity: Second Edition}}}},\
  Dover Books on Physics\ (\bibinfo  {publisher} {Dover Publications},\
  \bibinfo {year} {2004})\BibitemShut {NoStop}%
\bibitem [{\citenamefont {Wang}\ \emph {et~al.}(2018)\citenamefont {Wang},
  \citenamefont {Bretheau}, \citenamefont {Rodan-Legrain}, \citenamefont
  {Pisoni}, \citenamefont {Watanabe}, \citenamefont {Taniguchi},\ and\
  \citenamefont {Jarillo-Herrero}}]{PhysRevB.98.121411}%
  \BibitemOpen
  \bibfield  {author} {\bibinfo {author} {\bibfnamefont {J.~I.-J.}\
  \bibnamefont {Wang}}, \bibinfo {author} {\bibfnamefont {L.}~\bibnamefont
  {Bretheau}}, \bibinfo {author} {\bibfnamefont {D.}~\bibnamefont
  {Rodan-Legrain}}, \bibinfo {author} {\bibfnamefont {R.}~\bibnamefont
  {Pisoni}}, \bibinfo {author} {\bibfnamefont {K.}~\bibnamefont {Watanabe}},
  \bibinfo {author} {\bibfnamefont {T.}~\bibnamefont {Taniguchi}},\ and\
  \bibinfo {author} {\bibfnamefont {P.}~\bibnamefont {Jarillo-Herrero}},\
  }\bibfield  {title} {\bibinfo {title} {Tunneling spectroscopy of graphene
  nanodevices coupled to large-gap superconductors},\ }\href
  {https://doi.org/10.1103/PhysRevB.98.121411} {\bibfield  {journal} {\bibinfo
  {journal} {Phys. Rev. B}\ }\textbf {\bibinfo {volume} {98}},\ \bibinfo
  {pages} {121411(R)} (\bibinfo {year} {2018})}\BibitemShut {NoStop}%
\bibitem [{\citenamefont {Li}\ \emph {et~al.}(2020)\citenamefont {Li},
  \citenamefont {Leng}, \citenamefont {Fu}, \citenamefont {Watanabe},
  \citenamefont {Taniguchi}, \citenamefont {Liu}, \citenamefont {Liu},\ and\
  \citenamefont {Zhu}}]{PhysRevB.101.195405}%
  \BibitemOpen
  \bibfield  {author} {\bibinfo {author} {\bibfnamefont {J.}~\bibnamefont
  {Li}}, \bibinfo {author} {\bibfnamefont {H.-B.}\ \bibnamefont {Leng}},
  \bibinfo {author} {\bibfnamefont {H.}~\bibnamefont {Fu}}, \bibinfo {author}
  {\bibfnamefont {K.}~\bibnamefont {Watanabe}}, \bibinfo {author}
  {\bibfnamefont {T.}~\bibnamefont {Taniguchi}}, \bibinfo {author}
  {\bibfnamefont {X.}~\bibnamefont {Liu}}, \bibinfo {author} {\bibfnamefont
  {C.-X.}\ \bibnamefont {Liu}},\ and\ \bibinfo {author} {\bibfnamefont
  {J.}~\bibnamefont {Zhu}},\ }\bibfield  {title} {\bibinfo {title}
  {{Superconducting proximity effect in a transparent van der Waals
  superconductor-metal junction}},\ }\href
  {https://doi.org/10.1103/PhysRevB.101.195405} {\bibfield  {journal} {\bibinfo
   {journal} {Phys. Rev. B}\ }\textbf {\bibinfo {volume} {101}},\ \bibinfo
  {pages} {195405} (\bibinfo {year} {2020})}\BibitemShut {NoStop}%
\bibitem [{\citenamefont {Lee}\ and\ \citenamefont {Lee}(2018)}]{LeeLee2018}%
  \BibitemOpen
  \bibfield  {author} {\bibinfo {author} {\bibfnamefont {G.-H.}\ \bibnamefont
  {Lee}}\ and\ \bibinfo {author} {\bibfnamefont {H.-J.}\ \bibnamefont {Lee}},\
  }\bibfield  {title} {\bibinfo {title} {{Proximity coupling in
  superconductor-graphene heterostructures}},\ }\href
  {https://doi.org/10.1088/1361-6633/aaafe1} {\bibfield  {journal} {\bibinfo
  {journal} {Reports on Progress in Physics}\ }\textbf {\bibinfo {volume}
  {81}},\ \bibinfo {pages} {056502} (\bibinfo {year} {2018})}\BibitemShut
  {NoStop}%
\bibitem [{Note1()}]{Note1}%
  \BibitemOpen
  \bibinfo {note} {However, in special cases that involved numerical
  diagonalization we use even larger $\Delta _0 = 50\protect \tmspace
  +\thinmuskip {.1667em}\protect \mathrm {meV}$ just to reach convergence and
  cross-check analytical results.}\BibitemShut {Stop}%
\bibitem [{\citenamefont {Gmitra}\ \emph {et~al.}(2013)\citenamefont {Gmitra},
  \citenamefont {Kochan},\ and\ \citenamefont
  {Fabian}}]{Gmitra2013_SOC-in-H-Graphene}%
  \BibitemOpen
  \bibfield  {author} {\bibinfo {author} {\bibfnamefont {M.}~\bibnamefont
  {Gmitra}}, \bibinfo {author} {\bibfnamefont {D.}~\bibnamefont {Kochan}},\
  and\ \bibinfo {author} {\bibfnamefont {J.}~\bibnamefont {Fabian}},\
  }\bibfield  {title} {\bibinfo {title} {{Spin-Orbit Coupling in Hydrogenated
  Graphene}},\ }\href {https://doi.org/10.1103/PhysRevLett.110.246602}
  {\bibfield  {journal} {\bibinfo  {journal} {Physical Review Letters}\
  }\textbf {\bibinfo {volume} {110}},\ \bibinfo {pages} {246602} (\bibinfo
  {year} {2013})}\BibitemShut {NoStop}%
\bibitem [{\citenamefont {Irmer}\ \emph {et~al.}(2015)\citenamefont {Irmer},
  \citenamefont {Frank}, \citenamefont {Putz}, \citenamefont {Gmitra},
  \citenamefont {Kochan},\ and\ \citenamefont
  {Fabian}}]{Irmer2015-SOC-F-Graphene}%
  \BibitemOpen
  \bibfield  {author} {\bibinfo {author} {\bibfnamefont {S.}~\bibnamefont
  {Irmer}}, \bibinfo {author} {\bibfnamefont {T.}~\bibnamefont {Frank}},
  \bibinfo {author} {\bibfnamefont {S.}~\bibnamefont {Putz}}, \bibinfo {author}
  {\bibfnamefont {M.}~\bibnamefont {Gmitra}}, \bibinfo {author} {\bibfnamefont
  {D.}~\bibnamefont {Kochan}},\ and\ \bibinfo {author} {\bibfnamefont
  {J.}~\bibnamefont {Fabian}},\ }\bibfield  {title} {\bibinfo {title}
  {{Spin-orbit coupling in fluorinated graphene}},\ }\href
  {https://doi.org/10.1103/PhysRevB.91.115141} {\bibfield  {journal} {\bibinfo
  {journal} {Physical Review B}\ }\textbf {\bibinfo {volume} {91}},\ \bibinfo
  {pages} {115141} (\bibinfo {year} {2015})}\BibitemShut {NoStop}%
\bibitem [{\citenamefont {Zollner}\ \emph {et~al.}(2016)\citenamefont
  {Zollner}, \citenamefont {Frank}, \citenamefont {Irmer}, \citenamefont
  {Gmitra}, \citenamefont {Kochan},\ and\ \citenamefont
  {Fabian}}]{Zollner2016-SOC-Methyl}%
  \BibitemOpen
  \bibfield  {author} {\bibinfo {author} {\bibfnamefont {K.}~\bibnamefont
  {Zollner}}, \bibinfo {author} {\bibfnamefont {T.}~\bibnamefont {Frank}},
  \bibinfo {author} {\bibfnamefont {S.}~\bibnamefont {Irmer}}, \bibinfo
  {author} {\bibfnamefont {M.}~\bibnamefont {Gmitra}}, \bibinfo {author}
  {\bibfnamefont {D.}~\bibnamefont {Kochan}},\ and\ \bibinfo {author}
  {\bibfnamefont {J.}~\bibnamefont {Fabian}},\ }\bibfield  {title} {\bibinfo
  {title} {{Spin-orbit coupling in methyl functionalized graphene}},\ }\href
  {https://doi.org/10.1103/PhysRevB.93.045423} {\bibfield  {journal} {\bibinfo
  {journal} {Physical Review B}\ }\textbf {\bibinfo {volume} {93}},\ \bibinfo
  {pages} {045423} (\bibinfo {year} {2016})}\BibitemShut {NoStop}%
\bibitem [{\citenamefont {Frank}\ \emph {et~al.}(2017)\citenamefont {Frank},
  \citenamefont {Irmer}, \citenamefont {Gmitra}, \citenamefont {Kochan},\ and\
  \citenamefont {Fabian}}]{FrankPRB2017Copper}%
  \BibitemOpen
  \bibfield  {author} {\bibinfo {author} {\bibfnamefont {T.}~\bibnamefont
  {Frank}}, \bibinfo {author} {\bibfnamefont {S.}~\bibnamefont {Irmer}},
  \bibinfo {author} {\bibfnamefont {M.}~\bibnamefont {Gmitra}}, \bibinfo
  {author} {\bibfnamefont {D.}~\bibnamefont {Kochan}},\ and\ \bibinfo {author}
  {\bibfnamefont {J.}~\bibnamefont {Fabian}},\ }\bibfield  {title} {\bibinfo
  {title} {{Copper adatoms on graphene: Theory of orbital and spin-orbital
  effects}},\ }\href {https://doi.org/10.1103/PhysRevB.95.035402} {\bibfield
  {journal} {\bibinfo  {journal} {Physical Review B}\ }\textbf {\bibinfo
  {volume} {95}},\ \bibinfo {pages} {035402} (\bibinfo {year}
  {2017})}\BibitemShut {NoStop}%
\bibitem [{\citenamefont {Kochan}\ \emph {et~al.}(2017)\citenamefont {Kochan},
  \citenamefont {Irmer},\ and\ \citenamefont {Fabian}}]{Denis_SOC_Ham}%
  \BibitemOpen
  \bibfield  {author} {\bibinfo {author} {\bibfnamefont {D.}~\bibnamefont
  {Kochan}}, \bibinfo {author} {\bibfnamefont {S.}~\bibnamefont {Irmer}},\ and\
  \bibinfo {author} {\bibfnamefont {J.}~\bibnamefont {Fabian}},\ }\bibfield
  {title} {\bibinfo {title} {{Model spin-orbit coupling Hamiltonians for
  graphene systems}},\ }\href {https://doi.org/10.1103/PhysRevB.95.165415}
  {\bibfield  {journal} {\bibinfo  {journal} {Phys. Rev. B}\ }\textbf {\bibinfo
  {volume} {95}},\ \bibinfo {pages} {165415} (\bibinfo {year}
  {2017})}\BibitemShut {NoStop}%
\bibitem [{\citenamefont {Hewson}(1993)}]{hewson_1993}%
  \BibitemOpen
  \bibfield  {author} {\bibinfo {author} {\bibfnamefont {A.~C.}\ \bibnamefont
  {Hewson}},\ }\href {https://doi.org/10.1017/CBO9780511470752} {\emph
  {\bibinfo {title} {{The Kondo Problem to Heavy Fermions}}}},\ Cambridge
  Studies in Magnetism\ (\bibinfo  {publisher} {Cambridge University Press},\
  \bibinfo {year} {1993})\BibitemShut {NoStop}%
\bibitem [{\citenamefont {{Kochan D.; Gmitra M.; Fabian
  J.}}(2014)}]{KochanD2014}%
  \BibitemOpen
  \bibfield  {author} {\bibinfo {author} {\bibnamefont {{Kochan D.; Gmitra M.;
  Fabian J.}}},\ }\bibfield  {title} {\bibinfo {title} {{RESONANT SCATTERING
  OFF MAGNETIC IMPURITIES IN GRAPHENE: MECHANISM FOR ULTRAFAST SPIN
  RELAXATION}},\ }in\ \href
  {https://doi.org/https://doi.org/10.1142/9789814740371_0007} {\emph {\bibinfo
  {booktitle} {Symmetry, Spin Dynamics and the Properties of Nanostructures
  Lecture Notes of the 11th International School on Theoretical Physics 11th
  International School on Theoretical Physics Rzesz{\'{o}}w, Poland, 1 – 6
  September 2014}}},\ \bibinfo {editor} {edited by\ \bibinfo {editor}
  {\bibfnamefont {J.}~\bibnamefont {Barna{\'{s}}}}, \bibinfo {editor}
  {\bibfnamefont {V.}~\bibnamefont {Dugaev}},\ and\ \bibinfo {editor}
  {\bibfnamefont {A.}~\bibnamefont {Wal}}}\ (\bibinfo {year} {2014})\ pp.\
  \bibinfo {pages} {136--162}\BibitemShut {NoStop}%
\bibitem [{\citenamefont {Balatsky}\ \emph {et~al.}(2006)\citenamefont
  {Balatsky}, \citenamefont {Vekhter},\ and\ \citenamefont
  {Zhu}}]{Balatsky:RevModPhys2006}%
  \BibitemOpen
  \bibfield  {author} {\bibinfo {author} {\bibfnamefont {A.~V.}\ \bibnamefont
  {Balatsky}}, \bibinfo {author} {\bibfnamefont {I.}~\bibnamefont {Vekhter}},\
  and\ \bibinfo {author} {\bibfnamefont {J.-X.}\ \bibnamefont {Zhu}},\
  }\bibfield  {title} {\bibinfo {title} {{Impurity-induced states in
  conventional and unconventional superconductors}},\ }\href
  {https://doi.org/10.1103/RevModPhys.78.373} {\bibfield  {journal} {\bibinfo
  {journal} {Rev. Mod. Phys.}\ }\textbf {\bibinfo {volume} {78}},\ \bibinfo
  {pages} {373} (\bibinfo {year} {2006})}\BibitemShut {NoStop}%
\bibitem [{\citenamefont {Lifshitz}\ \emph {et~al.}(1988)\citenamefont
  {Lifshitz}, \citenamefont {Gredescul},\ and\ \citenamefont
  {Pastur}}]{Lifshitz:Book1988}%
  \BibitemOpen
  \bibfield  {author} {\bibinfo {author} {\bibfnamefont {I.~M.}\ \bibnamefont
  {Lifshitz}}, \bibinfo {author} {\bibfnamefont {S.~A.}\ \bibnamefont
  {Gredescul}},\ and\ \bibinfo {author} {\bibfnamefont {L.~A.}\ \bibnamefont
  {Pastur}},\ }\href@noop {} {\emph {\bibinfo {title} {Introduction to the
  Theory of Disordered Systems}}}\ (\bibinfo  {publisher} {Wiley-VCH, Berlin},\
  \bibinfo {year} {1988})\BibitemShut {NoStop}%
\bibitem [{Note2()}]{Note2}%
  \BibitemOpen
  \bibinfo {note} {As a comment, while in this toy model we assume no
  macroscopic spin polarization neither spin-orbit interaction in the
  unperturbed system we just employ the reduced Nambu formalism, however, one
  should keep in mind that for any solution with an energy $E$ the full
  Nambu-space approach will give as a solution also the energy
  $-E$.}\BibitemShut {Stop}%
\bibitem [{\citenamefont {Lifshitz}(1956)}]{Lifshitz1956}%
  \BibitemOpen
  \bibfield  {author} {\bibinfo {author} {\bibfnamefont {M.}~\bibnamefont
  {Lifshitz}},\ }\bibfield  {title} {\bibinfo {title} {{Some problems of the
  dynamic theory of non-ideal crystal lattices}},\ }\href
  {https://doi.org/10.1007/BF02746071} {\bibfield  {journal} {\bibinfo
  {journal} {Il Nuovo Cimento Series 10}\ }\textbf {\bibinfo {volume} {3}},\
  \bibinfo {pages} {716} (\bibinfo {year} {1956})}\BibitemShut {NoStop}%
\bibitem [{\citenamefont {Anderson}(1961)}]{AndersonPR:1961}%
  \BibitemOpen
  \bibfield  {author} {\bibinfo {author} {\bibfnamefont {P.~W.}\ \bibnamefont
  {Anderson}},\ }\bibfield  {title} {\bibinfo {title} {{Localized Magnetic
  States in Metals}},\ }\href {https://doi.org/10.1103/PhysRev.124.41}
  {\bibfield  {journal} {\bibinfo  {journal} {Phys. Rev.}\ }\textbf {\bibinfo
  {volume} {124}},\ \bibinfo {pages} {41} (\bibinfo {year} {1961})}\BibitemShut
  {NoStop}%
\bibitem [{\citenamefont {Lifshitz}(1964)}]{Lifshitz1964}%
  \BibitemOpen
  \bibfield  {author} {\bibinfo {author} {\bibfnamefont {I.~M.}\ \bibnamefont
  {Lifshitz}},\ }\bibfield  {title} {\bibinfo {title} {The energy spectrum of
  disordered systems},\ }\href {https://doi.org/10.1080/00018736400101061}
  {\bibfield  {journal} {\bibinfo  {journal} {Adv. Phys.}\ }\textbf {\bibinfo
  {volume} {13(52)}},\ \bibinfo {pages} {483} (\bibinfo {year}
  {1964})}\BibitemShut {NoStop}%
\bibitem [{\citenamefont {Elliott}\ \emph {et~al.}(1974)\citenamefont
  {Elliott}, \citenamefont {Krumhansl},\ and\ \citenamefont
  {Leath}}]{Elliott1974}%
  \BibitemOpen
  \bibfield  {author} {\bibinfo {author} {\bibfnamefont {R.~J.}\ \bibnamefont
  {Elliott}}, \bibinfo {author} {\bibfnamefont {J.~A.}\ \bibnamefont
  {Krumhansl}},\ and\ \bibinfo {author} {\bibfnamefont {P.~L.}\ \bibnamefont
  {Leath}},\ }\bibfield  {title} {\bibinfo {title} {{The theory and properties
  of randomly disordered crystals and related physical systems}},\ }\href
  {https://doi.org/10.1103/RevModPhys.46.465} {\bibfield  {journal} {\bibinfo
  {journal} {Reviews of Modern Physics}\ }\textbf {\bibinfo {volume} {46}},\
  \bibinfo {pages} {465} (\bibinfo {year} {1974})}\BibitemShut {NoStop}%
\bibitem [{Note3()}]{Note3}%
  \BibitemOpen
  \bibinfo {note} {\protect \emph {Private correspondence:} a very similar
  formula (unpublished) was obtained using a different perspective by
  Dr.~Tom\'{a}\v {s} Novotn\'{y}.}\BibitemShut {Stop}%
\bibitem [{\citenamefont {Uldemolins}\ \emph {et~al.}(2021)\citenamefont
  {Uldemolins}, \citenamefont {Mesaros},\ and\ \citenamefont
  {Simon}}]{Uldemolins_2021}%
  \BibitemOpen
  \bibfield  {author} {\bibinfo {author} {\bibfnamefont {M.}~\bibnamefont
  {Uldemolins}}, \bibinfo {author} {\bibfnamefont {A.}~\bibnamefont
  {Mesaros}},\ and\ \bibinfo {author} {\bibfnamefont {P.}~\bibnamefont
  {Simon}},\ }\bibfield  {title} {\bibinfo {title} {{Effect of Van Hove
  singularities on Shiba states in two-dimensional $s$-wave superconductors}},\
  }\href {https://doi.org/10.1103/PhysRevB.103.214514} {\bibfield  {journal}
  {\bibinfo  {journal} {Phys. Rev. B}\ }\textbf {\bibinfo {volume} {103}},\
  \bibinfo {pages} {214514} (\bibinfo {year} {2021})}\BibitemShut {NoStop}%
\bibitem [{\citenamefont {Kochan}\ \emph
  {et~al.}(2014{\natexlab{b}})\citenamefont {Kochan}, \citenamefont {Gmitra},\
  and\ \citenamefont {Fabian}}]{Kochan2014_PRL-SR-Graphene}%
  \BibitemOpen
  \bibfield  {author} {\bibinfo {author} {\bibfnamefont {D.}~\bibnamefont
  {Kochan}}, \bibinfo {author} {\bibfnamefont {M.}~\bibnamefont {Gmitra}},\
  and\ \bibinfo {author} {\bibfnamefont {J.}~\bibnamefont {Fabian}},\
  }\bibfield  {title} {\bibinfo {title} {{Spin Relaxation Mechanism in
  Graphene: Resonant Scattering by Magnetic Impurities}},\ }\href
  {https://doi.org/10.1103/PhysRevLett.112.116602} {\bibfield  {journal}
  {\bibinfo  {journal} {Physical Review Letters}\ }\textbf {\bibinfo {volume}
  {112}},\ \bibinfo {pages} {116602} (\bibinfo {year}
  {2014}{\natexlab{b}})}\BibitemShut {NoStop}%
\bibitem [{\citenamefont {Lopez-Bezanilla}\ and\ \citenamefont
  {Lado}(2019)}]{Lado:PhysRevMaterials2019}%
  \BibitemOpen
  \bibfield  {author} {\bibinfo {author} {\bibfnamefont {A.}~\bibnamefont
  {Lopez-Bezanilla}}\ and\ \bibinfo {author} {\bibfnamefont {J.~L.}\
  \bibnamefont {Lado}},\ }\bibfield  {title} {\bibinfo {title} {Defect-induced
  magnetism and yu-shiba-rusinov states in twisted bilayer graphene},\ }\href
  {https://doi.org/10.1103/PhysRevMaterials.3.084003} {\bibfield  {journal}
  {\bibinfo  {journal} {Phys. Rev. Materials}\ }\textbf {\bibinfo {volume}
  {3}},\ \bibinfo {pages} {084003} (\bibinfo {year} {2019})}\BibitemShut
  {NoStop}%
\bibitem [{\citenamefont {Han}\ and\ \citenamefont
  {Kawakami}(2011)}]{Han:PRL2011}%
  \BibitemOpen
  \bibfield  {author} {\bibinfo {author} {\bibfnamefont {W.}~\bibnamefont
  {Han}}\ and\ \bibinfo {author} {\bibfnamefont {R.~K.}\ \bibnamefont
  {Kawakami}},\ }\bibfield  {title} {\bibinfo {title} {{Spin Relaxation in
  Single-Layer and Bilayer Graphene}},\ }\href
  {https://doi.org/10.1103/PhysRevLett.107.047207} {\bibfield  {journal}
  {\bibinfo  {journal} {Phys. Rev. Lett.}\ }\textbf {\bibinfo {volume} {107}},\
  \bibinfo {pages} {047207} (\bibinfo {year} {2011})}\BibitemShut {NoStop}%
\bibitem [{\citenamefont {Yang}\ \emph {et~al.}(2011)\citenamefont {Yang},
  \citenamefont {Balakrishnan}, \citenamefont {Volmer}, \citenamefont {Avsar},
  \citenamefont {Jaiswal}, \citenamefont {Samm}, \citenamefont {Ali},
  \citenamefont {Pachoud}, \citenamefont {Zeng}, \citenamefont {Popinciuc},
  \citenamefont {G\"untherodt}, \citenamefont {Beschoten},\ and\ \citenamefont
  {\"Ozyilmaz}}]{Yang:PRL2011}%
  \BibitemOpen
  \bibfield  {author} {\bibinfo {author} {\bibfnamefont {T.-Y.}\ \bibnamefont
  {Yang}}, \bibinfo {author} {\bibfnamefont {J.}~\bibnamefont {Balakrishnan}},
  \bibinfo {author} {\bibfnamefont {F.}~\bibnamefont {Volmer}}, \bibinfo
  {author} {\bibfnamefont {A.}~\bibnamefont {Avsar}}, \bibinfo {author}
  {\bibfnamefont {M.}~\bibnamefont {Jaiswal}}, \bibinfo {author} {\bibfnamefont
  {J.}~\bibnamefont {Samm}}, \bibinfo {author} {\bibfnamefont {S.~R.}\
  \bibnamefont {Ali}}, \bibinfo {author} {\bibfnamefont {A.}~\bibnamefont
  {Pachoud}}, \bibinfo {author} {\bibfnamefont {M.}~\bibnamefont {Zeng}},
  \bibinfo {author} {\bibfnamefont {M.}~\bibnamefont {Popinciuc}}, \bibinfo
  {author} {\bibfnamefont {G.}~\bibnamefont {G\"untherodt}}, \bibinfo {author}
  {\bibfnamefont {B.}~\bibnamefont {Beschoten}},\ and\ \bibinfo {author}
  {\bibfnamefont {B.}~\bibnamefont {\"Ozyilmaz}},\ }\bibfield  {title}
  {\bibinfo {title} {{Observation of Long Spin-Relaxation Times in Bilayer
  Graphene at Room Temperature}},\ }\href
  {https://doi.org/10.1103/PhysRevLett.107.047206} {\bibfield  {journal}
  {\bibinfo  {journal} {Phys. Rev. Lett.}\ }\textbf {\bibinfo {volume} {107}},\
  \bibinfo {pages} {047206} (\bibinfo {year} {2011})}\BibitemShut {NoStop}%
\bibitem [{\citenamefont {Ingla-Ayn\'es}\ \emph {et~al.}(2015)\citenamefont
  {Ingla-Ayn\'es}, \citenamefont {Guimar\~aes}, \citenamefont {Meijerink},
  \citenamefont {Zomer},\ and\ \citenamefont {van Wees}}]{Ingla-Aynes:PRB2015}%
  \BibitemOpen
  \bibfield  {author} {\bibinfo {author} {\bibfnamefont {J.}~\bibnamefont
  {Ingla-Ayn\'es}}, \bibinfo {author} {\bibfnamefont {M.~H.~D.}\ \bibnamefont
  {Guimar\~aes}}, \bibinfo {author} {\bibfnamefont {R.~J.}\ \bibnamefont
  {Meijerink}}, \bibinfo {author} {\bibfnamefont {P.~J.}\ \bibnamefont
  {Zomer}},\ and\ \bibinfo {author} {\bibfnamefont {B.~J.}\ \bibnamefont {van
  Wees}},\ }\bibfield  {title} {\bibinfo {title}
  {{$24\ensuremath{-}\ensuremath{\mu}\mathrm{m}$ spin relaxation length in
  boron nitride encapsulated bilayer graphene}},\ }\href
  {https://doi.org/10.1103/PhysRevB.92.201410} {\bibfield  {journal} {\bibinfo
  {journal} {Phys. Rev. B}\ }\textbf {\bibinfo {volume} {92}},\ \bibinfo
  {pages} {201410(R)} (\bibinfo {year} {2015})}\BibitemShut {NoStop}%
\bibitem [{\citenamefont {Avsar}\ \emph {et~al.}(2016)\citenamefont {Avsar},
  \citenamefont {Vera-Marun}, \citenamefont {Tan}, \citenamefont {Koon},
  \citenamefont {Watanabe}, \citenamefont {Taniguchi}, \citenamefont {Adam},\
  and\ \citenamefont {{\"{O}}zyilmaz}}]{Avsar:NPG2016}%
  \BibitemOpen
  \bibfield  {author} {\bibinfo {author} {\bibfnamefont {A.}~\bibnamefont
  {Avsar}}, \bibinfo {author} {\bibfnamefont {I.~J.}\ \bibnamefont
  {Vera-Marun}}, \bibinfo {author} {\bibfnamefont {J.~Y.}\ \bibnamefont {Tan}},
  \bibinfo {author} {\bibfnamefont {G.~K.~W.}\ \bibnamefont {Koon}}, \bibinfo
  {author} {\bibfnamefont {K.}~\bibnamefont {Watanabe}}, \bibinfo {author}
  {\bibfnamefont {T.}~\bibnamefont {Taniguchi}}, \bibinfo {author}
  {\bibfnamefont {S.}~\bibnamefont {Adam}},\ and\ \bibinfo {author}
  {\bibfnamefont {B.}~\bibnamefont {{\"{O}}zyilmaz}},\ }\bibfield  {title}
  {\bibinfo {title} {{Electronic spin transport in dual-gated bilayer
  graphene}},\ }\href {https://doi.org/10.1038/am.2016.65} {\bibfield
  {journal} {\bibinfo  {journal} {NPG Asia Materials}\ }\textbf {\bibinfo
  {volume} {8}},\ \bibinfo {pages} {e274} (\bibinfo {year} {2016})}\BibitemShut
  {NoStop}%
\bibitem [{\citenamefont {Pereira}\ \emph {et~al.}(2006)\citenamefont
  {Pereira}, \citenamefont {Guinea}, \citenamefont {Lopes~dos Santos},
  \citenamefont {Peres},\ and\ \citenamefont {Castro~Neto}}]{Pereira:PRL2006}%
  \BibitemOpen
  \bibfield  {author} {\bibinfo {author} {\bibfnamefont {V.~M.}\ \bibnamefont
  {Pereira}}, \bibinfo {author} {\bibfnamefont {F.}~\bibnamefont {Guinea}},
  \bibinfo {author} {\bibfnamefont {J.~M.~B.}\ \bibnamefont {Lopes~dos
  Santos}}, \bibinfo {author} {\bibfnamefont {N.~M.~R.}\ \bibnamefont
  {Peres}},\ and\ \bibinfo {author} {\bibfnamefont {A.~H.}\ \bibnamefont
  {Castro~Neto}},\ }\bibfield  {title} {\bibinfo {title} {{Disorder Induced
  Localized States in Graphene}},\ }\href
  {https://doi.org/10.1103/PhysRevLett.96.036801} {\bibfield  {journal}
  {\bibinfo  {journal} {Phys. Rev. Lett.}\ }\textbf {\bibinfo {volume} {96}},\
  \bibinfo {pages} {036801} (\bibinfo {year} {2006})}\BibitemShut {NoStop}%
\bibitem [{\citenamefont {Castro}\ \emph {et~al.}(2010)\citenamefont {Castro},
  \citenamefont {L\'opez-Sancho},\ and\ \citenamefont
  {Vozmediano}}]{Castro:PRL2010}%
  \BibitemOpen
  \bibfield  {author} {\bibinfo {author} {\bibfnamefont {E.~V.}\ \bibnamefont
  {Castro}}, \bibinfo {author} {\bibfnamefont {M.~P.}\ \bibnamefont
  {L\'opez-Sancho}},\ and\ \bibinfo {author} {\bibfnamefont {M.~A.~H.}\
  \bibnamefont {Vozmediano}},\ }\bibfield  {title} {\bibinfo {title} {{New Type
  of Vacancy-Induced Localized States in Multilayer Graphene}},\ }\href
  {https://doi.org/10.1103/PhysRevLett.104.036802} {\bibfield  {journal}
  {\bibinfo  {journal} {Phys. Rev. Lett.}\ }\textbf {\bibinfo {volume} {104}},\
  \bibinfo {pages} {036802} (\bibinfo {year} {2010})}\BibitemShut {NoStop}%
\bibitem [{\citenamefont {Kulik}(1970)}]{Kulik1970}%
  \BibitemOpen
  \bibfield  {author} {\bibinfo {author} {\bibfnamefont {I.~O.}\ \bibnamefont
  {Kulik}},\ }\bibfield  {title} {\bibinfo {title} {{Macroscopic Quantization
  and the Proximity Effect in S-N-S Junctions}},\ }\href@noop {} {\bibfield
  {journal} {\bibinfo  {journal} {Zh. Eksp. Teor. Fiz}\ }\textbf {\bibinfo
  {volume} {30}},\ \bibinfo {pages} {1745} (\bibinfo {year}
  {1970})}\BibitemShut {NoStop}%
\bibitem [{\citenamefont {Costa}\ \emph {et~al.}(2018)\citenamefont {Costa},
  \citenamefont {Fabian},\ and\ \citenamefont {Kochan}}]{Costa2018}%
  \BibitemOpen
  \bibfield  {author} {\bibinfo {author} {\bibfnamefont {A.}~\bibnamefont
  {Costa}}, \bibinfo {author} {\bibfnamefont {J.}~\bibnamefont {Fabian}},\ and\
  \bibinfo {author} {\bibfnamefont {D.}~\bibnamefont {Kochan}},\ }\bibfield
  {title} {\bibinfo {title} {{Connection between zero-energy Yu-Shiba-Rusinov
  states and 0-{$\pi$} transitions in magnetic Josephson junctions}},\ }\href
  {https://doi.org/10.1103/PhysRevB.98.134511} {\bibfield  {journal} {\bibinfo
  {journal} {Physical Review B}\ }\textbf {\bibinfo {volume} {98}},\ \bibinfo
  {pages} {134511} (\bibinfo {year} {2018})}\BibitemShut {NoStop}%
\bibitem [{\citenamefont {M{\'{e}}nard}\ \emph {et~al.}(2015)\citenamefont
  {M{\'{e}}nard}, \citenamefont {Guissart}, \citenamefont {Brun}, \citenamefont
  {Pons}, \citenamefont {Stolyarov}, \citenamefont {Debontridder},
  \citenamefont {Leclerc}, \citenamefont {Janod}, \citenamefont {Cario},
  \citenamefont {Roditchev}, \citenamefont {Simon},\ and\ \citenamefont
  {Cren}}]{Menard2015a}%
  \BibitemOpen
  \bibfield  {author} {\bibinfo {author} {\bibfnamefont {G.~C.}\ \bibnamefont
  {M{\'{e}}nard}}, \bibinfo {author} {\bibfnamefont {S.}~\bibnamefont
  {Guissart}}, \bibinfo {author} {\bibfnamefont {C.}~\bibnamefont {Brun}},
  \bibinfo {author} {\bibfnamefont {S.}~\bibnamefont {Pons}}, \bibinfo {author}
  {\bibfnamefont {V.~S.}\ \bibnamefont {Stolyarov}}, \bibinfo {author}
  {\bibfnamefont {F.}~\bibnamefont {Debontridder}}, \bibinfo {author}
  {\bibfnamefont {M.~V.}\ \bibnamefont {Leclerc}}, \bibinfo {author}
  {\bibfnamefont {E.}~\bibnamefont {Janod}}, \bibinfo {author} {\bibfnamefont
  {L.}~\bibnamefont {Cario}}, \bibinfo {author} {\bibfnamefont
  {D.}~\bibnamefont {Roditchev}}, \bibinfo {author} {\bibfnamefont
  {P.}~\bibnamefont {Simon}},\ and\ \bibinfo {author} {\bibfnamefont
  {T.}~\bibnamefont {Cren}},\ }\bibfield  {title} {\bibinfo {title} {{Coherent
  long-range magnetic bound states in a superconductor}},\ }\href
  {https://doi.org/10.1038/nphys3508} {\bibfield  {journal} {\bibinfo
  {journal} {Nature Physics}\ }\textbf {\bibinfo {volume} {11}},\ \bibinfo
  {pages} {1013} (\bibinfo {year} {2015})}\BibitemShut {NoStop}%
\bibitem [{\citenamefont {Heinrich}\ \emph {et~al.}(2018)\citenamefont
  {Heinrich}, \citenamefont {Pascual},\ and\ \citenamefont
  {Franke}}]{Heinrich:ProgSurfScience2018}%
  \BibitemOpen
  \bibfield  {author} {\bibinfo {author} {\bibfnamefont {B.~W.}\ \bibnamefont
  {Heinrich}}, \bibinfo {author} {\bibfnamefont {J.~I.}\ \bibnamefont
  {Pascual}},\ and\ \bibinfo {author} {\bibfnamefont {K.~J.}\ \bibnamefont
  {Franke}},\ }\bibfield  {title} {\bibinfo {title} {{Single magnetic
  adsorbates on s-wave superconductors}},\ }\href
  {https://doi.org/https://doi.org/10.1016/j.progsurf.2018.01.001} {\bibfield
  {journal} {\bibinfo  {journal} {Progress in Surface Science}\ }\textbf
  {\bibinfo {volume} {93}},\ \bibinfo {pages} {1} (\bibinfo {year}
  {2018})}\BibitemShut {NoStop}%
\bibitem [{\citenamefont {Wang}\ \emph {et~al.}(2021)\citenamefont {Wang},
  \citenamefont {Wiebe}, \citenamefont {Zhong}, \citenamefont {Gu},\ and\
  \citenamefont {Wiesendanger}}]{WangWiesendanger:PRL2021}%
  \BibitemOpen
  \bibfield  {author} {\bibinfo {author} {\bibfnamefont {D.}~\bibnamefont
  {Wang}}, \bibinfo {author} {\bibfnamefont {J.}~\bibnamefont {Wiebe}},
  \bibinfo {author} {\bibfnamefont {R.}~\bibnamefont {Zhong}}, \bibinfo
  {author} {\bibfnamefont {G.}~\bibnamefont {Gu}},\ and\ \bibinfo {author}
  {\bibfnamefont {R.}~\bibnamefont {Wiesendanger}},\ }\bibfield  {title}
  {\bibinfo {title} {{Spin-Polarized Yu-Shiba-Rusinov States in an Iron-Based
  Superconductor}},\ }\href {https://doi.org/10.1103/PhysRevLett.126.076802}
  {\bibfield  {journal} {\bibinfo  {journal} {Phys. Rev. Lett.}\ }\textbf
  {\bibinfo {volume} {126}},\ \bibinfo {pages} {076802} (\bibinfo {year}
  {2021})}\BibitemShut {NoStop}%
\bibitem [{\citenamefont {K{\"u}ster}\ \emph {et~al.}(2021)\citenamefont
  {K{\"u}ster}, \citenamefont {Montero}, \citenamefont {Guimar{\~a}es},
  \citenamefont {Brinker}, \citenamefont {Lounis}, \citenamefont {Parkin},\
  and\ \citenamefont {Sessi}}]{Kuster:NatCom2021}%
  \BibitemOpen
  \bibfield  {author} {\bibinfo {author} {\bibfnamefont {F.}~\bibnamefont
  {K{\"u}ster}}, \bibinfo {author} {\bibfnamefont {A.~M.}\ \bibnamefont
  {Montero}}, \bibinfo {author} {\bibfnamefont {F.~S.~M.}\ \bibnamefont
  {Guimar{\~a}es}}, \bibinfo {author} {\bibfnamefont {S.}~\bibnamefont
  {Brinker}}, \bibinfo {author} {\bibfnamefont {S.}~\bibnamefont {Lounis}},
  \bibinfo {author} {\bibfnamefont {S.~S.~P.}\ \bibnamefont {Parkin}},\ and\
  \bibinfo {author} {\bibfnamefont {P.}~\bibnamefont {Sessi}},\ }\bibfield
  {title} {\bibinfo {title} {{Correlating Josephson supercurrents and Shiba
  states in quantum spins unconventionally coupled to superconductors}},\
  }\href {https://doi.org/10.1038/s41467-021-21347-5} {\bibfield  {journal}
  {\bibinfo  {journal} {Nature Communications}\ }\textbf {\bibinfo {volume}
  {12}},\ \bibinfo {pages} {1108} (\bibinfo {year} {2021})}\BibitemShut
  {NoStop}%
\bibitem [{\citenamefont {Sakurai}(1970)}]{Sakurai:ProgTheorPhys1970}%
  \BibitemOpen
  \bibfield  {author} {\bibinfo {author} {\bibfnamefont {A.}~\bibnamefont
  {Sakurai}},\ }\bibfield  {title} {\bibinfo {title} {{Comments on
  Superconductors with Magnetic Impurities}},\ }\href
  {https://doi.org/10.1143/PTP.44.1472} {\bibfield  {journal} {\bibinfo
  {journal} {Progress of Theoretical Physics}\ }\textbf {\bibinfo {volume}
  {44}},\ \bibinfo {pages} {1472} (\bibinfo {year} {1970})}\BibitemShut
  {NoStop}%
\bibitem [{\citenamefont {Sau}\ and\ \citenamefont
  {Demler}(2013)}]{SauDemler:PRB2013}%
  \BibitemOpen
  \bibfield  {author} {\bibinfo {author} {\bibfnamefont {J.~D.}\ \bibnamefont
  {Sau}}\ and\ \bibinfo {author} {\bibfnamefont {E.}~\bibnamefont {Demler}},\
  }\bibfield  {title} {\bibinfo {title} {Bound states at impurities as a probe
  of topological superconductivity in nanowires},\ }\href
  {https://doi.org/10.1103/PhysRevB.88.205402} {\bibfield  {journal} {\bibinfo
  {journal} {Phys. Rev. B}\ }\textbf {\bibinfo {volume} {88}},\ \bibinfo
  {pages} {205402} (\bibinfo {year} {2013})}\BibitemShut {NoStop}%
\bibitem [{\citenamefont {Pientka}\ \emph {et~al.}(2015)\citenamefont
  {Pientka}, \citenamefont {Peng}, \citenamefont {Glazman},\ and\ \citenamefont
  {von Oppen}}]{Pientka:PhysScripta2015}%
  \BibitemOpen
  \bibfield  {author} {\bibinfo {author} {\bibfnamefont {F.}~\bibnamefont
  {Pientka}}, \bibinfo {author} {\bibfnamefont {Y.}~\bibnamefont {Peng}},
  \bibinfo {author} {\bibfnamefont {L.}~\bibnamefont {Glazman}},\ and\ \bibinfo
  {author} {\bibfnamefont {F.}~\bibnamefont {von Oppen}},\ }\bibfield  {title}
  {\bibinfo {title} {{Topological superconducting phase and Majorana bound
  states in Shiba chains}},\ }\href
  {https://doi.org/10.1088/0031-8949/2015/t164/014008} {\bibfield  {journal}
  {\bibinfo  {journal} {Physica Scripta}\ }\textbf {\bibinfo {volume} {T164}},\
  \bibinfo {pages} {014008} (\bibinfo {year} {2015})}\BibitemShut {NoStop}%
\bibitem [{\citenamefont {Kochan}\ \emph {et~al.}(2015)\citenamefont {Kochan},
  \citenamefont {Irmer}, \citenamefont {Gmitra},\ and\ \citenamefont
  {Fabian}}]{Kochan2015_PRL-SR-BLG}%
  \BibitemOpen
  \bibfield  {author} {\bibinfo {author} {\bibfnamefont {D.}~\bibnamefont
  {Kochan}}, \bibinfo {author} {\bibfnamefont {S.}~\bibnamefont {Irmer}},
  \bibinfo {author} {\bibfnamefont {M.}~\bibnamefont {Gmitra}},\ and\ \bibinfo
  {author} {\bibfnamefont {J.}~\bibnamefont {Fabian}},\ }\bibfield  {title}
  {\bibinfo {title} {{Resonant Scattering by Magnetic Impurities as a Model for
  Spin Relaxation in Bilayer Graphene}},\ }\href
  {https://doi.org/10.1103/PhysRevLett.115.196601} {\bibfield  {journal}
  {\bibinfo  {journal} {Physical Review Letters}\ }\textbf {\bibinfo {volume}
  {115}},\ \bibinfo {pages} {196601} (\bibinfo {year} {2015})}\BibitemShut
  {NoStop}%
\end{thebibliography}%


\begin{thebibliography}{26}%
\makeatletter
\providecommand \@ifxundefined [1]{%
 \@ifx{#1\undefined}
}%
\providecommand \@ifnum [1]{%
 \ifnum #1\expandafter \@firstoftwo
 \else \expandafter \@secondoftwo
 \fi
}%
\providecommand \@ifx [1]{%
 \ifx #1\expandafter \@firstoftwo
 \else \expandafter \@secondoftwo
 \fi
}%
\providecommand \natexlab [1]{#1}%
\providecommand \enquote  [1]{``#1''}%
\providecommand \bibnamefont  [1]{#1}%
\providecommand \bibfnamefont [1]{#1}%
\providecommand \citenamefont [1]{#1}%
\providecommand \href@noop [0]{\@secondoftwo}%
\providecommand \href [0]{\begingroup \@sanitize@url \@href}%
\providecommand \@href[1]{\@@startlink{#1}\@@href}%
\providecommand \@@href[1]{\endgroup#1\@@endlink}%
\providecommand \@sanitize@url [0]{\catcode `\\12\catcode `\$12\catcode
  `\&12\catcode `\#12\catcode `\^12\catcode `\_12\catcode `\%12\relax}%
\providecommand \@@startlink[1]{}%
\providecommand \@@endlink[0]{}%
\providecommand \url  [0]{\begingroup\@sanitize@url \@url }%
\providecommand \@url [1]{\endgroup\@href {#1}{\urlprefix }}%
\providecommand \urlprefix  [0]{URL }%
\providecommand \Eprint [0]{\href }%
\providecommand \doibase [0]{https://doi.org/}%
\providecommand \selectlanguage [0]{\@gobble}%
\providecommand \bibinfo  [0]{\@secondoftwo}%
\providecommand \bibfield  [0]{\@secondoftwo}%
\providecommand \translation [1]{[#1]}%
\providecommand \BibitemOpen [0]{}%
\providecommand \bibitemStop [0]{}%
\providecommand \bibitemNoStop [0]{.\EOS\space}%
\providecommand \EOS [0]{\spacefactor3000\relax}%
\providecommand \BibitemShut  [1]{\csname bibitem#1\endcsname}%
\let\auto@bib@innerbib\@empty
\bibitem [{\citenamefont {Groth}\ \emph {et~al.}(2014)\citenamefont {Groth},
  \citenamefont {Wimmer}, \citenamefont {Akhmerov},\ and\ \citenamefont
  {Waintal}}]{Kwant}%
  \BibitemOpen
  \bibfield  {author} {\bibinfo {author} {\bibfnamefont {C.~W.}\ \bibnamefont
  {Groth}}, \bibinfo {author} {\bibfnamefont {M.}~\bibnamefont {Wimmer}},
  \bibinfo {author} {\bibfnamefont {A.~R.}\ \bibnamefont {Akhmerov}},\ and\
  \bibinfo {author} {\bibfnamefont {X.}~\bibnamefont {Waintal}},\ }\bibfield
  {title} {\bibinfo {title} {{Kwant: a software package for quantum
  transport}},\ }\href {https://doi.org/10.1088/1367-2630/16/6/063065}
  {\bibfield  {journal} {\bibinfo  {journal} {New Journal of Physics}\ }\textbf
  {\bibinfo {volume} {16}},\ \bibinfo {pages} {063065} (\bibinfo {year}
  {2014})}\BibitemShut {NoStop}%
\bibitem [{\citenamefont {Kochan}\ \emph {et~al.}(2020)\citenamefont {Kochan},
  \citenamefont {Barth}, \citenamefont {Costa}, \citenamefont {Richter},\ and\
  \citenamefont {Fabian}}]{Kochan:PRL_sc_graphene_spin_relaxation}%
  \BibitemOpen
  \bibfield  {author} {\bibinfo {author} {\bibfnamefont {D.}~\bibnamefont
  {Kochan}}, \bibinfo {author} {\bibfnamefont {M.}~\bibnamefont {Barth}},
  \bibinfo {author} {\bibfnamefont {A.}~\bibnamefont {Costa}}, \bibinfo
  {author} {\bibfnamefont {K.}~\bibnamefont {Richter}},\ and\ \bibinfo {author}
  {\bibfnamefont {J.}~\bibnamefont {Fabian}},\ }\bibfield  {title} {\bibinfo
  {title} {{Spin Relaxation in $s$-Wave Superconductors in the Presence of
  Resonant Spin-Flip Scatterers}},\ }\href
  {https://doi.org/10.1103/PhysRevLett.125.087001} {\bibfield  {journal}
  {\bibinfo  {journal} {Phys. Rev. Lett.}\ }\textbf {\bibinfo {volume} {125}},\
  \bibinfo {pages} {087001} (\bibinfo {year} {2020})}\BibitemShut {NoStop}%
\bibitem [{\citenamefont {Bundesmann}\ \emph {et~al.}(2015)\citenamefont
  {Bundesmann}, \citenamefont {Kochan}, \citenamefont {Tkatschenko},
  \citenamefont {Fabian},\ and\ \citenamefont {Richter}}]{Bundesmann_SR_SOC}%
  \BibitemOpen
  \bibfield  {author} {\bibinfo {author} {\bibfnamefont {J.}~\bibnamefont
  {Bundesmann}}, \bibinfo {author} {\bibfnamefont {D.}~\bibnamefont {Kochan}},
  \bibinfo {author} {\bibfnamefont {F.}~\bibnamefont {Tkatschenko}}, \bibinfo
  {author} {\bibfnamefont {J.}~\bibnamefont {Fabian}},\ and\ \bibinfo {author}
  {\bibfnamefont {K.}~\bibnamefont {Richter}},\ }\bibfield  {title} {\bibinfo
  {title} {{Theory of spin-orbit-induced spin relaxation in functionalized
  graphene}},\ }\href {https://doi.org/10.1103/PhysRevB.92.081403} {\bibfield
  {journal} {\bibinfo  {journal} {Physical Review B}\ }\textbf {\bibinfo
  {volume} {92}},\ \bibinfo {pages} {081403(R)} (\bibinfo {year}
  {2015})}\BibitemShut {NoStop}%
\bibitem [{\citenamefont {Katoch}\ \emph {et~al.}(2018)\citenamefont {Katoch},
  \citenamefont {Zhu}, \citenamefont {Kochan}, \citenamefont {Singh},
  \citenamefont {Fabian},\ and\ \citenamefont {Kawakami}}]{KatochPRL2018}%
  \BibitemOpen
  \bibfield  {author} {\bibinfo {author} {\bibfnamefont {J.}~\bibnamefont
  {Katoch}}, \bibinfo {author} {\bibfnamefont {T.}~\bibnamefont {Zhu}},
  \bibinfo {author} {\bibfnamefont {D.}~\bibnamefont {Kochan}}, \bibinfo
  {author} {\bibfnamefont {S.}~\bibnamefont {Singh}}, \bibinfo {author}
  {\bibfnamefont {J.}~\bibnamefont {Fabian}},\ and\ \bibinfo {author}
  {\bibfnamefont {R.~K.~R.}\ \bibnamefont {Kawakami}},\ }\bibfield  {title}
  {\bibinfo {title} {{Transport Spectroscopy of Sublattice-Resolved Resonant
  Scattering in Hydrogen-Doped Bilayer Graphene}},\ }\href
  {https://doi.org/10.1103/PhysRevLett.121.136801} {\bibfield  {journal}
  {\bibinfo  {journal} {Physical Review Letters}\ }\textbf {\bibinfo {volume}
  {121}},\ \bibinfo {pages} {136801} (\bibinfo {year} {2018})}\BibitemShut
  {NoStop}%
\bibitem [{\citenamefont {Kochan}\ \emph {et~al.}(2015)\citenamefont {Kochan},
  \citenamefont {Irmer}, \citenamefont {Gmitra},\ and\ \citenamefont
  {Fabian}}]{Kochan2015_PRL-SR-BLG}%
  \BibitemOpen
  \bibfield  {author} {\bibinfo {author} {\bibfnamefont {D.}~\bibnamefont
  {Kochan}}, \bibinfo {author} {\bibfnamefont {S.}~\bibnamefont {Irmer}},
  \bibinfo {author} {\bibfnamefont {M.}~\bibnamefont {Gmitra}},\ and\ \bibinfo
  {author} {\bibfnamefont {J.}~\bibnamefont {Fabian}},\ }\bibfield  {title}
  {\bibinfo {title} {{Resonant Scattering by Magnetic Impurities as a Model for
  Spin Relaxation in Bilayer Graphene}},\ }\href
  {https://doi.org/10.1103/PhysRevLett.115.196601} {\bibfield  {journal}
  {\bibinfo  {journal} {Physical Review Letters}\ }\textbf {\bibinfo {volume}
  {115}},\ \bibinfo {pages} {196601} (\bibinfo {year} {2015})}\BibitemShut
  {NoStop}%
\bibitem [{\citenamefont {Mashkoori}\ \emph {et~al.}(2017)\citenamefont
  {Mashkoori}, \citenamefont {Bj{\"o}rnson},\ and\ \citenamefont
  {Black-Schaffer}}]{Mashkoori2017ImpurityBS}%
  \BibitemOpen
  \bibfield  {author} {\bibinfo {author} {\bibfnamefont {M.}~\bibnamefont
  {Mashkoori}}, \bibinfo {author} {\bibfnamefont {K.}~\bibnamefont
  {Bj{\"o}rnson}},\ and\ \bibinfo {author} {\bibfnamefont {A.}~\bibnamefont
  {Black-Schaffer}},\ }\bibfield  {title} {\bibinfo {title} {Impurity bound
  states in fully gapped d-wave superconductors with subdominant order
  parameters},\ }\href@noop {} {\bibfield  {journal} {\bibinfo  {journal}
  {Scientific Reports}\ }\textbf {\bibinfo {volume} {7}} (\bibinfo {year}
  {2017})}\BibitemShut {NoStop}%
\bibitem [{\citenamefont {Eaton}\ \emph {et~al.}(2020)\citenamefont {Eaton},
  \citenamefont {Bateman}, \citenamefont {Hauberg},\ and\ \citenamefont
  {Wehbring}}]{Octave}%
  \BibitemOpen
  \bibfield  {author} {\bibinfo {author} {\bibfnamefont {J.~W.}\ \bibnamefont
  {Eaton}}, \bibinfo {author} {\bibfnamefont {D.}~\bibnamefont {Bateman}},
  \bibinfo {author} {\bibfnamefont {S.}~\bibnamefont {Hauberg}},\ and\ \bibinfo
  {author} {\bibfnamefont {R.}~\bibnamefont {Wehbring}},\ }\href
  {https://www.gnu.org/software/octave/doc/v6.1.0/} {\emph {\bibinfo {title}
  {{GNU Octave} version 6.1.0 manual: a high-level interactive language for
  numerical computations}}} (\bibinfo {year} {2020})\BibitemShut {NoStop}%
\bibitem [{\citenamefont {{Andreev}}(1966)}]{Andreev_1966}%
  \BibitemOpen
  \bibfield  {author} {\bibinfo {author} {\bibfnamefont {A.~F.}\ \bibnamefont
  {{Andreev}}},\ }\bibfield  {title} {\bibinfo {title} {{Electron Spectrum of
  the Intermediate State of Superconductors}},\ }\href@noop {} {\bibfield
  {journal} {\bibinfo  {journal} {Soviet Journal of Experimental and
  Theoretical Physics}\ }\textbf {\bibinfo {volume} {22}},\ \bibinfo {pages}
  {455} (\bibinfo {year} {1966})}\BibitemShut {NoStop}%
\bibitem [{\citenamefont {Kulik}\ and\ \citenamefont
  {Omel'yanchuk}(1977)}]{Kulik_1977}%
  \BibitemOpen
  \bibfield  {author} {\bibinfo {author} {\bibfnamefont {I.~O.}\ \bibnamefont
  {Kulik}}\ and\ \bibinfo {author} {\bibfnamefont {A.~N.}\ \bibnamefont
  {Omel'yanchuk}},\ }\bibfield  {title} {\bibinfo {title} {Properties of
  superconducting microbridges in the pure limit},\ }\bibfield  {journal}
  {\bibinfo  {journal} {Sov. J. Low Temp. Phys. (Engl. Transl.); (United
  States)}\ }\textbf {\bibinfo {volume} {3}},\ \href
  {https://www.osti.gov/biblio/6927703} {} (\bibinfo {year} {1977})\BibitemShut
  {NoStop}%
\bibitem [{\citenamefont {Sauls}(2018)}]{Sauls_2018}%
  \BibitemOpen
  \bibfield  {author} {\bibinfo {author} {\bibfnamefont {J.~A.}\ \bibnamefont
  {Sauls}},\ }\bibfield  {title} {\bibinfo {title} {Andreev bound states and
  their signatures},\ }\href {https://doi.org/10.1098/rsta.2018.0140}
  {\bibfield  {journal} {\bibinfo  {journal} {Philosophical Transactions of the
  Royal Society A: Mathematical, Physical and Engineering Sciences}\ }\textbf
  {\bibinfo {volume} {376}},\ \bibinfo {pages} {20180140} (\bibinfo {year}
  {2018})}\BibitemShut {NoStop}%
\bibitem [{\citenamefont {Josephson}(1962)}]{JOSEPHSON1962}%
  \BibitemOpen
  \bibfield  {author} {\bibinfo {author} {\bibfnamefont {B.}~\bibnamefont
  {Josephson}},\ }\bibfield  {title} {\bibinfo {title} {Possible new effects in
  superconductive tunnelling},\ }\href
  {https://doi.org/https://doi.org/10.1016/0031-9163(62)91369-0} {\bibfield
  {journal} {\bibinfo  {journal} {Physics Letters}\ }\textbf {\bibinfo {volume}
  {1}},\ \bibinfo {pages} {251} (\bibinfo {year} {1962})}\BibitemShut {NoStop}%
\bibitem [{\citenamefont {Josephson}(1974)}]{JOSEPHSON1974}%
  \BibitemOpen
  \bibfield  {author} {\bibinfo {author} {\bibfnamefont {B.~D.}\ \bibnamefont
  {Josephson}},\ }\bibfield  {title} {\bibinfo {title} {The discovery of
  tunnelling supercurrents},\ }\href
  {https://doi.org/10.1103/RevModPhys.46.251} {\bibfield  {journal} {\bibinfo
  {journal} {Rev. Mod. Phys.}\ }\textbf {\bibinfo {volume} {46}},\ \bibinfo
  {pages} {251} (\bibinfo {year} {1974})}\BibitemShut {NoStop}%
\bibitem [{\citenamefont {Titov}\ and\ \citenamefont
  {Beenakker}(2006)}]{Titov:PhysRevB2006}%
  \BibitemOpen
  \bibfield  {author} {\bibinfo {author} {\bibfnamefont {M.}~\bibnamefont
  {Titov}}\ and\ \bibinfo {author} {\bibfnamefont {C.~W.~J.}\ \bibnamefont
  {Beenakker}},\ }\bibfield  {title} {\bibinfo {title} {Josephson effect in
  ballistic graphene},\ }\href {https://doi.org/10.1103/PhysRevB.74.041401}
  {\bibfield  {journal} {\bibinfo  {journal} {Phys. Rev. B}\ }\textbf {\bibinfo
  {volume} {74}},\ \bibinfo {pages} {041401(R)} (\bibinfo {year}
  {2006})}\BibitemShut {NoStop}%
\bibitem [{\citenamefont {Mu\~noz}\ \emph {et~al.}(2012)\citenamefont
  {Mu\~noz}, \citenamefont {Covaci},\ and\ \citenamefont
  {Peeters}}]{Munoz:PhysRevB2012}%
  \BibitemOpen
  \bibfield  {author} {\bibinfo {author} {\bibfnamefont {W.~A.}\ \bibnamefont
  {Mu\~noz}}, \bibinfo {author} {\bibfnamefont {L.}~\bibnamefont {Covaci}},\
  and\ \bibinfo {author} {\bibfnamefont {F.~M.}\ \bibnamefont {Peeters}},\
  }\bibfield  {title} {\bibinfo {title} {Tight-binding study of bilayer
  graphene josephson junctions},\ }\href
  {https://doi.org/10.1103/PhysRevB.86.184505} {\bibfield  {journal} {\bibinfo
  {journal} {Phys. Rev. B}\ }\textbf {\bibinfo {volume} {86}},\ \bibinfo
  {pages} {184505} (\bibinfo {year} {2012})}\BibitemShut {NoStop}%
\bibitem [{\citenamefont {Alidoust}\ \emph {et~al.}(2019)\citenamefont
  {Alidoust}, \citenamefont {Willatzen},\ and\ \citenamefont
  {Jauho}}]{Alidoust:PhysRevB2019}%
  \BibitemOpen
  \bibfield  {author} {\bibinfo {author} {\bibfnamefont {M.}~\bibnamefont
  {Alidoust}}, \bibinfo {author} {\bibfnamefont {M.}~\bibnamefont
  {Willatzen}},\ and\ \bibinfo {author} {\bibfnamefont {A.-P.}\ \bibnamefont
  {Jauho}},\ }\bibfield  {title} {\bibinfo {title} {Symmetry of superconducting
  correlations in displaced bilayers of graphene},\ }\href
  {https://doi.org/10.1103/PhysRevB.99.155413} {\bibfield  {journal} {\bibinfo
  {journal} {Phys. Rev. B}\ }\textbf {\bibinfo {volume} {99}},\ \bibinfo
  {pages} {155413} (\bibinfo {year} {2019})}\BibitemShut {NoStop}%
\bibitem [{\citenamefont {Alidoust}\ \emph {et~al.}(2020)\citenamefont
  {Alidoust}, \citenamefont {Jauho},\ and\ \citenamefont
  {Akola}}]{Alidoust:PhysRevResearch2020}%
  \BibitemOpen
  \bibfield  {author} {\bibinfo {author} {\bibfnamefont {M.}~\bibnamefont
  {Alidoust}}, \bibinfo {author} {\bibfnamefont {A.-P.}\ \bibnamefont
  {Jauho}},\ and\ \bibinfo {author} {\bibfnamefont {J.}~\bibnamefont {Akola}},\
  }\bibfield  {title} {\bibinfo {title} {Josephson effect in graphene bilayers
  with adjustable relative displacement},\ }\href
  {https://doi.org/10.1103/PhysRevResearch.2.032074} {\bibfield  {journal}
  {\bibinfo  {journal} {Phys. Rev. Research}\ }\textbf {\bibinfo {volume}
  {2}},\ \bibinfo {pages} {032074(R)} (\bibinfo {year} {2020})}\BibitemShut
  {NoStop}%
\bibitem [{\citenamefont {Sriram}\ \emph {et~al.}(2019)\citenamefont {Sriram},
  \citenamefont {Kalantre}, \citenamefont {Gharavi}, \citenamefont {Baugh},\
  and\ \citenamefont {Muralidharan}}]{ABS_spectral_density_Muralidharan}%
  \BibitemOpen
  \bibfield  {author} {\bibinfo {author} {\bibfnamefont {P.}~\bibnamefont
  {Sriram}}, \bibinfo {author} {\bibfnamefont {S.~S.}\ \bibnamefont
  {Kalantre}}, \bibinfo {author} {\bibfnamefont {K.}~\bibnamefont {Gharavi}},
  \bibinfo {author} {\bibfnamefont {J.}~\bibnamefont {Baugh}},\ and\ \bibinfo
  {author} {\bibfnamefont {B.}~\bibnamefont {Muralidharan}},\ }\bibfield
  {title} {\bibinfo {title} {{Supercurrent interference in semiconductor
  nanowire Josephson junctions}},\ }\href
  {https://doi.org/10.1103/PhysRevB.100.155431} {\bibfield  {journal} {\bibinfo
   {journal} {Phys. Rev. B}\ }\textbf {\bibinfo {volume} {100}},\ \bibinfo
  {pages} {155431} (\bibinfo {year} {2019})}\BibitemShut {NoStop}%
\bibitem [{\citenamefont {Furusaki}(1994)}]{FURUSAKI1994214}%
  \BibitemOpen
  \bibfield  {author} {\bibinfo {author} {\bibfnamefont {A.}~\bibnamefont
  {Furusaki}},\ }\bibfield  {title} {\bibinfo {title} {{DC Josephson effect in
  dirty SNS junctions: Numerical study}},\ }\href
  {https://doi.org/https://doi.org/10.1016/0921-4526(94)90061-2} {\bibfield
  {journal} {\bibinfo  {journal} {Physica B: Condensed Matter}\ }\textbf
  {\bibinfo {volume} {203}},\ \bibinfo {pages} {214} (\bibinfo {year}
  {1994})}\BibitemShut {NoStop}%
\bibitem [{\citenamefont {Ostroukh}\ \emph {et~al.}(2016)\citenamefont
  {Ostroukh}, \citenamefont {Baxevanis}, \citenamefont {Akhmerov},\ and\
  \citenamefont {Beenakker}}]{Ostroukh_paper}%
  \BibitemOpen
  \bibfield  {author} {\bibinfo {author} {\bibfnamefont {V.~P.}\ \bibnamefont
  {Ostroukh}}, \bibinfo {author} {\bibfnamefont {B.}~\bibnamefont {Baxevanis}},
  \bibinfo {author} {\bibfnamefont {A.~R.}\ \bibnamefont {Akhmerov}},\ and\
  \bibinfo {author} {\bibfnamefont {C.~W.~J.}\ \bibnamefont {Beenakker}},\
  }\bibfield  {title} {\bibinfo {title} {{Two-dimensional Josephson vortex
  lattice and anomalously slow decay of the Fraunhofer oscillations in a
  ballistic SNS junction with a warped Fermi surface}},\ }\href
  {https://doi.org/10.1103/PhysRevB.94.094514} {\bibfield  {journal} {\bibinfo
  {journal} {Phys. Rev. B}\ }\textbf {\bibinfo {volume} {94}},\ \bibinfo
  {pages} {094514} (\bibinfo {year} {2016})}\BibitemShut {NoStop}%
\bibitem [{\citenamefont {Zuo}\ \emph {et~al.}(2017)\citenamefont {Zuo},
  \citenamefont {Mourik}, \citenamefont {Szombati}, \citenamefont {Nijholt},
  \citenamefont {van Woerkom}, \citenamefont {Geresdi}, \citenamefont {Chen},
  \citenamefont {Ostroukh}, \citenamefont {Akhmerov}, \citenamefont {Plissard},
  \citenamefont {Car}, \citenamefont {Bakkers}, \citenamefont {Pikulin},
  \citenamefont {Kouwenhoven},\ and\ \citenamefont
  {Frolov}}]{Akhmerov_supercurrent}%
  \BibitemOpen
  \bibfield  {author} {\bibinfo {author} {\bibfnamefont {K.}~\bibnamefont
  {Zuo}}, \bibinfo {author} {\bibfnamefont {V.}~\bibnamefont {Mourik}},
  \bibinfo {author} {\bibfnamefont {D.~B.}\ \bibnamefont {Szombati}}, \bibinfo
  {author} {\bibfnamefont {B.}~\bibnamefont {Nijholt}}, \bibinfo {author}
  {\bibfnamefont {D.~J.}\ \bibnamefont {van Woerkom}}, \bibinfo {author}
  {\bibfnamefont {A.}~\bibnamefont {Geresdi}}, \bibinfo {author} {\bibfnamefont
  {J.}~\bibnamefont {Chen}}, \bibinfo {author} {\bibfnamefont {V.~P.}\
  \bibnamefont {Ostroukh}}, \bibinfo {author} {\bibfnamefont {A.~R.}\
  \bibnamefont {Akhmerov}}, \bibinfo {author} {\bibfnamefont {S.~R.}\
  \bibnamefont {Plissard}}, \bibinfo {author} {\bibfnamefont {D.}~\bibnamefont
  {Car}}, \bibinfo {author} {\bibfnamefont {E.~P. A.~M.}\ \bibnamefont
  {Bakkers}}, \bibinfo {author} {\bibfnamefont {D.~I.}\ \bibnamefont
  {Pikulin}}, \bibinfo {author} {\bibfnamefont {L.~P.}\ \bibnamefont
  {Kouwenhoven}},\ and\ \bibinfo {author} {\bibfnamefont {S.~M.}\ \bibnamefont
  {Frolov}},\ }\bibfield  {title} {\bibinfo {title} {{Supercurrent Interference
  in Few-Mode Nanowire Josephson Junctions}},\ }\href
  {https://doi.org/10.1103/PhysRevLett.119.187704} {\bibfield  {journal}
  {\bibinfo  {journal} {Phys. Rev. Lett.}\ }\textbf {\bibinfo {volume} {119}},\
  \bibinfo {pages} {187704} (\bibinfo {year} {2017})}\BibitemShut {NoStop}%
\bibitem [{\citenamefont {McClure}(1957)}]{McClure:PR1957}%
  \BibitemOpen
  \bibfield  {author} {\bibinfo {author} {\bibfnamefont {J.~W.}\ \bibnamefont
  {McClure}},\ }\bibfield  {title} {\bibinfo {title} {{Band Structure of
  Graphite and de Haas-van Alphen Effect}},\ }\href
  {https://doi.org/10.1103/PhysRev.108.612} {\bibfield  {journal} {\bibinfo
  {journal} {Phys. Rev.}\ }\textbf {\bibinfo {volume} {108}},\ \bibinfo {pages}
  {612} (\bibinfo {year} {1957})}\BibitemShut {NoStop}%
\bibitem [{\citenamefont {Slonczewski}\ and\ \citenamefont
  {Weiss}(1958)}]{Slonczewski:PR1958}%
  \BibitemOpen
  \bibfield  {author} {\bibinfo {author} {\bibfnamefont {J.~C.}\ \bibnamefont
  {Slonczewski}}\ and\ \bibinfo {author} {\bibfnamefont {P.~R.}\ \bibnamefont
  {Weiss}},\ }\bibfield  {title} {\bibinfo {title} {{Band Structure of
  Graphite}},\ }\href {https://doi.org/10.1103/PhysRev.109.272} {\bibfield
  {journal} {\bibinfo  {journal} {Phys. Rev.}\ }\textbf {\bibinfo {volume}
  {109}},\ \bibinfo {pages} {272} (\bibinfo {year} {1958})}\BibitemShut
  {NoStop}%
\bibitem [{\citenamefont {Konschuh}\ \emph {et~al.}(2012)\citenamefont
  {Konschuh}, \citenamefont {Gmitra}, \citenamefont {Kochan},\ and\
  \citenamefont {Fabian}}]{KonschuhPRB:2012}%
  \BibitemOpen
  \bibfield  {author} {\bibinfo {author} {\bibfnamefont {S.}~\bibnamefont
  {Konschuh}}, \bibinfo {author} {\bibfnamefont {M.}~\bibnamefont {Gmitra}},
  \bibinfo {author} {\bibfnamefont {D.}~\bibnamefont {Kochan}},\ and\ \bibinfo
  {author} {\bibfnamefont {J.}~\bibnamefont {Fabian}},\ }\bibfield  {title}
  {\bibinfo {title} {Theory of spin-orbit coupling in bilayer graphene},\
  }\href {https://doi.org/10.1103/PhysRevB.85.115423} {\bibfield  {journal}
  {\bibinfo  {journal} {Phys. Rev. B}\ }\textbf {\bibinfo {volume} {85}},\
  \bibinfo {pages} {115423} (\bibinfo {year} {2012})}\BibitemShut {NoStop}%
\bibitem [{\citenamefont {McCann}\ and\ \citenamefont
  {Koshino}(2013)}]{McCann_2013}%
  \BibitemOpen
  \bibfield  {author} {\bibinfo {author} {\bibfnamefont {E.}~\bibnamefont
  {McCann}}\ and\ \bibinfo {author} {\bibfnamefont {M.}~\bibnamefont
  {Koshino}},\ }\bibfield  {title} {\bibinfo {title} {The electronic properties
  of bilayer graphene},\ }\href {https://doi.org/10.1088/0034-4885/76/5/056503}
  {\bibfield  {journal} {\bibinfo  {journal} {Reports on Progress in Physics}\
  }\textbf {\bibinfo {volume} {76}},\ \bibinfo {pages} {056503} (\bibinfo
  {year} {2013})}\BibitemShut {NoStop}%
\bibitem [{\citenamefont {Beenakker}(1991)}]{Beenakker_1991}%
  \BibitemOpen
  \bibfield  {author} {\bibinfo {author} {\bibfnamefont {C.~W.~J.}\
  \bibnamefont {Beenakker}},\ }\bibfield  {title} {\bibinfo {title} {{Universal
  limit of critical-current fluctuations in mesoscopic Josephson junctions}},\
  }\href {https://doi.org/10.1103/PhysRevLett.67.3836} {\bibfield  {journal}
  {\bibinfo  {journal} {Phys. Rev. Lett.}\ }\textbf {\bibinfo {volume} {67}},\
  \bibinfo {pages} {3836} (\bibinfo {year} {1991})}\BibitemShut {NoStop}%
\bibitem [{\citenamefont {van Heck}\ \emph {et~al.}(2014)\citenamefont {van
  Heck}, \citenamefont {Mi},\ and\ \citenamefont
  {Akhmerov}}]{Akhmerov_ABS_smatrix}%
  \BibitemOpen
  \bibfield  {author} {\bibinfo {author} {\bibfnamefont {B.}~\bibnamefont {van
  Heck}}, \bibinfo {author} {\bibfnamefont {S.}~\bibnamefont {Mi}},\ and\
  \bibinfo {author} {\bibfnamefont {A.~R.}\ \bibnamefont {Akhmerov}},\
  }\bibfield  {title} {\bibinfo {title} {{Single fermion manipulation via
  superconducting phase differences in multiterminal Josephson junctions}},\
  }\href {https://doi.org/10.1103/PhysRevB.90.155450} {\bibfield  {journal}
  {\bibinfo  {journal} {Phys. Rev. B}\ }\textbf {\bibinfo {volume} {90}},\
  \bibinfo {pages} {155450} (\bibinfo {year} {2014})}\BibitemShut {NoStop}%
\end{thebibliography}%


\end{document}


\title{SUPPLEMENTAL MATERIAL \\[0.1 cm] Spin relaxation, Josephson effect and Yu-Shiba-Rusinov states in superconducting bilayer graphene}

\author{Michael Barth}%
 	\affiliation{Institute for Theoretical Physics, University of Regensburg, 93040 Regensburg, Germany}
 	
\author{Jacob Fuchs}%
 	\affiliation{Institute for Theoretical Physics, University of Regensburg, 93040 Regensburg, Germany}
        
\author{Denis Kochan}%
 	\affiliation{Institute for Theoretical Physics, University of Regensburg, 93040 Regensburg, Germany\\
 	Institute of Physics, Slovak Academy of Sciences, 84511 Bratislava, Slovakia}


\maketitle

The following Supplemental Material provides some technical details as well as additional data for calculations that probe the effects of certain system parameters. 

\section{Numerical implementation}\label{Sec:NumericalImpl}

\subsection{Quasi-particle spin relaxation rates}

For the numerical calculation of the temperature and doping dependencies of spin relaxation in the superconducting BLG we employ the Python package \textsc{Kwant}~\cite{Kwant}. More precisely, we use the same implementation scheme as developed 
and described in details in Ref.~\cite{Kochan:PRL_sc_graphene_spin_relaxation} and its supplemental material. 
The starting point is to bring the model Hamiltonian, Eq.~(\ref{Eq:Hamiltonian}) in the Bogoliubov-de~Gennes form
and place it on the hexagonal lattice. Numerically, this is implemented by constructing two separate, but identical lattices in real space---one for the spin up $\gamma_\uparrow$ and the second for the spin down $\gamma_\downarrow$ quasi-particles. Each spin lattice contains two leads and single scattering region (with a length $L$ and width $W$) consisting of the same BLG carbons and adatoms. The latter are mutually coupled as dictated by the orbital hoppings entering 
$H_{BLG}$ and $V_o$ (the orbital part of the adatom Hamiltonian $H_{ada}$). Such a setup allows us to study spin conserving dynamics. In order to account for spin-flips we need to connect the two spin lattices. This is done by $V_s$---the spin part of $H_{ada}$---that locally inter-couples spin up and down scattering regions. Thus injecting a spin up quasi-particle, say from the left lead, we can look at the probability to get a spin down quasi-particle at the right/left 
lead (spin-flip transmission/reflection), as a function of chemical potential, quasi-particle energy, propagating direction, magnitude of the superconducting 
gap, etc. This can be read out from the elements of the S-matrix---a core functionality  built into \textsc{Kwant}. 

A priori, \textsc{Kwant} simulates quasi-1D transport---from lead to lead. However, by imposing additional ``periodic boundary conditions'' in the transverse 
direction (with respect to the leads and scattering regions) one can also access certain 2D spectral and transport characteristics \cite{Bundesmann_SR_SOC,Kochan:PRL_sc_graphene_spin_relaxation}.
Down-to-earth, visualizing a 1D system as a horizontal strip, one can additionally connect its upper and lower edges by the corresponding nearest and next-nearest neighbour hoppings just multiplied with $\mathrm{e}^{\pm i k_{\mathrm{trans}} W}$, what is effectively restoring a periodicity along the transverse direction. Varying $k_{\mathrm{trans}}W\in [0,2\pi)$ we affect the quantization condition for the transverse modes in the quasi-1D system, and thereby effectively simulate 2D transport.

The spin-relaxation rates follow from the generalized Fermi-golden rule, Eq.~(\ref{Eq:Yafet}), that uses the T-matrix corresponding to $H_{ada}$, see Eq.~(\ref{Eq:T-matrix}), and hence
takes into account multiple and virtually off-shell---by the in-gap states mediated---scattering effects. The way how to get from the S-matrix computed by \textsc{Kwant} to the T-matrix required for the relaxation rates is bridged by the fundamental scattering relation:
\begin{align}
S^\mathit{\mathrm{f}, \mathrm{i}}_{\sigma^\prime,\sigma} &\equiv
\langle \Psi_{\mathrm{f}\sigma^\prime}|S|\Psi_{\mathrm{i}\sigma}\rangle\\
&= \langle \Psi_{\mathrm{f}\sigma^\prime}|\Psi_{\mathrm{i}\sigma}\rangle
- 2\pi i\,\delta(E_{\mathrm{f}\sigma^\prime}-E_{\mathrm{i}\sigma})\,
\langle \Psi_{\mathrm{f}\sigma^\prime}|\mathbb{T}|\Psi_{\mathrm{i}\sigma}\rangle,\nonumber
\end{align}
where i and f label initial and final states and energies. 
Using it one can obtain the temperature dependent spin-relaxation rates of quasi-particles purely in terms of \textsc{Kwant} outputs: 
\begin{equation}\label{Eq:spin-flip-probability7}
\frac{1}{\tau_s^{\mathit{Kw}}}=\frac{\eta_{ada}\frac{4W}{\sqrt{3}a}}{\hbar}
\frac{
	\int\limits_{\Delta}^\infty \mathrm{d}E\,\left(-\frac{\partial g}{\partial E}\right)\,
	\sum\limits_{\mathrm{i,f},\sigma}\bigl|S^\mathrm{f, i}_{-\sigma,\sigma}(E)\bigr|^2
}{
	\int\limits_{\Delta}^\infty \mathrm{d}E\,\left(-\frac{\partial g}{\partial E}\right)\sum\limits_{\mathrm{i},\sigma} \langle\Psi_{\mathrm{i},\sigma}|\Psi_{\mathrm{i},\sigma}\rangle_{\mathit{luc}}\,},
\end{equation}
where as before $\eta_{ada}$ is the adatom concentration in the scattering region per carbon, $g(E,T)$ is the Fermi-Dirac-distribution, and $a$ and $W$ are lattice constant and the system spatial width. 
Summation over $\mathrm{f}$ and $\mathrm{i}$ runs over all outgoing (final) and incomming (initial) leads' modes, $\Psi_{\mathrm{f},\sigma}$ and $\Psi_{\mathrm{i},\sigma}$, at given chemical potential $\mu$, while the sum over $\sigma$ counts the spin projections. In the denominator, $\langle\Psi_{\mathrm{i},\sigma}|\Psi_{\mathrm{i},\sigma}\rangle_{\mathit{luc}}$ 
represents an integral of the incoming mode density $|\Psi_{\mathrm{i},\sigma}|^2$ over the corresponding \textit{lead-unit-cell} that possesses spatial width 
$W$ and length $a$ [this is given by the normalization of the scattering modes as implemented in \textsc{Kwant}]. In practice, the energy integration is substituted by a Riemann summation until a cutoff energy of $25k_{\mathrm{B}}T$, since for low
temperatures (below the superconducting $T_c$) the derivative of $g(E)$ is very narrowly peaked around $E=0$ and falls off rapidly on a distance of a few
$k_{\mathrm{B}}T$'s.
The elementary derivation of Eq.~(\ref{Eq:spin-flip-probability7}) from Eq.~(\ref{Eq:Yafet}) is given in the supplemental material of Ref.~\cite{Kochan:PRL_sc_graphene_spin_relaxation}. It is worth to emphasize that ${1}/{\tau_s^{\mathit{Kw}}}$ is a function of $\mu$ that specifies the number of scattering modes, and $T$ that determines the size of the superconducting gap $\Delta$ in $H_{BLG}$, as well as the Fermi level thermal smearing.

In the following we will compare spin-relaxation rates for two different adatom positions in superconducting BLG---the dimer and non-dimer sites---as those exhibit different resonant behaviour \cite{KatochPRL2018} and different doping dependencies of spin relaxation \cite{Kochan2015_PRL-SR-BLG} already in the normal-phase.
To simulate the single adatom behaviour we place one adatom inside the scattering region. The exact impurity position inside the scattering region is not important due to periodic boundary conditions in the transverse direction.

\subsection{Numerical calculation: Yu-Shiba-Rusinov states}

In order to show consistency between the numerical and the analytical treatments we will compare later in Sec.~\ref{Sec:ResultYSR} the YSR spectra by implementing both approaches. Here we will shortly explain our 
numerical strategy.

In principle one could get access to the eigenenergies numerically by employing the Kernel polynomial method \cite{Mashkoori2017ImpurityBS}, where the local density of states can be computed in the vicinity of the adatom in a finite tight-binding system. 
Looking for the spectral maxima in the density of states while varying the energies inside the superconducting gap, one can find energies corresponding to the bound states. 
The drawback of this method is, that we do not get the corresponding eigenstates. 
The spatial profiles of these are important for understanding the behaviour of the spin relaxation on the adatom positions. 
Therefore we compute
eigenenergies and eigenstates by direct diagonalization of a finite BLG system with a single adatom using the package GNU Octave \cite{Octave}. 
The system size was fixed to a width of $W = 601a$ and a length of $L = 601a$. The other knob that affects computational demands is the size of the pairing gap.
The smaller the value of $\Delta_0$, the larger system size is needed to overcome the intrinsic length scale given by the superconducting correlation length that is inversely proportional to $\Delta_0$.
So, simulating the YSR spectra for small pairing gaps becomes computationally more demanding, and therefore we chose $\Delta_0 = 50\,\mathrm{meV}$ as a meaningful compromise between the computational effort and the physics we can learn from that. Of course, there are not limitations on the size of the gap when using the analytical Green's function method.

\subsection{BLG-based Josephson junctions with resonant magnetic impurities}

The measurements of the spin-relaxation rates in the superconducting phase along with YSR spectroscopy can be experimentally quite challenging. For that reason we also 
study transport signatures of the resonant magnetic impurities chemisorbed on the dimer and non-dimer sites in BLG-based Josephson junctions.
Such systems are experimentally easier to investigate, and as will be shown, they allow us to predict clear differences discriminating between the dimer and non-dimer resonant magnetic impurities.
More precisely, we will focus on the Andreev bound state (ABS) spectra~\cite{Andreev_1966,Kulik_1977,Sauls_2018} and the corresponding critical currents in the BLG-based Josephson junctions~\cite{JOSEPHSON1962,JOSEPHSON1974}. 
Josephson junctions built from single layer and BLG have been already studied theoretically \cite{Titov:PhysRevB2006,Munoz:PhysRevB2012}, and such systems are 
predicted to realize a variety of interesting physical phenomena. For example, studying the supercurrent flow in the junction as a function of the graphene bilayer displacement 
one might get inside into the superconducting pairing mechanisms \cite{Alidoust:PhysRevB2019,Alidoust:PhysRevResearch2020}.

\begin{figure}
  \centering
  \includegraphics[width=0.9\columnwidth]{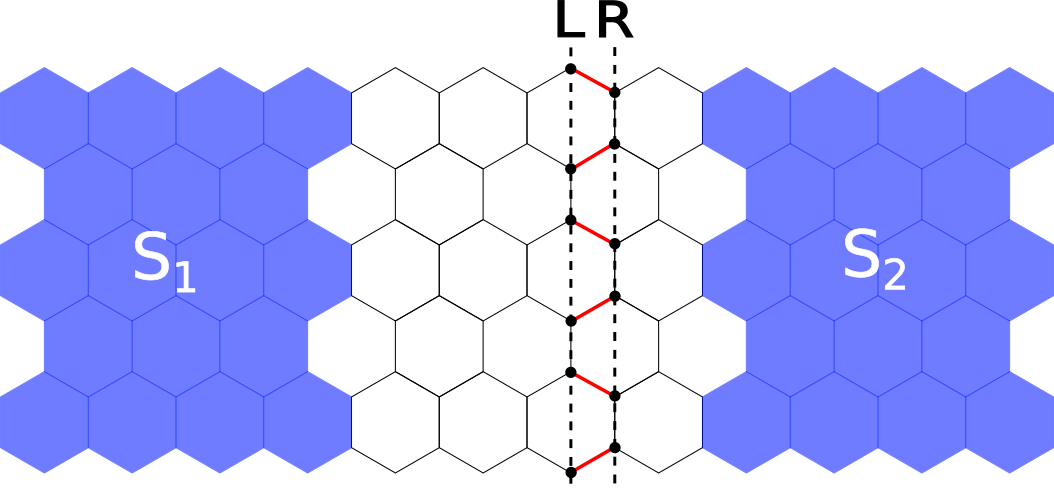}
  \caption{Schematic top view of a BLG-based Josephson junction consisting of two superconducting leads (blue) and the normal region (white) hosting magnetic impurities (sitting in the center, not shown), to show the junction geometry we draw only the top BLG lattice. 
  Sites lying on $L$ and $R$ vertical lines and the red bonds connecting them mark the needed entries of Eq.~(\ref{eq:furusaki}) for the evaluation of the
  Josephson current.}
  \label{fig:BLG_JJ}
\end{figure}

Figure~\ref{fig:BLG_JJ} shows a schematic top view of the BLG-based Josephson junction with two superconducting leads $\mathrm{S}_1$ and $\mathrm{S}_2$ 
possessing a phase difference $\phi = \varphi_1-\varphi_2$ and the normal BLG region hosting resonant magnetic impurities. 

The first task is to know the effect of resonant magnetic impurities on the corresponding ABS spectra---especially at doping levels where the resonant scattering 
off the magnetic impurities dominates. In order to get the corresponding ABS eigenenergies as functions of $\phi$ we calculate the spectral density~\cite{ABS_spectral_density_Muralidharan} for the whole tight-binding system and look for peaks inside of the superconducting gap. 
The spectral density is defined as
\begin{equation}\label{eq:spectral_density}
d(E) = \frac{1}{2\pi} \mathrm{Tr}[A(E)] = \frac{1}{2\pi} \mathrm{Tr}[i(G^r(E)-G^a(E))],
\end{equation}
where $G^r(E) = \left(E\mathbb{I} + i\eta - H  - \Sigma\right)^{-1}$ is the retarded and $G^a(E) = G^r(E)^\dagger$ the advanced Green's function associated with 
the tight-binding Hamiltonian $H=H_{BLG}+H_{ada}$ and the self-energy matrix $\Sigma$---taking into account the presence of left and right superconducting leads.
All these entries can be computed by using the built-in \textsc{Kwant} functionalities. Nevertheless, we want to emphasize a subtle point, a necessity of correct sorting of
the entries of $H$ and $\Sigma$ what we shortly discuss later.

We remark that the YSR states discussed above are conceptually different from the ABS that are inherent to the superconductor-normal-superconductor (SNS) Josephson junctions. Despite both live inside the superconducting gap, the first are bound to impurities in the bulk superconductor, while the later are bound in a 
normal region of the junction and carry a current. 
In this work we are using two different methods when calculating these spectra. In contrast to the computation of the YSR eigenenergies and eigenstates where we used a direct diagonalization, we employ for ABS energies the spectral density method (since we are not interested in the underlying wave-functions etc.). One can get the ABS spectra also by diagonalizing a finite SNS slab with the periodic boundary conditions across the junction. However, this requires an inspection of the convergence with respect to the system length. In practice, to get well-converged results it is necessary to use relatively large systems, what increases computational efforts. 
In turn the spectral density method is not depending on the extension of the superconducting regions since it allows to use semi-infinite leads that are naturally entering into the self-energies. Nevertheless, as far as we were able to examine, both methods give consistent results and can be used for mutual consistency cross-checks.

To calculate the supercurrent of the BLG Josephson junction we employ the Green's function technique devised by Furusaki \cite{FURUSAKI1994214}, which was further technically fine-tuned, e.g.,~for the \textsc{Kwant} implementation in Refs.~\cite{Ostroukh_paper,Akhmerov_supercurrent}. We use the part of the code provided in supplemental repository of Ref.~\cite{Akhmerov_supercurrent}, which we have readopted properly for the BLG systems.
The idea is to cut from the normal part of the Josephson junction two neighbouring regions (we denote them as $L$ and $R$ in Fig.~\ref{fig:BLG_JJ}) and calculate 
a sum of the bond currents flowing between their sites, what is encompassed in the following expression: 
\begin{equation}\label{eq:furusaki}
I_{LR}(\phi)= \frac{ek_{\mathrm{B}}T}{\hbar}\sum_{n=0}^{\infty}
\sum_{\substack{i\in R \\ j\in L}}\mathrm{Im} \left( H_{ji}G^r_{ij}(i\omega_n) - H_{ij}G^r_{ji}(i\omega_n) \right).
\end{equation}
Therein $\omega_n = \frac{k_{\mathrm{B}}T}{\hbar} (2n+1) \pi$ are fermionic Matsubara frequencies, and $H_{ji}$ and $G^r_{ij}$
are, correspondingly, the matrix elements of Hamiltonian $H=H_{BLG}+H_{ada}$ and the associated retarded Green's function $G^r$, taken between the atomic orbitals 
residing at lattice sites $j\in L$ and $i\in R$---this is constructed by \textsc{Kwant} routines once implementing $H$ and the leads' geometry. 
Summation over $j$ and $i$ run only over those $L$ and $R$ sites that are mutually coupled by (nearest neighbour) hoppings forming $H$. For simplicity we also set in these calculations $\gamma_3$ and $\gamma_4$ in $H_{BLG}$ to zero.
By using Eq.~(\ref{eq:furusaki}) one has direct access to the current-phase relation of the Josephson junction, which allows us to compute the critical current 
$I_c = \max_{\phi} |I_{LR}(\phi)|$. 
In the calculations presented below we use the zigzag terminated BLG ribbon with the hard wall boundary conditions that possesses a width of the normal region of $W=40\,a$ and three different lengths, namely $L=10\,a$, $L=20\,a$ and $L=60\,a$. Moreover, we assume the superconducting leads are connected to the normal region by transparent barriers. Leads are formed from the superconducting BLG with the pairing gap $\Delta_0=1$\,meV, and possess the same widths and chemical potentials as the normal part of the junction.

\begin{figure*}
  \centering
  \includegraphics[width=1.9\columnwidth]{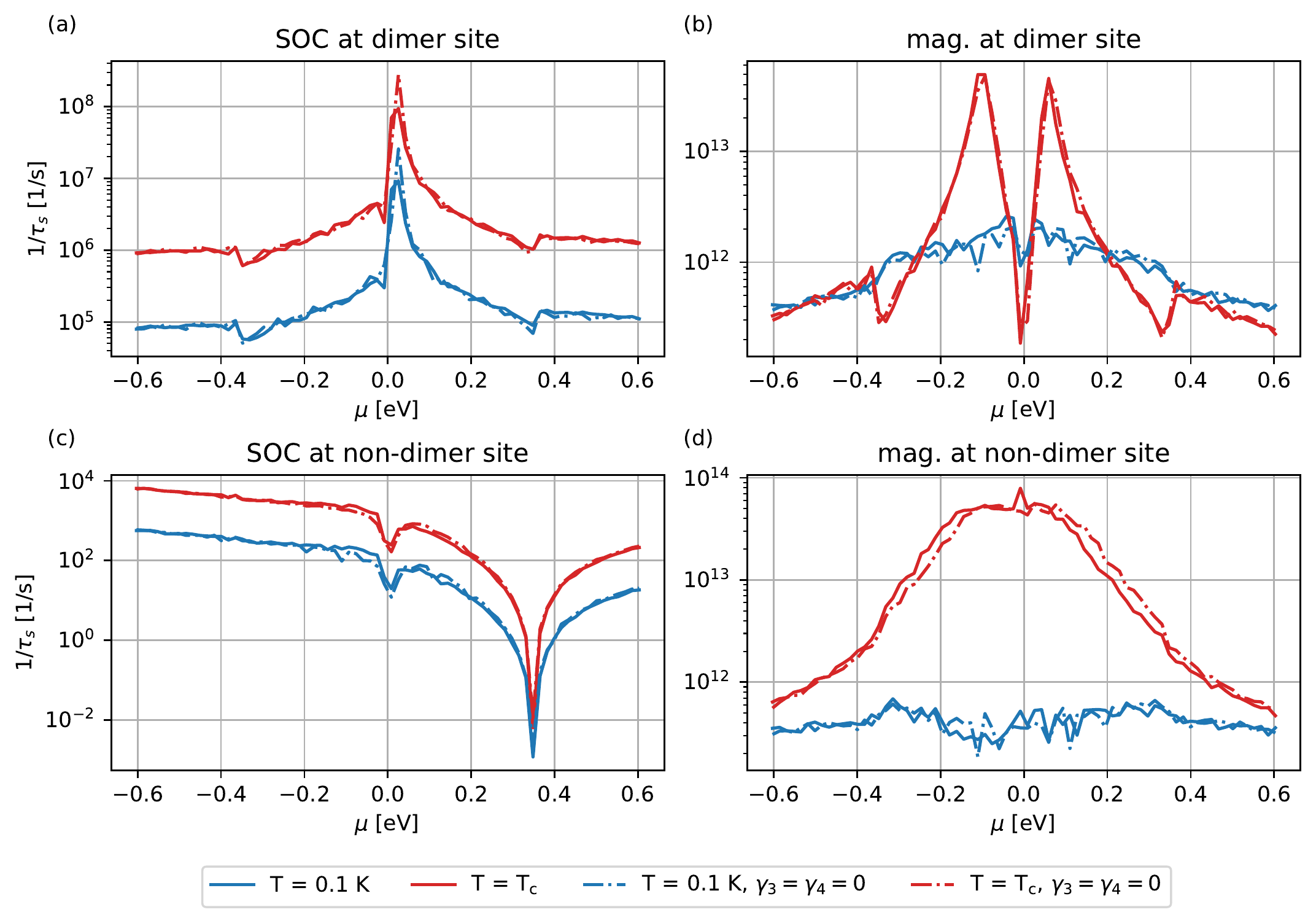}
  \caption{Quasi-particle spin-relaxation rates for different interlayer hopping contributions. The system size was again fixed to $\mathrm{W}=131a$ and $\mathrm{L=4a}$, resulting in~$\eta_{ada}~=~0.0413~\mathrm{\%}$. The phase averaging was also performed for 20 equally spaced values of $k_{\mathrm{trans}}$ in the interval $[0;2\pi]$.}
  \label{fig:Rates_mag_soc_compare_diff_hoppings}
\end{figure*}

\section{Comparison of spin-relaxation rates including $\gamma_3$ and $\gamma_4$ interlayer hoppings}\label{app:Comparison}

To describe the host system one can use the tight-binding Hamiltonian in the conventional McClure-Slonczewski-Weiss parameterization~\cite{McClure:PR1957,Slonczewski:PR1958,KonschuhPRB:2012,McCann_2013}:
\begin{align}\label{EqSM:HamBLG}
    H_{BLG}&=
    -\sum_{m, n, j, \sigma} (\gamma_0 \delta_{\langle mn\rangle}+\mu\delta_{mn})
    c^\dagger_{j,m,\sigma} c^{\phantom{\dagger}}_{j,n,\sigma} \nonumber \\ 
         &+\Delta\sum_{m, j} c^\dagger_{j,m,\uparrow} c^\dagger_{j,m,\downarrow}+\rm{h.c.} \nonumber \\ 
         & + \gamma_1 \sum_{m, \sigma} (c^\dagger_{B1,m,\sigma} c^{\phantom{\dagger}}_{A2,m,\sigma} +\rm{h.c.}) \\
         & + \gamma_3 \sum_{m, \sigma} (c^\dagger_{A1,m,\sigma} c^{\phantom{\dagger}}_{B2,m,\sigma} +\rm{h.c.})  \nonumber \\ 
         & + \gamma_4 \sum_{m, \sigma} (c^\dagger_{A1,m,\sigma} c^{\phantom{\dagger}}_{A2,m,\sigma} + c^\dagger_{B1,m,\sigma} c^{\phantom{\dagger}}_{B2,m,\sigma} + \rm{h.c.}), \nonumber
\end{align}

For simplicity of analytic calculations of YSR energies and to connect them with the doping dependencies of spin-relaxation rates, we used a simplified Hamiltonian $H_{BLG}$ setting $\gamma_3$ and $\gamma_4$ to zero.  
In order to see the influence of $\gamma_3$ and $\gamma_4$ interlayer hoppings on the spin-relaxation rates we compare in Fig.~\ref{fig:Rates_mag_soc_compare_diff_hoppings} the outcomes of the full and simplified model for the temperatures $T=T_c$ and $T=0.1~\mathrm{K}$.
We fixed the parameters to $\gamma_3 = 0.28\,\mathrm{eV}$ and $\gamma_4 = -0.14\,\mathrm{eV}$.
There appear only marginal differences at small values of $\mu$, but overall the spin-relaxation does not seem to be affected strongly by the interlayer hoppings $\gamma_3$ and $\gamma_4$. So we believe that the simplified treatment is well acknowledged.

\subsection{Comparison of ABS spectrum and curent-phase relation}

\begin{figure}
  \centering
  \includegraphics[width=\columnwidth]{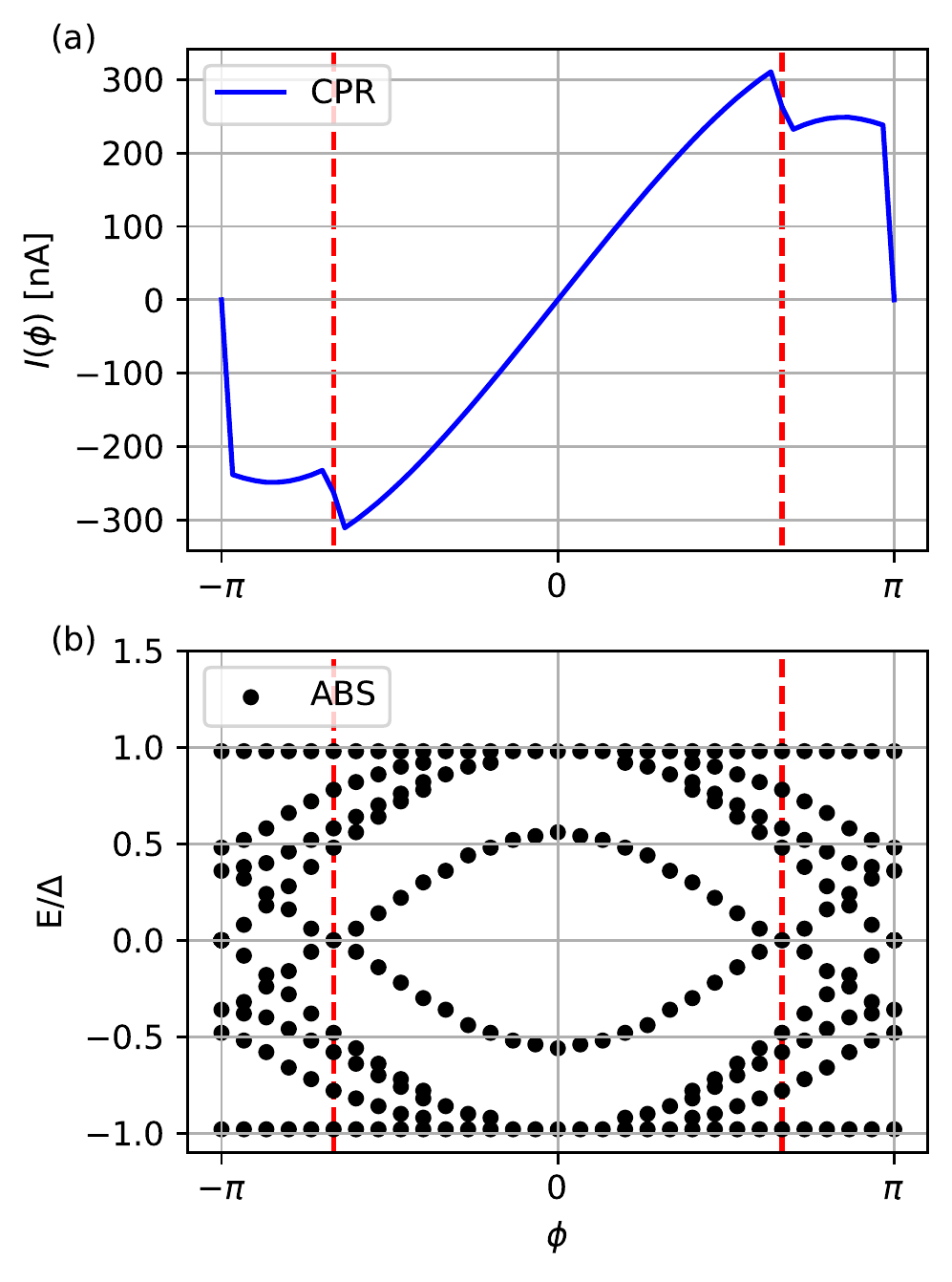}
  \caption{Panel~(a) shows CPR for a BLG-based Josephson junction with a single magnetic impurity (hydrogen) at a non-dimer site for a chemical potential 
  $\mu = -16\,\mathrm{meV}$. Panel~(b) displays corresponding ABS spectrum. The red dashed lines indicate phases $\phi\simeq\pm 2$ at which the lowest ABS branches 
  cross at zero and the current $I(\phi)$ experiences jumps.}
  \label{fig:compare_CPR_ABS}
\end{figure}

Signatures of the ABS bands affected by the magnetic impurities interacting via the exchange interaction reflect in the current-phase relation (CPR).
Figure~\ref{fig:compare_CPR_ABS}~(a) shows an example CPR for a junction with a hydrogen adatom chemisorbed at a non-dimer site at a doping level $\mu=-16\,\mathrm{meV}$ computed according to Eq.~(\ref{eq:furusaki}). At a phase difference of roughly $\phi\simeq\pm 2$ (marked by the dashed red vertical lines) 
the CPR exhibits a kink, where the magnitude of the current drops suddenly. The corresponding ABS spectrum obtained from Eq.~(\ref{eq:spectral_density}) is 
shown in Fig.~\ref{fig:compare_CPR_ABS}~(b). This comparison between two different methods serves as a consistency check. The current is proportional 
to the derivative of the ground-state energy of the ABS spectrum with respect to the phase difference, i.e., 
$I(\phi) \propto \sum_{\mathrm{ABS}}\partial E^{\mathrm{ABS}}(\phi) / \partial \phi$ with $E^{\mathrm{ABS}}(\phi) < 0$, and therefore the jump in $I(\phi)$ should originate from the sudden change in the slope of the ABS bands, i.e.,~a point of the zero energy crossing.

\section{Sorting of tight-binding Hamiltonian: technical info}\label{app:TechnicalAppendix}

For BLG the lattice consists of four sublattices, A1, B1, A2 and B2, and by default, \textsc{Kwant} is returning a tight-binding Hamiltonian $H$ that is sorted according to atomic sites grouped within the same sublattice, i.e.,~firstly, exhausting 
sites belonging to the sublattice A1, then B1 etc. The entries of the self energy matrix $\Sigma$ are in turn ordered differently: the first are entries containing the sites interfacing the L lead, and the last are entries containing the sites interfacing the R lead.

In order to make proper matrix manipulations in the expressions containing both $H$ and $\Sigma$, we have to ensure 
the proper entry orderings. In our code we reshuffled the entries of $H$. \textsc{Kwant} offers an option to change the site ordering in $H$. 

\section{Consistency check: ABS spectra from the S-matrix} \label{app:ABS_smatrix}

\begin{figure}
  \centering
  \includegraphics[width=0.9\linewidth]{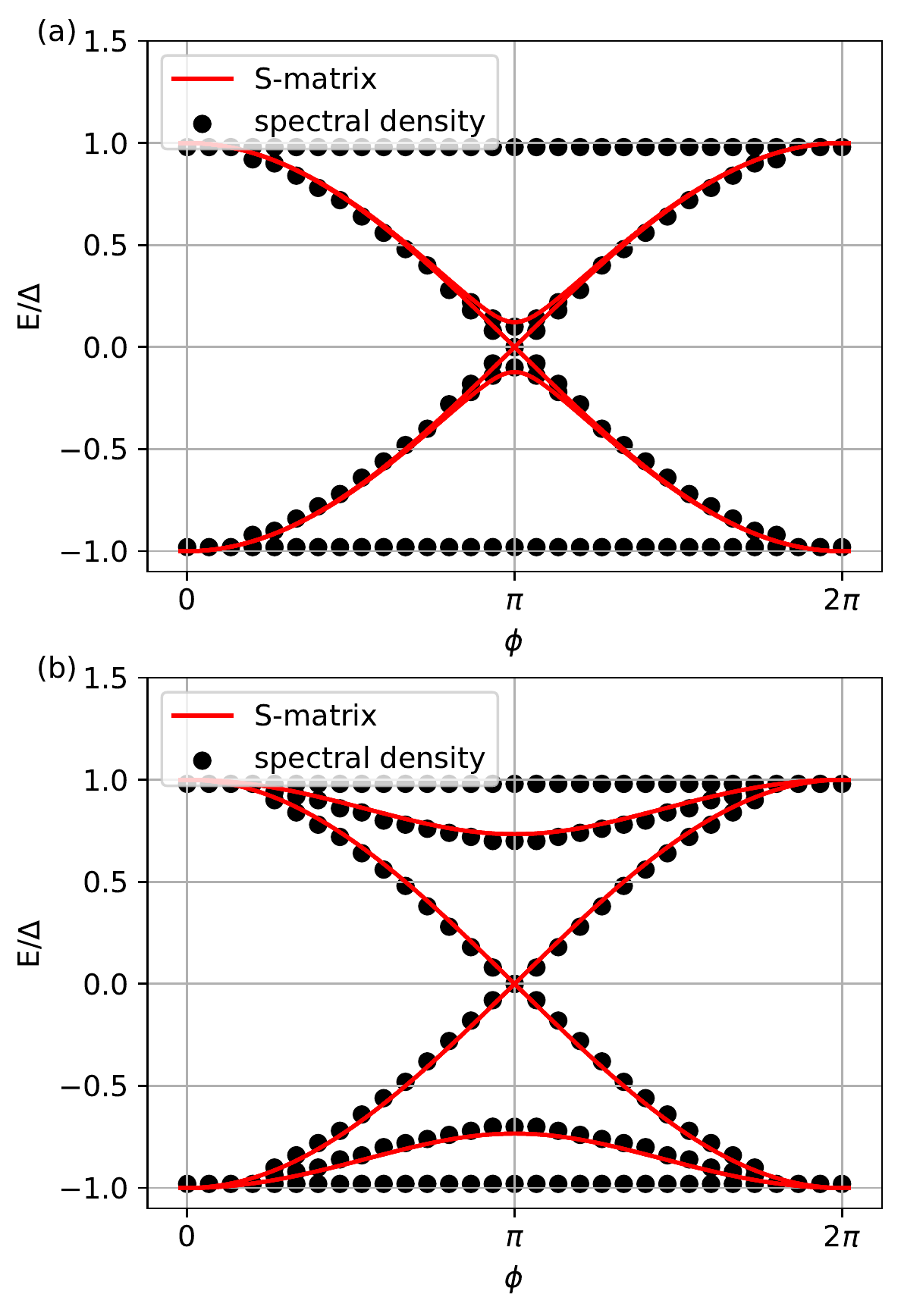}
  \caption{Comparison of the ABS spectra obtained by the spectral density (dots) and the S-matrix (red lines) methods for a short ballistic junction with a length $L=10a$ and width $W=40a$. The magnetic exchange energy $J$ is put to zero. In panel (a) the impurity is occupying a dimer site with $\mu = -100\,\mathrm{meV}$, while in (b) it sits at the non-dimer site and $\mu = 150\,\mathrm{meV}$.}
  \label{fig:compare_smatrix_spectraldensity}
\end{figure}

The ABS spectrum can be in some cases determined by computing the S-matrix of the normal system~\cite{Beenakker_1991}. 
This method is applicable if scattering occurs away from the NS interfaces, i.e.,~inside the normal region.
The BLG Josephson junction with a chemisorbed adatom is therefore a very suitable system for probing this method. 
According to~\cite{Akhmerov_ABS_smatrix} the ABS energies $\epsilon_{ABS}$ inside of the gap can be computed by solving the eigenvalue problem
\begin{equation}
    A\Psi^e_{in} = \frac{|\epsilon_{ABS}|}{\Delta} e^{i\chi}\Psi^e_{in},
\end{equation}
where $A = \left(r_As-s^Tr_A^T\right)/2$ is a normal matrix, meaning $AA^\dagger=A^\dagger A$.
This property of $A$ allows one to compute complex eigenvalues $|\epsilon_{ABS}|\,e^{i\chi}$ by a singular value decomposition.
In the definition of $A$, $s$ stands for the electron scattering matrix in the normal region, and $r_A$ is defined as the diagonal matrix assuming the incoming and outgoing leads' modes are connected by time-reversal symmetry:
\begin{equation}
    r_A =
    \begin{pmatrix}
        i\exp(i\varphi_L) \mathbb{I}_{N_L} & 0 \\
        0 & i\exp(i\varphi_R) \mathbb{I}_{N_R}
    \end{pmatrix},
\end{equation}
where $N_L$ and $N_R$ are the numbers of modes in the left and right leads, and $\varphi_{L}$ and $\varphi_{R}$ are the corresponding superconducting phases.
As already stated in the main text, this method is only applicable to our system if we put magnetic exchange $J=0$.
The short junction regime, which we are considering, leads to quantum interference effects triggered by the superconducting pairing due to the leads and the magnetic exchange due to the 
adatom, these are not included in the S-matrix method.
Only in the long junction limit, when the interference quenches the two techniques give consistent results.
Nevertheless, we are comparing the results of ABS spectra by two methods in Fig.~\ref{fig:compare_smatrix_spectraldensity} for the nonmagnetic impurity at dimer and non-dimer site 
for two representative values of $\mu$.
The ABS spectra obtained by both methods agree perfectly. If $J=0$, the impurity is a trivial scattering center that gaps out the ABS spectrum.
With an increased width $W$, the scattering is further reduced and the gap in the ABS spectrum is decreasing as expected.

\bibliography{library_sm}
